\documentclass[notitlepage,12pt]{article}%
\usepackage{latexsym,epsfig,amsmath}
\usepackage{epsfig,graphicx}
\usepackage{amsfonts}
\usepackage{tabularx}
\usepackage{rotating}
\usepackage{harvard}
\usepackage{hyperref}
\usepackage{amsmath}
\usepackage{amssymb}
\usepackage{graphicx}
\usepackage{setspace}
\usepackage[flushleft,online,para]{threeparttable}
\usepackage{floatrow}
\usepackage{placeins}%
\providecommand{\U}[1]{\protect\rule{.1in}{.1in}}
\renewcommand{\baselinestretch}{1.5}
\begin{document}

\title{Estimation with Aggregate Shocks}
\author{Jinyong Hahn\thanks{UCLA, Department of Economics, 8283 Bunche Hall, Mail
Stop: 147703, Los Angeles, CA 90095, hahn@econ.ucla.edu}\\UCLA
\and Guido Kuersteiner\thanks{University of Maryland, Department of Economics,
Tydings Hall 3145, College Park, MD, 20742, kuersteiner@econ.umd.edu}\\University of Maryland
\and Maurizio Mazzocco\thanks{UCLA, Department of Economics, 8283 Bunche Hall, Mail
Stop: 147703, Los Angeles, CA 90095, mmazzocc@econ.ucla.edu}\\UCLA}
\maketitle

\begin{abstract}
Aggregate shocks affect most households' and firms' decisions. Using three
stylized models we show that inference based on cross-sectional data alone
generally fails to correctly account for decision making of rational agents
facing aggregate uncertainty. We propose an econometric framework that
overcomes these problems by explicitly parameterizing the agents' inference
problem relative to aggregate shocks. Our framework and examples illustrate
that the cross-sectional and time-series aspects of the model are often
interdependent. Therefore, estimation of model parameters in the presence of
aggregate shocks requires the combined use of cross-sectional and time series
data. We provide easy-to-use formulas for test statistics and confidence
intervals that account for the interaction between the cross-sectional and
time-series variation. Lastly, we perform Monte Carlo simulations that
highlight the properties of the proposed method and the risks of not properly
accounting for the presence of aggregate shocks.

\end{abstract}

\newpage

\section{Introduction}

An extensive body of economic research suggests that aggregate shocks have
important effects on households' and firms' decisions. Consider for instance
the oil shock that hit developed countries in 1973. A large literature has
provided evidence that this aggregate shock triggered a recession in the
United States, where the demand and supply of non-durable and durable goods
declined, inflation grew, the unemployment rate increased, and real wages dropped.

The profession has generally adopted one of the following three strategies to
deal with aggregate shocks. The most common strategy is to assume that
aggregate shocks have no effect on households' and firms' decisions, and hence
that aggregate shocks can be ignored. Almost all papers estimating discrete
choice dynamic models or dynamic games are based on this premise. Examples
include Keane and Wolpin (1997), Bajari, Bankard, and Levin (2007), and
Eckstein and Lifshitz (2011). The second approach is to add time dummies to
the model in an attempt to capture the effect of aggregate shocks on the
estimation of the parameters of interest, as was done for instance in Runkle
(1991) and Shea (1995). The last strategy is to fully specify how aggregate
shocks affect individual decisions jointly with the rest of the structure of
the economic problem. We are aware of only one paper that uses this strategy,
Lee and Wolpin (2010).

The previous discussion reveals that there is no generally agreed upon
econometric framework for estimation and statistical inference in models where
aggregate shocks have an effect on individual decisions. This paper makes two
main contributions related to this deficiency. We first provide a general
econometric framework that can be used to evaluate the effect of aggregate
shocks on estimation and statistical inference and apply it to three examples.
The examples reveal which issues may arise if aggregate shocks are a feature
of the data, but the researcher does not properly account for them. The
examples also provide important insights on which econometric method can be
employed in the estimation of model parameters when aggregate shocks are
present. Using those insights, we propose a method based on a combination of
cross-sectional variables and a long time-series of aggregate variables. There are no available formulas that can be used for statistical inference when those two data
sources are combined. The second contribution of this paper is to provide
simple-to-use formulas for test statistics and confidence intervals that can
be employed when our proposed method is used.

We proceed in four steps. In Section \ref{section-model}, we introduce the
generic identification problem by examining a general class of models with the
following two features. First, each model in this class is composed of two
submodels. The first submodel includes all the cross-sectional features,
whereas the second submodel is composed of all the time-series aspects. As a
consequence, the parameters of the model can also be divided into two groups:
the parameters that characterize the cross-sectional submodel and the
parameters that enter the time-series submodel. The second feature is that the
two submodels are linked by a vector of aggregates shocks and by the
parameters that govern their dynamics. Individual decision making thus depends
on aggregate shocks.

Given the interplay between the two submodels, aggregate shocks have
complicated effects on the estimation of the parameters of interest. To better
understand those effects, in the second step, we present three examples of the
general framework that illustrate the complexities generated by the existence
of the aggregate shocks.

In Section \ref{portfolio-example}, we consider as a first example a simple
model of portfolio choice with aggregate shocks. The simplicity of the model
enables us to clearly illustrate the effect of aggregates shocks on the
estimation of model parameters and on their asymptotic distribution. Using the
example, we first show that, if the econometrician does not account for
uncertainty generated by aggregate shocks, the estimates of model parameters
are biased and inconsistent. Our results also illustrate that the inclusion of
time dummies generally does not correctly account for the existence of
aggregate shocks.\footnote{In the Euler equation context, Chamberlain (1984)
considers a special example characterized by a nonstationary aggregate
environment and time-varying nonstochastic preference shocks. Under this
special environment, he shows that, when aggregate shocks are present but
disregarded, the estimated parameters can be inconsistent even when time
dummies are included. In this paper, we show that the presence of aggregate
shocks produces inconsistent estimates if those shocks are ignores, even when
time dummies are employed, in very general and realistic contexts and not only
in the very special case adopted by Chamberlain (1984).} We then provide some
insight on the sign of the bias. When aggregate uncertainty is ignored, agents
in the estimated model appear more risk averse than they are. This is a way
for the misspecified model to account for the uncertainty in the data that is
not properly modeled. As a consequence, the main parameter in the portfolio
model, the coefficient of risk aversion, is biased upward. Lastly, we show
that a method based on a combination of cross-sectional and time-series
variables produces unbiased and consistent estimates of the model parameters.

In Section \ref{Estimation of Production Functions}, as a second example, we
study the estimation of firms' production functions when aggregate shocks
affect firms' decisions. This example shows that there are exceptional cases
where model parameters can be consistently estimated using only repeated
cross-sections if time dummies are skillfully used rather than simply added as
time intercepts. Specifically, our analysis indicates that the method proposed
by Olley and Pakes (1996) fails to produce consistent estimates if aggregate
shocks are present. It also indicates that the production functions can be
consistently estimated if their method is modified with the proper inclusion
of time dummies. The results of Section
\ref{Estimation of Production Functions} are of independent interest since
aggregate shocks have significant effects in most markets and the estimation
of firms' production functions is an important topic in industrial
organization, see for instance Levinsohn and Petrin (2003) and Ackerberg,
Caves, and Frazer (2015).

In Section \ref{GE-Model} we present as our last example a general equilibrium
model of education and labor supply decisions. The portfolio example has the
quality of being simple. But, because of its simplicity, it generates a
one-directional relationship between the time-series and cross-sectional
submodels: the parameters of the cross-sectional model can be consistently
estimated only if the parameters of the time-series model are known, but the
time-series parameters can be consistently estimated without knowledge of
cross-sectional parameters. However, this is not generally the case. In many situations, the link between the two submodels is bi-directional.
The advantage of the general-equilibrium example is that it produces a
bi-directional relationship we can use to illustrate the complexity of the
effect of aggregate shocks on the estimation of the model parameters and on
their asymptotic distribution. The general equilibrium example also
illustrates how our method based on cross-sectional and time-series variables
can be used to generate consistent estimates when the link between the two
sub-models is bi-directional.

The examples make clear that in general consistent estimation of parameters in
models with aggregate shocks is not feasible with only cross-sectional or time
series data. They also clarify that a method based on the combination of
cross-sectional variables and a long time-series of aggregate variables
generates consistent estimates. Since there is no existing formula for the
computation of the standard errors when those two data sources are
combined\textbf{,} as the third step, in Section \ref{Standard Errors} we
provide easy-to-use algorithms that can be employed to obtain test statistics
and confidence intervals for parameters estimated using the proposed method.
The underlying asymptotic theory, which is presented in the companion paper
Hahn, Kuersteiner, and Mazzocco (2016), is highly technical due to the
complicated interactions that exists between the two submodels. It is
therefore surprising that the formulas necessary to perform inference take
simple forms that are easy to adopt. We conclude the section by illustrating,
using the portfolio choice model and the general equilibrium model, how the
formulas can be computed in specific cases.

Finally, to evaluate our econometric framework, we perform a Monte Carlo
experiment for the general equilibrium model. The Monte Carlo results indicate
that our method performs well when the length of the time-series is
sufficiently large. In that case, the parameter estimates are statistically
close to the true values and the coverage probabilities are statistically
close to the nominal levels. To document biases that may arise from ignoring
aggregate shocks and using only cross-sectional variation, we also estimate
the model's parameters under the incorrect assumption that the economy is not
affected by aggregate shocks. Our results show that this form of
misspecification can generate extremely large biases for the parameters that
require both cross-sectional and longitudinal variation to be consistently
estimated. For instance, we find that a parameter that is of considerable
interest to economists, the coefficient of risk aversion, is between five and
six times larger than the true value if aggregate shocks are ignored. This
result is consistent with the intuition provided by the portfolio choice
model. If aggregate shocks are ignored by the econometrician, agents in the
model are estimated to be more risk averse than they are to account for the
high degree of uncertainty present in the data.

In addition to the econometric literature that deals with inferential issues,
our paper also contributes to a growing literature whose objective is the
estimation of general equilibrium models. Some examples of papers in this
literature are Heckman and Sedlacek (1985), Heckman, Lochner, and Taber
(1998), Lee (2005), Lee and Wolpin (2006), Gemici and Wiswall (2011),
Gillingham, Iskhakov, Munk-Nielsen, Rust, and Schjerning (2015). Aggregate
shocks are a natural feature of general equilibrium models. Without them those
models have the unpleasant implication that all aggregate variables can be
fully explained by observables and, hence, that errors have no effects on
those variables. Our general econometric framework makes this point clear by
highlighting the impact of aggregate shocks on parameter estimation and the
variation required in the data to estimate those models. More importantly, our
results provide easy-to-use formulas that can be employed to perform
statistical inference in a general equilibrium context.

A separate discussion is required for the paper by Lee and Wolpin (2006). That
paper is the only one that estimates a model that fully specifies how
aggregate shocks affect individual decisions. Using that approach, the authors
can obtain consistent estimates of the parameters of interest. Their paper is
primarily focused on the estimation of a specific empirical model. They do not
address the broader question of which statistical assumptions and what type of
data requirements are needed more generally to obtain consistent estimators
when aggregate shocks are present, which is the focus of this paper. Moreover,
as we argue later on, in Lee and Wolpin's (2010) paper there are issues with
statistical inference and efficiency.

\section{The General Identification Problem\label{section-model}}

This section introduces the identification problem generated by the existence
of aggregate shocks in general terms. We consider a class of models with three
main features. First, the model can be divided into two parts. The first part
encompasses all the aspects of the model that can be analyzed using
cross-sectional variables and will be denoted with the term cross-sectional
submodel. The second part includes aspects whose examination requires
time-series variables and will be denoted with the term time-series submodel.
Second, the two submodels are linked by the presence of a vector of aggregate
shocks $\nu_{t}$ and by the parameters that govern their dynamics. The vector
of aggregate shocks may not be observed. If that is the case, it is treated as
a set of parameters to be estimated. Lastly, the parameters of the model can
be consistently estimated only if a combination of cross-sectional and
time-series data are available, which is the case for many interesting models
with aggregate shocks.

We now formally introduce the general model. It consists of two distinct
vectors of variables $y_{i,t}$ and $z_{s}$. The first vector $y_{i,t}$
includes all the variables that characterize the cross-sectional submodel,
where $i$ describes an individual decision-maker, a household or a firm, and
$t$ a time period in the cross-section.\footnote{Even if the time subscript
$t$ is not necessary in this subsection, we keep it here for notational
consistency because later we consider the case where longitudinal data are
collected.} The second vector $z_{s}$ is composed of all the variables
associated with the time-series model. Accordingly, the parameters of the
general model can be divided into two sets, $\beta$ and $\rho$. The first set
of parameters $\beta$ characterizes the cross-sectional submodel, in the sense
that, if the second set $\rho$ was known, $\beta$ and $\nu_{t}$ can be
consistently estimated using exclusively variation in the cross-sectional
variables $y_{i,t}$. Similarly, the vector $\rho$ characterizes the
time-series submodel meaning that, if $\beta$ were known, those parameters can
be consistently estimated using exclusively the time series variables $z_{s}$.
There are two functions that relate the cross-sectional and time-series
variables to the parameters. The function $f\left(  \left.  y_{i,t}\right\vert
\beta,\nu_{t},\rho\right)  $ restricts the behavior of the cross-sectional
variables conditional on a particular value of the parameters. Analogously,
the function $g\left(  \left.  z_{s}\right\vert \beta,\rho\right)  $ describes
the behavior of the time-series variables for a given value of the parameters.
An example is a situation in which (i) the variables $y_{i,t}$ for
$i=1,\ldots,n$ are i.i.d. given the aggregate shock $\nu_{t}$, (ii) the
variables $z_{s}$ correspond to $\left(  \nu_{s},\nu_{s-1}\right)  $, (iii)
the cross-sectional function $f\left(  \left.  y_{i,t}\right\vert \beta
,\nu_{t},\rho\right)  $ denotes the log likelihood of $y_{i,t}$ given the
aggregate shock $\nu_{t}$, and (iv) the time-series function $g\left(  \left.
z_{s}\right\vert \beta,\rho\right)  =g\left(  \left.  \nu_{s}\right\vert
\nu_{s-1},\rho\right)  $ is the log of the conditional probability density
function of the aggregate shock $\nu_{s}$ given $\nu_{s-1}$. In this special
case the time-series function $g$ does not depend on the cross-sectional
parameters $\beta$.

We assume that our cross-sectional data consist of $\left\{  y_{i,t}%
,\;i=1,\ldots,n\right\}  $, and our time series data consist of $\left\{
z_{s},\;s=\tau_{0}+1,\ldots,\tau_{0}+\tau\right\}  $. For simplicity, we
assume that $\tau_{0}=0$ in this section.

The parameters of the general model can be estimated by maximizing a
well-specified objective function. Since in our case the general framework is
composed of two submodels, a natural approach is to estimate the parameters of
interest by maximizing two separate objective functions, one for the
cross-sectional model and one for the time-series model. We denote these
criterion functions by $F_{n}\left(  \beta,\nu_{t},\rho\right)  $ and
$G_{\tau}\left(  \beta,\rho\right)  $. In the case of maximum likelihood these
functions are simply $F_{n}\left(  \beta,\nu_{t},\rho\right)  =\frac{1}{n}%
\sum_{i=1}^{n}f\left(  \left.  y_{i,t}\right\vert \beta,\nu_{t},\rho\right)  $
and $G_{\tau}\left(  \beta,\rho\right)  =\frac{1}{\tau}\sum_{s=1}^{\tau
}g\left(  \left.  z_{s}\right\vert \beta,\rho\right)  $. Another scenario
where separate criterion functions arise naturally is when $f$ and $g$
represent moment conditions. The use of two separate objective functions is
helpful in our context because it enables us to discuss which issues arise if
only cross-sectional variables or only time-series variables are used in the
estimation. Moreover, considering the two components separately adds
flexibility since data are not required for all variables in the same period.

In this paper, we consider the class of models for which identification of the
parameters requires the joint use of cross-sectional and time-series data.
Specifically, for any fixed and feasible value of $\rho$ the maximum of the
objective function $F$ over the parameters $\beta$ and the aggregate shocks
$\nu$ remains unchanged and independent of the value of $\rho$. The parameter
$\rho$ can therefore not be identified using only cross-sectional variation.
Similarly, the objective function $G$ of the time-series model evaluated at
the time-series parameters and aggregate shocks takes the same value for any
feasible set of cross-sectional parameters. Consequently, the parameters
$\beta$ cannot be identified using only time-series variation. In our class of
models, however, all the parameters of interest can be consistently estimated
if cross-sectional data are combined with time-series data.

In the next three sections, we use three examples to illustrate the effects of
aggregate shocks on the estimation of model parameters and the method we
propose to address the issues generated by the presence of those
shocks.\footnote{Our models assume rational expectations. We do not consider
examples that incorporate model uncertainty, i.e., the possibility that agents
need to learn or estimate model parameters when making decisions. We restrict
our attention to rational expectation models because there is only a limited
number of papers that consider self-confirming equilibria or robust control.
See Cho, Sargent, and Williams (2002) or Hansen, Sargent, and Tallarini
(1999). This is, however, an important topic that we leave for future
research.}

\section{Example 1: Portfolio Choice\label{portfolio-example}}

We start with a simple portfolio choice example that clearly illustrates the
perils of ignoring aggregate shocks. Using this example, we make the following
points. First, the presence of aggregate shocks generally produces estimates
that are biased and inconsistent unless the econometrician properly accounts
for the uncertainty generated by the aggregate shocks. Second, the use of time
dummies generally does not solve the problems generated by the existence of
aggregate shocks. Third, if the researcher does not account for the aggregate
shocks, the parameter estimates will adjust to make the model consistent with
the aggregate uncertainty that is present in the data but not modeled, hence
the bias. For instance, in a model with risk averse agents such as our
portfolio example, ignoring the aggregate shocks produces estimates of the
risk aversion parameter that are upward biased.

Consider an economy that, in each period $t$, is populated by $n$ households.
These households are born at the beginning of period $t$, live for one period,
and are replaced in the next period by $n$ new families. The households living
in consecutive periods do not overlap and, hence, make independent decisions.
Each household is endowed with deterministic income and has preferences over a
non-durable consumption good $c_{i,t}$. The preferences can be represented by
Constant Absolute Risk Aversion (CARA) utility functions which take the
following form: $U\left(  c_{i,t}\right)  =-e^{-\delta c_{i,t}}$. For
simplicity, we normalize income to be equal to 1.

During the period in which households are alive, they can invest a share of
their income in a risky asset with return $u_{i,t}$. The remaining share is
automatically invested in a risk-free asset with a return $r$ that does not
change over time. At the end of the period, the return on the investment is
realized and households consume the quantity of the non-durable good they can
purchase with their realized income. The return on the risky asset depends on
aggregate shocks. Specifically, it takes the following form: $u_{i,t}=\nu
_{t}+\epsilon_{i,t}$, where $\nu_{t}$ is the aggregate shock and
$\epsilon_{i,t}$ is an i.i.d. idiosyncratic shock. The idiosyncratic shock,
and hence the heterogeneity in the return on the risky asset, can be
interpreted as differences across households in transaction costs, in
information on the profitability of different stocks, or in marginal tax
rates. We assume that $\nu_{t}\sim N\left(  \mu,\sigma_{\nu}^{2}\right)  $,
$\epsilon_{i,t}\sim N\left(  0,\sigma_{\epsilon}^{2}\right)  $, and hence that
$u_{i,t}\sim N\left(  \mu,\sigma^{2}\right)  $, where $\sigma^{2}=\sigma_{\nu
}^{2}+\sigma_{\epsilon}^{2}$.

Household $i$ living in period $t$ chooses the fraction of income to be
allocated to the risk-free asset $\alpha_{i,t}$ by maximizing its life-time
expected utility:
\begin{gather}
\max_{\alpha_{i,t}}\ E\left[  -e^{-\delta c_{i,t}}\right] \nonumber\\
s.t.\quad c_{i,t}=\alpha_{i,t}\left(  1+r\right)  +\left(  1-\alpha
_{i,t}\right)  \left(  1+u_{i,t}\right)  , \label{portfolio problem}%
\end{gather}
where the expectation is taken with respect to the return on the risky asset.
It can be shown\footnote{This is shown in the Appendix, which is available
upon request.} that the household's optimal choice of $\alpha_{i,t}$ is given
by
\begin{equation}
\alpha_{i,t}^{\ast}=\alpha=\dfrac{\delta\sigma^{2}+r-\mu}{\delta\sigma^{2}}.
\label{delta-port-choice}%
\end{equation}
We will assume that the econometrician is mainly interested in estimating the
risk aversion parameter $\delta$.

We now consider an estimator that takes the form of a population analog of
(\ref{delta-port-choice}), and study the impact of aggregate shocks on the
estimator's consistency when an econometrician works only with cross-sectional
data. Our analysis reveals that such an estimator is inconsistent because
cross-sectional data do not contain information about aggregate uncertainty.
It also makes explicit the dependence of the estimator on the probability
distribution of the aggregate shock and thus points to the following method
for consistently estimating $\delta$. First, using time series variation, the
parameters pertaining to aggregate uncertainty are consistently estimated.
Second, those estimates are plugged into the cross-sectional model to estimate
the remaining parameters.\footnote{Our model is a stylized version of many
models considered in a large literature interested in estimating the parameter
$\delta$ using cross-sectional variation. Estimators are often based on moment
conditions derived from first order conditions (FOC) related to optimal
investment and consumption decisions. Such estimators have similar problems,
which we discuss in Appendix \ref{Euler-appendix}. The appendix is available
upon request.}

Without loss of generality, we assume that the cross-sectional data are
observed in period $t=1$. Econometricians observe data on the return of the
risky asset $u_{i,t}$ and on the return of the risk-free asset $r$. We assume
that they also observe a noisy measure of the share of resources invested in
the risk-free asset $\alpha_{i,t}=\alpha+e_{i,t}$, where $e_{i,t}$ is a
measurement error with zero mean and variance $\sigma_{e}^{2}$. The vector of
cross-sectional variables $y_{i}$ is therefore composed of $u_{i1}$ and
$\alpha_{i1}$ and the vector of cross-sectional parameters $\beta$ is composed
of $\delta$, $\sigma_{\epsilon}^{2}$, and $\sigma_{e}^{2}$. The vector of
time-series variables includes only the aggregate shock, i.e. $z_{t}=\nu_{t}$,
and the vector of time-series variables parameters is composed of $\mu$ and
$\sigma_{\nu}^{2}$. Since, $\nu_{t}$ corresponds to the aggregate return of
the risky asset, we assume that $\nu_{t}$ is observed.

Consider an econometrician who ignores the existence of the aggregate shocks,
by assuming that the aggregate return is fixed at $\mu$ for all $t$, and uses
only cross-sectional variation. Recall that $\mu=E\left[  u_{i1}\right]  $,
$\sigma^{2}=\operatorname{Var}\left(  u_{i1}\right)  $, and $\alpha=E\left[
\alpha_{i1}\right]  $. That econometrician will therefore estimate those
parameters using the following method-of-moments estimators:
\[
\hat{\mu}=\frac{1}{n}\sum_{i=1}^{n}u_{i1}=\bar{u},\qquad\hat{\sigma}^{2}%
=\frac{1}{n}\sum_{i=1}^{n}\left(  u_{i1}-\bar{u}\right)  ^{2},\qquad
\text{and}\qquad\hat{\alpha}=\frac{1}{n}\sum_{i=1}^{n}\alpha_{i1}.
\]
Econometricians can then use equation (\ref{delta-port-choice}) to write the
risk aversion parameter as $\delta=\left.  \left(  \mu-r\right)  \right/
\left(  \sigma^{2}\left(  1-\alpha\right)  \right)  $ and estimate it with the
sample analog $\hat{\delta}=\left.  \left(  \hat{\mu}-r\right)  \right/
\left(  \hat{\sigma}^{2}\left(  1-\hat{\alpha}\right)  \right)  $.

In the presence of the aggregate shocks $\nu_{t}$, however, the
method-of-moments estimators take the following form:
\begin{align*}
\hat{\mu}  &  =\frac{1}{n}\sum_{i=1}^{n}u_{i1}=\nu_{1}+\frac{1}{n}\sum
_{i=1}^{n}\epsilon_{i1}=\nu_{1}+o_{p}\left(  1\right)  ,\\
\hat{\sigma}^{2}  &  =\frac{1}{n}\sum_{i=1}^{n}\left(  u_{i1}-\bar{u}\right)
^{2}=\frac{1}{n}\sum_{i=1}^{n}\left(  \epsilon_{i1}-\bar{\epsilon}\right)
^{2}=\sigma_{\epsilon}^{2}+o_{p}\left(  1\right)  ,\\
\hat{\alpha}  &  =\alpha+\frac{1}{n}\sum_{i=1}^{n}e_{i1}=\alpha+o_{p}\left(
1\right)  ,
\end{align*}
which implies that $\delta$ will be estimated to be
\begin{equation}
\hat{\delta}=\frac{\nu_{1}+o_{p}\left(  1\right)  -r}{\left(  \sigma
_{\epsilon}^{2}+o_{p}\left(  1\right)  \right)  \left(  1-\alpha+o_{p}\left(
1\right)  \right)  }=\frac{\nu_{1}-r}{\sigma_{\epsilon}^{2}\left(
1-\alpha\right)  }+o_{p}\left(  1\right)  . \label{biased-estimator}%
\end{equation}

Using Equation (\ref{biased-estimator}), we can study the properties of
estimator $\hat{\delta}$. Without aggregate shocks, we would have $\nu_{1}%
=\mu$, $\sigma_{\nu}^{2}=0$, $\sigma_{\epsilon}^{2}=\sigma^{2}$ and,
therefore, $\hat{\delta}$ would converge to $\delta$, a nonstochastic
constant, as $n$ grows to infinity. It is therefore a consistent estimator of
the risk aversion parameter. However, in the presence of the aggregate shock,
the proposed estimator has different properties. We consider first the case in
which econometricians condition on the realization of the aggregate shock
$\nu$ or, equivalently, assumes that the realization of the aggregate shock is
known. In this case, the estimator $\hat{\delta}$ is inconsistent with
probability 1, since it converges to $\frac{\nu_{1}-r}{\sigma_{\epsilon}%
^{2}\left(  1-\alpha\right)  }$ and not to the true value $\frac{\mu
-r}{\left(  \sigma_{\nu}^{2}+\sigma_{\epsilon}^{2}\right)  \left(
1-\alpha\right)  }$.

As discussed in the introduction, a common practice to account for the effect
of aggregate shocks is to include time dummies in the model. The portfolio
example clarifies that the addition of time dummies does not solve the problem
generated by the presence of aggregate shocks. The inclusion of time dummies
is equivalent to the assumption that the realization of the aggregate shock is
known or that econometricians condition on the realization of $\nu$. But the
previous result indicates that, using exclusively cross-sectional data, the
estimator $\hat{\delta}$ is biased even if the realizations of the aggregate
shocks are known. To provide the intuition behind this result, note that, if
aggregate shocks affect individual behavior, the decisions recorded in the
data account for the uncertainty generated by the variation in $\nu$. Even if
econometricians assume that the realizations of the aggregate shocks are
known, the only way the portfolio model can rationalize the degree of
uncertainty displayed by the data is by making the agents more risk averse
than they actually are. Hence, the bias and inconsistency described above.

We now consider the case in which econometricians do not condition on the
realization of the aggregate shock. As $n$ grows to infinity, $\hat{\delta}$
converges to a random variable with a mean that is different from the true
value of the risk aversion parameter. The estimator will therefore be biased
and inconsistent. To see this, remember that $\nu_{1}\sim N\left(  \mu
,\sigma_{\nu}^{2}\right)  $. As a consequence, the unconditional asymptotic
distribution of $\hat{\delta}$ takes the following form:
\[
\hat{\delta}\rightarrow N\left(  \frac{\mu-r}{\sigma_{\epsilon}^{2}\left(
1-\alpha\right)  },\left(  \frac{1}{\sigma_{\epsilon}^{2}\left(
1-\alpha\right)  }\right)  ^{2}\sigma_{\nu}^{2}\right)  =N\left(
\delta+\delta\dfrac{\sigma_{\nu}^{2}}{\sigma_{\epsilon}^{2}},\frac{\sigma
_{\nu}^{2}}{\left(  \sigma_{\epsilon}^{2}\left(  \alpha-1\right)  \right)
^{2}}\right)  ,
\]
which is centered at $\delta+\delta\dfrac{\sigma_{\nu}^{2}}{\sigma_{\epsilon
}^{2}}$ and not at $\delta$, hence the \textquotedblleft
bias\textquotedblright. The intuition behind the bias is the same as for the
case in which the realization of the aggregate shock is known. But when
econometricians do not condition on $\nu$, it is straightforward to sign the
bias. The bias is equal to $\delta\dfrac{\sigma_{\nu}^{2}}{\sigma_{\epsilon
}^{2}}$ and always positive, which is consistent with the intuition described
above according to which ignoring aggregate shocks generates estimates of the
risk aversion parameter that are too high. The formula of the bias also
enables one to reach the intuitive conclusion that its size increases when the
magnitude of the aggregate uncertainty ($\sigma_{\nu}^{2}$) is large relative
to the magnitude of the micro-level uncertainty ($\sigma_{\epsilon}^{2}$)
\footnote{When the realization of $\nu$ is assumed to be known, one can only
sign the expected bias, where the expectation is taken over the realization of
the aggregate shock, since the bias depends on the actual realization of the
shock. The expected bias is always positive and increasing in $\sigma_{\nu
}^{2}$ as our intuition indicates.}

We are not the first to consider a case in which the estimator converges to a
random variable. Andrews (2005) and more recently Kuersteiner and Prucha
(2013) discuss similar scenarios. Our example is remarkable because the nature
of the asymptotic randomness is such that the estimator is not even
asymptotically unbiased. This is not the case in Andrews (2005) or Kuersteiner
and Prucha (2013), where in spite of the asymptotic randomness the estimator
is unbiased.\footnote{Kuersteiner and Prucha (2013) also consider cases where
the estimator is random and inconsistent. However, in their case this happens
for different reasons: the endogeneity of the factors. The inconsistency
considered here occurs even when the factors are strictly exogenous.}

As mentioned above, there is a simple statistical explanation for our result:
cross-sectional variation is not sufficient for the consistent estimation of
the risk aversion parameter if aggregate shocks affect individual decisions.
To make this point transparent, observe that, conditional on the aggregate
shock, the assumptions of this section imply that the cross-sectional variable
$y_{i}=\left(  u_{i1},\alpha_{i1}\right)  $ have the following distribution
\begin{equation}
\left.  y_{i}\right\vert \nu_{1}\sim N\left(  \left[
\begin{array}
[c]{c}%
\nu_{1}\\
\dfrac{\delta\left(  \sigma_{\nu}^{2}+\sigma_{\epsilon}^{2}\right)  +r-\mu
}{\delta\left(  \sigma_{\nu}^{2}+\sigma_{\epsilon}^{2}\right)  }%
\end{array}
\right]  ,\left[
\begin{array}
[c]{cc}%
\sigma_{\epsilon}^{2} & 0\\
0 & \sigma_{e}^{2}%
\end{array}
\right]  \right)  , \label{distribution of y}%
\end{equation}
Using (\ref{distribution of y}), it is straightforward to see that any
arbitrary choice of the time-series parameters $\rho=\left(  \mu,\sigma_{\nu
}^{2}\right)  $ maximize the cross-sectional likelihood, as long as one
chooses $\delta$ that satisfies the following equation:
\[
\dfrac{\delta\left(  \sigma_{\nu}^{2}+\sigma_{\epsilon}^{2}\right)  +r-\mu
}{\delta\left(  \sigma_{\nu}^{2}+\sigma_{\epsilon}^{2}\right)  }=\alpha.
\]
Consequently, the cross-sectional parameters $\mu$ and $\sigma_{\nu}^{2}$
cannot be consistently estimated by maximizing the cross-sectional likelihood
and, hence, $\delta$ cannot be consistently estimated using only
cross-sectional data.

We can now describe the method we propose in this paper as a general solution
to the issues introduced by the presence of aggregate shocks. The method,
which generates unbiased estimates of the model parameters, relies on the
combined use of cross-sectional and time-series variables. Specifically, under
the assumption that the realizations of the aggregate shocks are observed, the
researcher can consistently estimate the parameters that characterize the
distribution of those shocks $\mu$ and $\sigma_{\nu}^{2}$ using a time-series
of aggregate data $\left\{  z_{t}\right\}  $.\footnote{The assumption that the
realizations of aggregate shocks are observed is made to simplify the
discussion and can be easily relaxed. In Section \ref{GE-Model}, we apply the
proposed estimation method to a general equilibrium example in which the
realizations of the aggregate shocks are not observed.} The risk aversion
parameter $\delta$ and the remaining two parameters $\sigma_{\epsilon}^{2}$
and $\sigma_{e}^{2}$ can then be consistently estimated using cross-sectional
variables, by replacing the consistent estimators of $\mu$ and $\sigma_{\nu
}^{2}$ in the correctly specified cross-section likelihood derived in equation
(\ref{distribution of y}).

The example presented in this section is a simplified version of the general
class of models introduced in Section \ref{section-model}. The variables and
parameters of the time-series submodel affect the cross-sectional submodel,
but the cross-sectional variables and parameters have no impact on the
time-series submodel. As a consequence, the time-series parameters can be
consistently estimated without knowing the cross-sectional parameters. The
recursive feature of the example is due to the exogenously specified price
process and the partial equilibrium nature of the model. In more complicated
situations, such as general equilibrium models, where aggregate shocks are a
natural feature, the relationship between the two submodels is generally
bi-directional. But before considering an example of the general case, we
study a situation in which the effect of aggregate shocks can be accounted for
with the proper use of time dummies.

\section{Example 2: Estimation of Production
Functions\label{Estimation of Production Functions}}

In the previous section, we presented an example that illustrates the
complicated nature of identification in the presence of aggregate shocks. The
example highlights that generally there is no simple method for estimating the
class of models considered in this paper. Estimation requires a careful
examination of the interplay between the cross-sectional and time-series
submodels. In this section, we consider an example showing that there are
exceptions to this general rule. In the case we analyze, the researcher is
interested in only a subset of the parameters, and its identification can be
achieved using only cross-sectional data even if aggregate shocks affect
individual decisions, provided that time dummies are skillfully employed. We
will show that the naive practice of introducing additive time dummies is not
sufficient to deal with the effects generated by aggregate shocks. But the
solution is simpler than the general approach we adopted to identify the
parameters of the portfolio model.

The example we consider here is a simplified version of the problem studied by
Olley and Pakes (1996) and deals with an important topic in industrial
organization: the estimation of firms' production functions. A
profit-maximizing firm $j$ produces a product $Y_{j,t}$ in period $t,$
employing a production function that depends on the logarithm of labor
$l_{j,t}$, the logarithm of capital $k_{j,t}$, and a productivity shock
$\omega_{j,t}$. By denoting the logarithm of $Y_{j,t}$ by $y_{j,t}$, the
production function takes the following form:
\begin{equation}
y_{j,t}=\beta_{0}+\beta_{l}l_{j,t}+\beta_{k}k_{j,t}+\omega_{j,t}+\eta_{j,t},
\label{production function}%
\end{equation}
where $\eta_{i,t}$ is a measurement error. The firm chooses the amount of
labor to use in production and the new investment in capital $i_{j,t}$ by
maximizing a dynamic profit function subject to the constraints that in each
period capital accumulates according to the following equation:\footnote{For
details of the profit function, see Olley and Pakes (1996).}
\[
k_{j,t+1}=\left(  1-\delta\right)  k_{j,t}+i_{j,t},
\]
where $\delta$ is the rate at which capital depreciates. In the model proposed
by Olley and Pakes (1996), firms are heterogeneous in their age and can choose
to exit the market. In this section, we will abstract from age heterogeneity
and exit decisions because they make the model more complicated without adding
more insight on the effect of aggregate shocks on the estimation of production functions.

A crucial feature of the model proposed by Olley and Pakes (1996) and of our
example is that the optimal investment decision in period $t$ is a function of
the current stock of capital and of the productivity shock, i.e.
\begin{equation}
i_{j,t}=i_{t}\left(  \omega_{j,t},k_{j,t}\right)  .
\label{investment decision}%
\end{equation}
Olley and Pakes (1996) do not allow for aggregate shocks, but in this example
we consider a situation in which the productivity shock at $t$ is the sum of
an aggregate shock $\nu_{t}$ drawn from a distribution $F\left(  \nu\left|
\rho\right.  \right)  $ and of an i.i.d. idiosyncratic shock $\varepsilon
_{j,t}$, i.e.
\begin{equation}
\omega_{j,t}=\nu_{t}+\varepsilon_{j,t}.
\end{equation}
One example of aggregate shock affecting the productivity of a firm is the
arrival of technological innovations in the economy. We will assume that the
firm observes the realization of the aggregate shock and, separately, of the
i.i.d. shock.

We first review the estimation method proposed by Olley and Pakes (1996) for
the production function (\ref{production function}) when aggregate shocks are
not present. We then discuss how that method has to be modified with the
appropriate use of time dummies if aggregate shocks affect firms' decisions.

The main problem in the estimation of the production function
(\ref{production function}) is that the productivity shock is correlated with
labor and capital, but not observed by the econometrician. To deal with that
issue, Olley and Pakes (1996) use the result that the investment decision
(\ref{investment decision}) is strictly increasing in the productivity shock
for every value of capital to invert the corresponding function, solve for the
productivity shock, and obtain
\begin{equation}
\omega_{j,t}=h_{t}\left(  i_{j,t},k_{j,t}\right)  .
\label{productivity shock production function}%
\end{equation}
One can then replace the productivity shock in the production function using
equation (\ref{productivity shock production function}) to obtain
\begin{equation}
y_{j,t}=\beta_{l}l_{j,t}+\phi_{t}\left(  i_{j,t},k_{j,t}\right)  +\eta_{j,t},
\label{production function no productity shock}%
\end{equation}
where
\begin{equation}
\phi_{t}\left(  i_{j,t},k_{j,t}\right)  =\beta_{0}+\beta_{k}k_{j,t}%
+h_{t}\left(  i_{j,t},k_{j,t}\right)  . \label{definition of phi}%
\end{equation}
The parameter $\beta_{l}$ and the function $\phi_{t}$ can then be estimated by
regressing, period by period, $y_{j,t}$ on $l_{j,t}$ and a flexible polynomial
(i.e., a nonparametric approximation) in $i_{j,t}$ and $k_{j,t}$ or,
similarly, by interacting time dummies with the polynomial in $i_{j,t}$ and
$k_{j,t}$.\footnote{Given our simplifying assumptions that there are no exit
decisions and age heterogeneity, without aggregate shocks, the function $\phi$
is independent of time. We use the more general notation that allows for time
dependence to highlight where the estimation approach developed in Olley and
Pakes (1996) fails when aggregate shocks are present.} The parameter
$\beta_{l}$ is therefore identified by
\begin{equation}
\beta_{l}=\frac{E\left[  \left(  l_{j,t}-E\left[  \left.  l_{j,t}\right\vert
i_{j,t},k_{j,t}\right]  \right)  \left(  y_{j,t}-E\left[  \left.
y_{j,t}\right\vert i_{j,t},k_{j,t}\right]  \right)  \right]  }{E\left[
\left(  l_{j,t}-E\left[  \left.  l_{j,t}\right\vert i_{j,t},k_{j,t}\right]
\right)  ^{2}\right]  }. \label{betal}%
\end{equation}
To identify the parameter on the logarithm of capital $\beta_{k}$ observe that
the production function (\ref{production function}) implies the following:
\begin{equation}
E\left[  \left.  y_{i,t+1}-\beta_{l}l_{j,t+1}\right\vert k_{j,t+1}\right]
=\beta_{0}+\beta_{k}k_{j,t+1}+E\left[  \left.  \omega_{j,t+1}\right\vert
\omega_{j,t}\right]  =\beta_{0}+\beta_{k}k_{j,t+1}+g\left(  \omega
_{j,t}\right)  , \label{expectation second step}%
\end{equation}
where the first equality follows from $k_{j,t+1}$ being determined conditional
on $\omega_{j,t}$. Note that, in the absence of aggregate shocks, the function
$g\left(  .\right)  $ is independent of time. The shock $\omega_{j,t}%
=h_{t}\left(  i_{j,t},k_{j,t}\right)  $ is not observed, but using equations
(\ref{productivity shock production function}) and (\ref{definition of phi}),
it can be written in the following form:
\begin{equation}
\omega_{j,t}=\phi_{t}\left(  i_{j,t},k_{j,t}\right)  -\beta_{0}-\beta
_{k}k_{j,t}, \label{omega equation}%
\end{equation}
where $\phi_{t}$ is known from the first-step estimation. Substituting for
$\omega_{j,t}$ into the function $g\left(  .\right)  $ in equation
(\ref{expectation second step}) and letting $\xi_{j,t+1}=\omega_{j,t+1}%
-E\left[  \left.  \omega_{j,t+1}\right\vert \omega_{j,t}\right]  $, equation
(\ref{expectation second step}) can be written as follows:
\begin{equation}
y_{i,t+1}-\beta_{l}l_{j,t+1}=\beta_{k}k_{j,t+1}+g\left(  \phi_{t}-\beta
_{k}k_{j,t}\right)  +\xi_{j,t+1}+\eta_{j,t}. \label{equation for second step}%
\end{equation}
where $\beta_{0}$ has been included in the function $g\left(  \cdot\right)  $.
The parameter $\beta_{k}$ can then be estimated by using the estimates of
$\beta_{l}$ and $\phi_{t}$ obtained in the first step and by minimizing the
sum of squared residuals in the previous equation, employing a kernel or a
series estimator for the function $g$.

We now consider the case in which aggregate shocks affect the firm's decisions
and analyze how the model parameters can be identified using only
cross-sectional variation. The introduction of aggregate shocks changes the
estimation method in two main ways. First, the investment decision is affected
by the aggregate shock and takes the following form:
\[
i_{j,t}=i_{t}\left(  \nu_{t},\varepsilon_{j,t},k_{j,t}\right)  .
\]
where $\nu_{t}$ and $\varepsilon_{j,t}$ enter as independent arguments because
the firm observes them separately. Second, all expectations are conditional on
the realization of the aggregate shock since in the cross-section there is no
variation in that shock and only its realization is relevant.

If the investment function is strictly increasing in the productivity shock
$\omega_{j,t}$ for all capital levels, it is also strictly increasing in
$\nu_{t}$ and $\varepsilon_{j,t}$ for all $k_{j,t}$, because $\omega_{j,t}%
=\nu_{t}+\varepsilon_{j,t}$. Using this result, we can invert $i_{t}\left(
\cdot\right)  $ to derive $\varepsilon_{j,t}$ as a function of the aggregate
shock, investment, and the stock of capital, i.e.
\[
\varepsilon_{j,t}=h_{t}\left(  \nu_{t},i_{j,t},k_{j,t}\right)  .
\]

The production function can therefore be rewritten in the following form:
\begin{align}
y_{j,t}  &  =\beta_{0}+\beta_{l}l_{j,t}+\beta_{k}k_{j,t}+\nu_{t}%
+\varepsilon_{j,t}+\eta_{j,t}\label{production function with aggregate shocks}%
\\
&  =\beta_{l}l_{j,t}+\left[  \beta_{0}+\beta_{k}k_{j,t}+\nu_{t}+h_{t}\left(
\nu_{t},i_{j,t},k_{j,t}\right)  \right]  +\eta_{j,t}\nonumber\\
&  =\beta_{l}l_{j,t}+\bar{\phi}_{t}\left(  \nu_{t},i_{j,t},k_{j,t}\right)
+\eta_{j,t}\nonumber\\
&  =\beta_{l}l_{j,t}+\phi_{t}\left(  i_{j,t},k_{j,t}\right)  +\eta
_{j,t}.\nonumber
\end{align}
where we have included the aggregate shock in the function $\phi_{t}$.
Analogously to the case of no aggregate shocks, $\beta_{l}$ can be
consistently estimated by regressing period by period $y_{j,t}$ on $l_{j,t}$
and a polynomial in $i_{j,t}$ and $k_{j,t}$ or, similarly, by interacting the
polynomial with time dummies.

Note that the estimation of $\beta_{l}$ is not affected by the uncertainty
generated by the aggregate shocks since that uncertainty is captured by the
time subscript in the function $\phi_{t}$ and the method developed by Olley
and Pakes (1996) already requires the estimation of a different function
$\phi$ for each period. The parameter $\beta_{l}$ is therefore identified by
\begin{equation}
\beta_{l}=\frac{E\left[  \left(  l_{j,t}-E\left[  \left.  l_{j,t}\right\vert
i_{j,t},k_{j,t},\nu_{t}\right]  \right)  \left(  y_{j,t}-E\left[  \left.
y_{j,t}\right\vert i_{j,t},k_{j,t},\nu_{t}\right]  \right)  \right]
}{E\left[  \left(  l_{j,t}-E\left[  \left.  l_{j,t}\right\vert i_{j,t}%
,k_{j,t},\nu_{t}\right]  \right)  ^{2}\right]  }. \label{betal-alt}%
\end{equation}
Observe that the expectation operator in the previous equation is in principle
defined with respect to a probability distribution function that includes the
randomness of the aggregate shock $\nu_{t}$. But, when one uses
cross-sectional variation, $\nu_{t}$ is fixed at its realized value. As a
consequence, the distribution is only affected by the randomness of
$\varepsilon_{it}$.

For the estimation of $\beta_{k}$, note that, under the assumption that the
$\nu_{t}$'s are independent of the $\varepsilon_{j,t}$'s,
\begin{align}
&  E\left[  \left.  y_{i,t+1}-\beta_{l}l_{j,t+1}\right\vert k_{j,t+1}%
,i_{j,t},k_{j,t},\nu_{t+1},\nu_{t},\varepsilon_{j,t}\right]
\label{second step expectation as}\\
&  =\beta_{0}+\beta_{k}k_{j,t+1}+E\left[  \left.  \nu_{t+1}+\varepsilon
_{j,t+1}\right\vert k_{j,t},\nu_{t+1},\nu_{t},\varepsilon_{j,t}\right]
\nonumber\\
&  =\beta_{0}+\beta_{k}k_{j,t+1}+\nu_{t+1}+E\left[  \left.  \varepsilon
_{j,t+1}\right\vert \varepsilon_{j,t}\right] \nonumber\\
&  =\beta_{0}+\beta_{k}k_{j,t+1}+\nu_{t+1}+g\left(  \varepsilon_{j,t}\right)
\nonumber
\end{align}
where the first equality follows from $k_{j,t+1}$ being known if $i_{j,t}$,
$k_{j,t}$, $\nu_{t}$, and $\varepsilon_{j,t}$ are known.

The only variable of equation (\ref{second step expectation as}) that is not
observed is $\varepsilon_{j,t}$. But remember that
\[
\varepsilon_{j,t}=h_{t}\left(  \nu_{t},i_{j,t},k_{j,t}\right)  =\phi
_{t}\left(  \nu_{t},i_{j,t},k_{j,t}\right)  -\beta_{0}-\beta_{k}k_{j,t}%
-\nu_{t}.
\]
We can therefore use the above expression to substitute for $\varepsilon
_{j,t}$ in equation (\ref{second step expectation as}) and obtain
\begin{align*}
&  E\left[  \left.  y_{i,t+1}-\beta_{l}l_{j,t+1}\right\vert k_{j,t+1}%
,i_{j,t},k_{j,t},\nu_{t+1},\nu_{t}\right] \\
&  =\beta_{0}+\beta_{k}k_{j,t+1}+\nu_{t+1}+g_{t}\left(  \phi_{t}\left(
\nu_{t},i_{j,t},k_{j,t}\right)  -\beta_{0}-\beta_{k}k_{j,t}-\nu_{t}\right) \\
&  =\beta_{k}k_{j,t+1}+g_{t,t+1}\left(  \phi_{t}-\beta_{k}k_{j,t}\right)  ,
\end{align*}
where in the last equality $\beta_{0}$, $\nu_{t}$, and $\nu_{t+1}$ have been
included in the function $g_{t,t+1}\left(  \cdot\right)  $. Hence, if one
defines $\xi_{j,t+1}=\varepsilon_{j,t+1}-E\left[  \left.  \varepsilon
_{j,t+1}\right\vert \nu_{t},\varepsilon_{j,t}\right]  $, the previous equation
can be written in the following form:
\begin{equation}
y_{i,t+1}-\beta_{l}l_{j,t+1}=\beta_{k}k_{j,t+1}+g_{t,t+1}\left(  \phi
_{t}-\beta_{k}k_{j,t}\right)  +\xi_{j,t+1}+\eta_{j,t+1}.
\label{second step regresion with as}%
\end{equation}
The inclusion of the aggregate shocks in the function $g\left(  \cdot\right)
$ implies that that function varies with time when aggregate shocks are
present. This is in contrast with the case considered in Olley and Pakes
(1996) where aggregate shocks are ignored and, hence, the function $g\left(
\cdot\right)  $ is independent of time.

Given equation (\ref{second step regresion with as}), if one attempts to
estimate $\beta_{k}$ using equation (\ref{equation for second step}), repeated
cross-sections and the method developed for the case with no aggregate shocks,
the estimated coefficient will generally be biased because the econometrician
does not account for the aggregate shocks and their correlation with the
firm's choice of capital. There is, however, a small variation of the method
proposed earlier that produces unbiased estimates of $\beta_{k}$, as long as
$\varepsilon_{j,t}$ is independent of $\eta_{j,t}$. The econometrician should
regress \textit{period by period} $y_{j,t}$ on $l_{j,t}$ and a nonparametric
function of $i_{j,t}$ and $k_{j,t}$ or, in practice, on a flexible polynomial
of $i_{j,t}$ and $k_{j,t}$ interacted with time dummies. It is this atypical
use of time dummies that enables the econometrician to account for the effect
of aggregate shocks on firms' decisions.

We conclude by drawing attention to two features of the production function
example that make it possible to use time dummies to deal with the effect of
the aggregate shocks. To do that, it is useful to cast the example in terms of
the cross-sectional and time-series submodels. The cross-sectional submodel
includes the variables $y_{j}$, $l_{j}$, $k_{j}$, and $i_{j}$, the parameters
$\beta_{0}$, $\beta_{l}$, and $\beta_{k}$, and the non-parametric functions
$\phi_{t}$ and $g_{t,t+1}$. The time-series submodel includes the aggregate
shocks $\nu_{t}$ and the parameters $\rho$ that define their distribution
function. The decomposition in the two submodels highlights two features of
the example. First, the time-series submodel affects the cross-sectional
counterpart only through the functions $\phi_{t}$ and $g_{t,t+1}$. Second, to
consistently estimate the production function parameters $\beta_{l}$ and
$\beta_{k}$, the functions $\phi_{t}$ and $g_{t,t+1}$ must be known to control
for the correlation between labor and capital on one side and the productivity
shocks on the other. But it is irrelevant how the aggregate shocks and the
corresponding parameters enter those functions. These two features imply that,
if the econometrician is only interested in estimating the production function
parameters $\beta_{l}$ and $\beta_{k}$, he can achieve this by simply
estimating the cross-sectional submodel. This is possible as long as the
functions $\phi_{t}$ and $g_{t,t+1}$ are allowed to vary in a non-parametric
way over time to deal with the existence of the aggregate shocks. The cleaver
use of time-dummies, therefore, solves all the issues raised by their
presence. However, if the econometrician is interested in estimating the
entire model, which includes the parameters that describe the distribution of
the aggregate shocks, he has to rely on the general approach based on the
combination of cross-sectional and time-series variables.

\section{Example 3: A General Equilibrium Model\label{GE-Model}}

In this section, we consider as a third example a general equilibrium model of
education and labor supply decisions in which aggregate shocks influence
individual choices. This example provides additional insights into the effects
of aggregate shocks on the estimation of model parameters. Differently from
the portfolio and production function examples, it considers a case in which
the relationship between the cross-sectional and time-series models is
bi-directional: the cross-sectional parameters cannot be identified from
cross-sectional data without knowledge of the time-series parameters and the
time-series parameters cannot be identified from time series data without
knowing the cross-sectional parameters. In addition, it confirms the results
obtained in the portfolio example. Disregarding the uncertainty generated by
the aggregate shocks can produce large biases in parameters that are important
for economists and policy makers. We show theoretically that ignoring the
presence of aggregate shocks generally produces estimates of the risk aversion
parameter that are severely biased. Monte Carlo simulations confirm biases in
that parameter as large as five, six times the size of the true parameter.

In principle, we could have used as a general example a model proposed in the
general equilibrium literature such as the model developed in Lee and Wolpin
(2006). We decided against this alternative because in those models the effect
of the aggregate shocks on the estimation of the model parameters and the
relationship between the cross-sectional and time-series submodels are
complicated and therefore difficult to describe. Instead, we decided to
develop a model that is sufficiently general to generate an interesting
relationship between the shocks and the estimation of the parameters of
interest and between the two submodels, but at the same time sufficiently
stylized for these relationships to be easy to describe and understand. 

In the model we develop, aggregate shocks affect the education decisions of
young individuals and their subsequent labor supply decisions when of
working-age. Specifically, we consider an economy in which in each period
$t\in T$ a young and a working-age generation overlap. Each generation is
composed of a continuum of individuals with measure $N_{t}$.\footnote{In the
rest of the Section we use interchangeably the word 'measure' and the more
intuitive but less precise word 'number' to refer to $N_{t}$ or similar
objects.} Each individual is endowed with preferences over a non-durable
consumption good and leisure. The preferences of individual $i$ are
represented by a Cobb-Douglas utility function $U^{i}\left(  c,l\right)
=\left.  \left(  c^{\sigma}l^{1-\sigma}\right)  ^{1-\gamma_{i}}\right/
\left(  1-\gamma_{i}\right)  $, where the risk aversion parameter $\gamma_{i}$
is a function of the observable variables $x_{i,t}$, the unobservable
variables $\xi_{i,t}$, and a vector of parameters $\mu$, i.e. $\gamma
_{i}=\gamma\left(  \left.  x_{i,t},\xi_{i,t}\right\vert \mu\right)  $. Future
utilities are discounted using a discount factor $\delta$.

Both young and working-age individuals are endowed with a number of hours
$\mathcal{T}$ that can be allocated to leisure or to a productive activity.
Young individuals are also endowed with an exogenous income $y_{i,t}$. In each
period, the economy is hit by an aggregate shock $\nu_{t}$ whose conditional
probability $P\left(  \left.  \nu_{t+1}\right\vert \nu_{t}\right)  $ is
determined by $\log\nu_{t+1}=\varrho\log\nu_{t}+\eta_{t}$. We assume that
$\eta_{t}$ is normally distributed with mean $0$ and variance $\omega^{2}$.
The aggregate shock affects the labor market in a way that will be established
later on.

In each $t$, young individuals choose the type of education to acquire. They
can choose either a flexible type of education $F$ or a rigid type of
education $R$. Working-age individuals with flexible education are affected
less by adverse aggregate shocks, but they have lower expected wages. The two
types of education have identical cost $C_{e}<y_{i,t}$ and need the same
amount of time to acquire $\mathcal{T}_{e}<\mathcal{T}$. Since young
individuals have typically limited financial wealth, we assume that there is
no saving decision when young and that any transfer from parents or relatives
is included in non-labor income $y_{i,t}$. We also abstract from student loans
and assume that all young individuals can afford to buy one of the two types
of education. As a consequence, a young individual will consume the part of
income $y_{i,t}$ that is not spent on education.

At each $t$, working-age individuals draw a wage offer $w_{i,t}^{F}$ if they
have chosen the flexible education when young and a wage offer $w_{i,t}^{R}$
otherwise. They also draw a productivity shock $\varepsilon^{S}_{i,t}$, for
$S=F,R$, which determines how productive their hours of work are in case they
choose to supply labor. We assume that the productivity shock is unknown to
the individuals when young. Given the wage offer and the productivity shock,
working-age individuals choose how much to work $h_{i,t}$ and how much to
consume. If a working-age individual decides to supply $h_{i,t}$ hours of
work, the effective amount of labor hours supplied is given by $\exp\left(
\varepsilon_{i,t}^{F}\right)  h_{i,t}$ for the flexible type of education $F$
and by $\exp\left(  \varepsilon_{i,t}^{R}\right)  h_{i,t}$ for the rigid type
of education $R$. We assume that $\varepsilon_{i,t}^{S}$ is normally
distributed with mean $\mu_{\varepsilon}^{S}$ and variance $\sigma_{S}^{2}$,
for $S=F,R$, and that $\sigma_{F}^{2}<\sigma_{R}^{2}$. To simplify the
analysis we normalize $E\left[  \exp\left(  \varepsilon_{i,t}^{S}\right)
\right]  =1$, for $S=F,R$.

The economy is populated by two types of firms to whom the working-age
individuals supply labor. The first type of firm employs only workers with
education $F$, whereas the second type of firm employs only workers with
education $R$. Both use the same type of capital $K$, which is assumed to be
fixed over periods. The labor demand functions of the two types of firms are
assumed to take the following form:
\[
\log H_{t}^{D,F}=\alpha_{0}+\alpha_{1}\log w_{t}^{F},
\]
and
\[
\log H_{t}^{D,R}=\alpha_{0}+\alpha_{1}\log w_{t}^{R}+\log{\nu_{t}},
\]
where $H^{D,S}$ is the total demand for \textit{effective labor}, with
$S=F,R$, $\alpha_{0}>0$, and $\alpha_{1}<0$. We assume that the two labor
demands have identical slopes for simplicity. These two labor demand functions
enable us to account for the common insight that workers with more flexible
education are affected less by aggregate shocks such as business cycle shocks.
The wage for each education group is determined by the equilibrium in the
corresponding labor market. It will therefore generally depend on the
aggregate shock.

We conclude the description of the model by pointing out that there is only
one source of uncertainty in the economy, the aggregate shock, and two sources
of heterogeneity across individuals, the risk aversion parameter and the
productivity shock.

The problem solved in period $t$ by individual $i$ of the young generation is
to choose consumption, leisure, and the type of education that satisfy:
\begin{gather}
\max_{c_{i,t},l_{i,t},c_{i,t+1},l_{i,t+1},S}\dfrac{\left(  c_{i,t}^{\sigma
}l_{i,t}^{1-\sigma}\right)  ^{1-\gamma_{i}}}{1-\gamma_{i}}+\delta\int%
\dfrac{\left(  c_{i,t+1}^{\sigma}l_{i,t+1}^{1-\sigma}\right)  ^{1-\gamma_{i}}%
}{1-\gamma_{i}}dP\left(  \left.  \nu_{t+1}\right\vert \nu_{t}\right)
\label{young-individual's-problem}\\
s.t.\ \ \ c_{i,t}=y_{i,t}-C_{e}\quad\text{and}\quad l_{i,t}=\mathcal{T}%
-\mathcal{T}_{e}\nonumber\\
c_{i,t+1}=w_{i,t+1}^{S}\left(  \nu_{t+1}\right)  \exp\left(  \varepsilon
_{i,t+1}^{S}\right)  \left(  \mathcal{T}-l_{i,t+1}\right)  \quad\text{for
every }\nu_{t+1}.\nonumber
\end{gather}
Here, $w_{i,t+1}^{S}\left(  \nu_{t+1}\right)  $ denotes the wage rate of
individual $i$ in the second period, which depends on the realization of the
aggregate shock $\nu_{t+1}$ and the education choice $S=F,R$. The wage rate is
per unit of the effective amount of labor hours supplied and is determined in equilibrium.

The problem solved by a working-age individual takes a simpler form.
Conditional on the realization of the aggregate shock $\nu_{t}$ and on the
type of education $S$ chosen when young, individual $i$ of the working-age
generation chooses consumption and leisure that solve the following problem:
\begin{gather}
\max_{c_{i,t},l_{i,t}}\dfrac{\left(  c_{i,t}^{\sigma}l_{i,t}^{1-\sigma
}\right)  ^{1-\gamma_{i}}}{1-\gamma_{i}}%
\label{working-age-individual's-problem}\\
s.t.\ c_{i,t}=w_{i,t}^{S}\left(  \nu_{t}\right)  \exp\left(  \varepsilon
_{i,t}^{S}\right)  \left(  \mathcal{T}-l_{i,t}\right)  .\nonumber
\end{gather}

We now solve the model starting from the problem of a working-age individual.
Using the first order conditions of problem
(\ref{working-age-individual's-problem}) the optimal choice of consumption,
leisure, and hence labor supply for a working-age individual takes the
following form:
\begin{align}
c_{i,t}^{\ast}  &  =\sigma w_{t}\left(  \nu_{t},S\right)  \exp\left(
\varepsilon_{i,t}^{S}\right)  \mathcal{T},\label{working-age-consumption}\\
l_{i,t}^{\ast}  &  =\left(  1-\sigma\right)  \mathcal{T}%
,\label{working-age-leisure}\\
h_{i,t}^{\ast}  &  =\mathcal{T}-l_{i,t}\left(  \nu_{i,t}\right)
=\sigma\mathcal{T}.\nonumber
\end{align}
The supply of effective labor is therefore equal to $\sigma\exp\left(
\varepsilon_{i,t}^{S}\right)  \mathcal{T}$. Given the optimal choice of
consumption and leisure, conditional on the aggregate shock, the value
function of a working-age individual with education $S$ can be written as
follows:
\[
V_{i,t}\left(  S,\nu_{t},\varepsilon_{i,t}\right)  =\dfrac{\left[  \left(
\sigma w_{i,t}^{S}\left(  \nu_{t}\right)  \exp\left(  \varepsilon_{i,t}%
^{S}\right)  \mathcal{T}\right)  ^{\sigma}\left(  \left(  1-\sigma\right)
\mathcal{T}\right)  ^{1-\sigma}\right]  ^{1-\gamma_{i}}}{1-\gamma_{i}}%
,\quad\text{S = F, R}.
\]

Given the value functions of a working-age individual, we can now characterize
the education choice of a young individual. This individual will choose
education $F$ if the expectation taken over the next period aggregate shocks
of the corresponding value function is greater than the analogous expectation
for education $R$:
\begin{equation}
E\left[  \left.  V_{i,t}\left(  F,\nu_{t+1},\varepsilon_{i,t+1}\right)
\right\vert \nu_{t}\right]  \geq E\left[  \left.  V_{i,t}\left(  R,\nu
_{t+1},\varepsilon_{i,t+1}\right)  \right\vert \nu_{t}\right]  .
\label{sector-choice}%
\end{equation}
To simplify the discussion, we assume that $\varepsilon_{i,t+1}$ is
independent of $\gamma_{i}$, thereby eliminating sample selection issues in
the wage equations.

Before we can determine which variables and parameters affect the education
choice, we have to derive the equilibrium in the labor market. It can be shown
that the labor market equilibrium is characterized by the following two wage
equations:\footnote{See Appendix \ref{sec-proof-wage-f-wage-R}, which is
available upon request.}
\begin{align}
\log w_{i,t}^{F}  &  =\dfrac{\log n_{t}^{F}+\log\sigma+\log\mathcal{T}%
-\alpha_{0}}{\alpha_{1}}+\varepsilon_{i,t}^{F},\label{wage-F}\\
\log w_{i,t}^{R}  &  =\dfrac{\log n_{t}^{R}+\log\sigma+\log\mathcal{T}%
-\alpha_{0}-\log\nu_{t}}{\alpha_{1}}+\varepsilon_{i,t}^{R}, \label{wage-R}%
\end{align}
where $w_{i,t}^{F}$ and $w_{i,t}^{R}$ are the individual wages observed in
sectors $F$ and $R$ and $n_{t}^{F}$ and $n_{t}^{R}$ are the measures of
individuals that choose education $F$ and $R$. We can now replace the
equilibrium wages inside inequality (\ref{sector-choice}) and analyze the
education decision of a young individual. It can be shown that a young
individual chooses the flexible type of education at time $t$ if the following
inequality is satisfied:\footnote{Details are given in the Appendix, which is
available upon request.}
\begin{equation}
\gamma_{i}\geq1-\frac{\log\left(  \frac{n_{t+1}^{F}}{n_{t+1}^{R}}\right)
+\frac{\sigma_{R}^{2}-\sigma_{F}^{2}}{2}+\varrho\log\nu_{t}}{\frac
{\sigma\left(  \sigma_{R}^{2}-\sigma_{F}^{2}+\omega^{2}\right)  }{2\alpha_{1}%
}}. \label{lem-sector-choice gamma geq 1}%
\end{equation}
This inequality provides some insight into the educational choice of young
individuals. Since $\alpha_{1}<0$, they are more likely to choose the flexible
education which insures them against aggregate shocks if the variance of the
aggregate shock is larger, if they are more risk averse, if the aggregate
shock at the time of the decision is lower as long as $\varrho>0$, and if the
elasticity of the wage for the rigid education with respect to the aggregate
shock is larger (the absolute value of $\alpha_{1}$ is lower).

Similarly to the first two examples, we can classify some of the variables and
some of the parameters as belonging to the cross-sectional submodel and the
remaining to the time-series submodel. The cross-sectional variables include
consumption $c_{i,t}$, leisure $l_{i,t}$, individual wages $w_{i,t}^{F}$ and
$w_{i,t}^{R}$, the variable determining the educational choice $D_{i,t}$, the
amount of time $\mathcal{T}$ an individual can divide between leisure and
productive activities, and the variables that enter the risk aversion
parameter $x_{i,t}$. The time-series variables are composed of the aggregate
shock $\nu_{t}$, the numbers of young individuals choosing the two types of
education $n^{F}$ and $n^{R}$, and the aggregate equilibrium wages in the two
sectors $w_{t}^{F}=E\left[  w_{it}^{F}\right]  $ and $w_{t}^{R}=E\left[
w_{it}^{F}\right]  $.\footnote{The expectation operator $E$ corresponds to the
expectation taken over the distribution of cross sectional variables.} We want
to stress the difference between individual wages and aggregate wages.
Individual wages are typically observed in panel data or repeated
cross-sections whose time dimension is generally short, whereas aggregate
wages are available in longer time-series of aggregate data. The
cross-sectional parameters consist of the relative taste for consumption
$\sigma$, the variances $\sigma_{F}^{2}$ and $\sigma_{R}^{2}$ of the
individual productivity shocks, the parameters defining the risk aversion
$\mu$, and the parameters of the wage equations $\alpha_{0}$ and $\alpha_{1}$,
whereas the time-series parameters include the two parameters governing the
evolution of the aggregate shock $\varrho$ and $\omega^{2}$, and the discount
factor $\delta$. The discount factor is notoriously difficult to estimate. For
this reason, in the rest of the section we will assume it is known.

We now employ the method proposed in this paper, which exploits a combination
of a long time-series of aggregate data and cross-sectional data, in the
estimation of the model parameters. We will highlight which parameters require
cross-sectional variables to be identified, which parameters require
time-series variables, and which parameters embody the bi-directional
relationship between the cross-sectional and time-series submodels.

We assume that the econometrician has access to two repeated cross-sections of
data for periods $t=1$ and $t=2$, which include i.i.d. observations on
educational choices $D_{i,t}$, wages $w_{i,t}^{S}$ with $S=F,R$, consumption
$c_{i,t}^{\ast}$, and leisure $l_{i,t}^{\ast}$. The econometrician has also
access to a time-series of aggregate data that spans $t=1,\ldots,\tau$. It
consists of the measures of people choosing the flexible and rigid educations
$n_{t}^{F}$ and $n_{t}^{R}$, and their corresponding aggregate wages
$w_{t}^{F}$ and $w_{t}^{R}$. For simplicity, we assume that the two
cross-sections consist of the same number of individuals $n$, and that the
first $\bar{n}_{1}$ and $\bar{n}_{2}$ individuals in the two cross sections
chose $S=F$.

The parameters $\alpha_{1}$, $\sigma$, $\sigma_{F}^{2}$, and $\sigma_{R}^{2}$
can be estimated using only the two cross-sections. Specifically, $\alpha_{1}$
can be consistently estimated using the wage equation for the flexible
education (\ref{wage-F}) for the periods $1$ and $2$ as the $\widehat{\alpha
}_{1}$ which solves
\begin{equation}
\frac{1}{\bar{n}_{1}}\sum_{i=1}^{\bar{n}_{1}}\log w_{i,1}^{F}-\frac{1}{\bar
{n}_{2}}\sum_{i=1}^{\bar{n}_{2}}\log w_{i,2}^{F}=\dfrac{1}{\widehat{\alpha
}_{1}}\left(  \log n_{1}^{F}-\log n_{2}^{F}\right)  .
\label{equation for estimation of a1}%
\end{equation}
Observe that this can be done because the productivity shock $\varepsilon_{t}$
and the risk aversion parameter $\gamma_{i}$ are assumed to be independent of
each other, which implies that there is no sample selectivity problem. The
parameter $\sigma$ can be consistently estimated employing the consumption and
leisure choices of the working-age individuals (\ref{working-age-consumption})
and (\ref{working-age-leisure}) for period $1$ as the $\widehat{\sigma}$ that
solves
\begin{equation}
\frac{1}{\bar{n}_{1}}\sum_{i=1}^{\bar{n}_{1}}\frac{c_{i,1}^{\ast}}%
{l_{i,1}^{\ast}}=w_{1}^{F}\frac{\widehat{\sigma}}{1-\widehat{\sigma}}.
\label{equation for estimation of sigma}%
\end{equation}
The variances of the productivity shocks for the two sectors $\sigma_{F}^{2}$
and $\sigma_{R}^{2}$ can be estimated using the wage equations for sectors $F$
and $R$ (\ref{wage-F}) and (\ref{wage-R}) as the sample variances of $\log
w_{i,t}^{F}$ and $\log w_{i,t}^{R}$.

The aggregate shocks and the parameters governing their evolution $\varrho$
and $\omega^{2}$ can then be estimated using the time-series of aggregate
data. Specifically, with $\alpha_{1}$ consistently estimated, the aggregate
shock in period $t$ can be consistently estimated for $t=1,\ldots,\tau$ using
the following equation:
\begin{equation}
\widehat{\log\nu_{t}}=\widehat{\alpha}_{1}\left(  \log w_{t}^{F}-\log
w_{t}^{R}\right)  -\left(  \log n_{t}^{F}-\log n_{t}^{R}\right)  ,
\label{equation for estimation of as}%
\end{equation}
which was derived by computing the difference between the equations defining
the equilibrium wages in sectors $R$ and $S$ and solving for $\log\nu_{t}%
$.\footnote{The equations defining the equilibrium wages are reported in the
Appendix as equations (\ref{equi-flex-wage}) and (\ref{equi-rigid-wage}).}
Observe that $\nu_{t}$ can only be estimated because $\alpha_{1}$ was
previously estimated using the cross-sections. The parameters $\varrho$ and
$\omega^{2}$ can then be consistently estimated by the time-series regression
of the equation that characterizes the evolution of the aggregate shocks:
\begin{equation}
\widehat{\log\nu_{t+1}}=\varrho\widehat{\log\nu_{t}}+\eta_{t}.
\label{equation for estimation of rho}%
\end{equation}

The only parameters left to estimate are the parameters $\mu$ defining the
individual risk aversion $\gamma_{i}$. They are the most interesting
parameters of the model because they incorporate the bi-directional
relationship between the cross-sectional and time-series submodels, as the
following discussion reveals. Specifically, if the distribution of $\gamma
_{i}$ is parametrically specified, the parameters $\mu$ can be consistently
estimated by MLE using cross-sectional variation on the educational choices
and the inequality that characterizes those choices
(\ref{lem-sector-choice gamma geq 1}). In the Monte Carlo exercise in Section
\ref{Section: Montecarlo Results}, we assume that $\log\gamma_{i}\sim N\left(
\mu,1\right)  $. Under this assumption, the distribution of risk aversion in
the population is characterized by only one parameter, its mean $\mu$. It can
be shown that in this case the probability that an individual chooses
education $F$ takes the following form:\footnote{For details see the Appendix,
which is available upon request.}
\[
1-\Phi\left(  \log\left(  1-\Theta_{t}\right)  -\mu\right)  .
\]
where
\[
\Theta_{t}\equiv\frac{\log\left(  \frac{n_{t+1}^{F}}{n_{t+1}^{R}}\right)
+\frac{\sigma_{R}^{2}-\sigma_{F}^{2}}{2}+\varrho\log\nu_{t}}{\frac
{\sigma\left(  \sigma_{R}^{2}-\sigma_{F}^{2}+\omega^{2}\right)  }{2\alpha_{1}%
}}.
\]
We can therefore estimate the mean of the distribution of risk aversion $\mu$
using a Probit maximum likelihood estimator, provided that $\nu_{t}$,
$\varrho$, $\omega^{2}$, $\sigma_{F}^{2}$, $\sigma_{R}^{2}$, $\sigma$, and
$\alpha_{1}$ and are known.\footnote{It is straightforward to relax the
distributional assumption on $\gamma_{i}$ and consider the more general case
where the risk aversion parameter $\gamma_{i}$ is a function of the observable
variables $x_{i,t}$, the unobservable variables $\xi_{i,t}$, and a vector of
parameters $\mu$, i.e. $\gamma_{i}=\gamma\left(  \left.  x_{i,t},\xi
_{i,t}\right\vert \mu\right)  $.} The cross-sectional parameter $\mu$ can
therefore be estimated only if the time-series parameters $\nu_{t}$, $\varrho
$, and $\omega^{2}$ have been previously estimated. But their estimation
requires the prior estimation of the cross-sectional parameter $\alpha_{1}$.
Hence, the bi-directional relationship between the cross-sectional and
time-series submodels.

To evaluate the effect of ignoring aggregate shocks when estimating the
parameters of the general equilibrium model, we now consider the case of an
econometrician who is unaware of the presence of aggregate shocks and, hence,
only uses cross-sectional variation for the identification and estimation of
the parameters of interest. The misspecification only changes the inequality
that characterizes the education choice (\ref{lem-sector-choice gamma geq 1}),
which in this case takes the following form:\footnote{For details see the
Appendix \ref{sec-misspecification}, which is available upon request.}
\begin{equation}
\gamma_{i}\geq1-\frac{\log\left(  \frac{n_{t+1}^{F}}{n_{t+1}^{R}}\right)
+\frac{\sigma_{R}^{2}-\sigma_{F}^{2}}{2}+\log\nu_{t+1}}{\frac{\sigma\left(
\sigma_{R}^{2}-\sigma_{F}^{2}\right)  }{2\alpha_{1}}}.
\label{education inequality misspecified}%
\end{equation}
As a consequence, under the misspecification and the assumption that
$\log\gamma_{i}\sim N\left(  \mu,1\right)  $, the probability that an
individual chooses education $F$ becomes
\[
1-\Phi\left(  \log\left(  1-\Theta_{t}^{\ast}\right)  -\mu\right)  ,
\]
where
\[
\Theta_{t}^{\ast}\equiv\frac{\log\left(  \frac{n_{t+1}^{F}}{n_{t+1}^{R}%
}\right)  +\frac{\sigma_{R}^{2}-\sigma_{F}^{2}}{2}+\log\nu_{t+1}}{\frac
{\sigma\left(  \sigma_{R}^{2}-\sigma_{F}^{2}\right)  }{2\alpha_{1}}}.
\]

Since this form of misspecification only changes the probability of choosing
education $F$, only estimation of the parameter $\mu$ is affected. To
understand its effect, we derive the estimation bias in closed form. In the
misspecified model, the probability that someone selects education $F$ can be
written as follows:
\[
1-\Phi\left(  \log\left(  1-\Theta_{t}^{\ast}\right)  -\mu\right)
=1-\Phi\left(  \log\left(  1-\Theta_{t}\right)  -\left(  \mu-\log\left(
1-\Theta_{t}^{\ast}\right)  +\log\left(  1-\Theta_{t}\right)  \right)
\right)  .
\]
Let $\hat{\mu}$ be the maximum likelihood estimator of the correctly specified
model. Then, the previous equation implies that the maximum likelihood
estimator $\widehat{\mu}_{\text{mis}}$ of the misspecified model satisfies the
following equation:
\[
\widehat{\mu}_{\text{mis}}=\widehat{\mu}+\log\left(  1-\Theta_{t}^{\ast
}\right)  -\log\left(  1-\Theta_{t}\right)  .
\]
The misspecification bias has therefore the following analytic form:
\begin{align}
&  \log\left(  1-\Theta_{t}^{\ast}\right)  -\log\left(  1-\Theta_{t}\right)
=\label{bias equation}\\
&  \log\left(  1-\frac{\log\left(  \frac{n_{t+1}^{F}}{n_{t+1}^{R}}\right)
+\frac{\sigma_{R}^{2}-\sigma_{F}^{2}}{2}+\log\nu_{t+1}}{\frac{\sigma\left(
\sigma_{R}^{2}-\sigma_{F}^{2}\right)  }{2\alpha_{1}}}\right)  -\log\left(
1-\frac{\log\left(  \frac{n_{t+1}^{F}}{n_{t+1}^{R}}\right)  +\frac{\sigma
_{R}^{2}-\sigma_{F}^{2}}{2}+\varrho\log\nu_{t}}{\frac{\sigma\left(  \sigma
_{R}^{2}-\sigma_{F}^{2}+\omega^{2}\right)  }{2\alpha_{1}}}\right)  .\nonumber
\end{align}
It shows that the magnitude of the bias depends on the size of the the
variance of the aggregate shocks $\omega^{2}$ and on the difference between
the expected aggregate shock in period $t+1$, $\varrho\log\nu_{t}$, and its
realization, $\log\nu_{t+1}$. Later in the paper, we will use particular
values for the model parameters to provide evidence on the magnitude of the
bias. Intuitively, ignoring the uncertainty generated by the aggregate shocks
should have the same effect as in the portfolio example of biasing upward the
estimated risk aversion parameter. Not accounting for the aggregate shocks is
equivalent to assuming that the agents face less uncertainty than they
actually experience when making the education decisions. Since the
individuals' decisions are based on the actual uncertainty, the only way the
model can explain those choices is by making people more risk averse. In the
general equilibrium model, this insight is not as straightforward to see as in
the portfolio example, since the bias depends also on the difference between
the current and next period aggregate shocks. For this reason we perform a
Monte Carlo exercise whose results are reported in Section
\ref{Section: Montecarlo Results}. They confirm the intuition regarding the
sign of the bias and show that its size can be extremely large. These insights
are not specific to the uncertainty generated by the aggregate shocks. They
apply equally to individual-specific shocks. If the econometrician disregards
the variation generated by those shocks, risk aversion will generally be
estimated to be larger than it actually is.

There is an alternative approach that can be used to estimate model parameters
when aggregate shocks affect behavior. The econometrician can use a single
panel data set in which the time-series dimension of the panel is sufficiently
long, instead of the repeated cross-sections combined with the time-series of
aggregate data. The general equilibrium model of this section is too
complicated to illustrate the limitations of the alternative panel-data
approach. Using a stylized linear panel model, however, one can show that,
when the alternative approach is used, the effective sample size of the data
is not $n\times T$ but $T$\textbf{,} with the cross-section generally playing
a minor role\footnote{A detailed exposition of the model and derivation are in
Appendix \ref{section-long-panel}, which is available upon request.}. The
reason is that the asymptotic theory for the alternative \textquotedblleft
long panel\textquotedblright\ approach requires, analogously to the
time-series analysis, the time dimension $T$ to go to infinity. A large
cross-section $n$ does not compensate for the lack of a long time-series in
the panel. This is in contrast to the textbook panel analysis, which indicates
that the effective sample size is $n\times T$. Since in practice almost all
panel data sets have limited time-series dimensions, using the alternative
panel approach would therefore lead to imprecise estimates relative to our
proposed method.

It is also important to point out that the practice of computing standard
errors under the assumption that the time-series parameters are known does not
solve the large-$T$ problem illustrated by our panel example. Under that
assumption, the standard errors for the cross sectional parameters are
incorrect and too small because they do not account for the noise introduced
by the estimation of the time series parameters. Lee and Wolpin (2006) use
such a procedure (see also their footnote 37). Their standard errors therefore
underestimate the true standard errors.\footnote{Donghoon Lee kindly confirmed
this in private communication.}

The econometric method proposed in this paper for the estimation of models
with aggregate shocks requires the combined use of cross-sectional data with
long time-series of aggregate data. There are no formulas available for the
computation of standard errors and confidence intervals that account for
jointly estimated time series and cross-sectional coefficients based on those
combined data sources. In the next section, we provide such formulas. They are
based on a new and complex asymptotic theory that we develop in the companion
paper Hahn, Kuersteiner, and Mazzocco (2016). Surprisingly, in spite of the
complexity of the theory, the formulas are straightforward and easy to use.

\section{Standard Errors\label{Standard Errors}}

The asymptotic theory underlying estimators obtained from the combination of
the two data sources considered in this paper is complex. It is based on a new
central limit theorem that requires a novel martingale representation. Given
its complexity, the theory is presented in a separate paper (Hahn, Kuersteiner
and Mazzocco (2016)). However, the mechanical implementation of test
statistics and confidence intervals is surprisingly straightforward. In this
Section, we first provide a step-by-step description of how those statistics
can be calculated. We then explain how they can be employed in concrete cases
using as examples the portfolio choice and the general equilibrium models
analyzed in the previous sections.

The computation starts with the explicit characterization of the
\textquotedblleft moments\textquotedblright\ that identify the cross-sectional
parameters $\beta$ and the time-series parameters $\rho$. In the most general
case, the aggregate shocks are unknown and must be estimated jointly with the
other model parameters using cross-sectional data, as illustrated in the
general equilibrium example. The shocks can therefore be treated as
cross-sectional parameters. This is accounted for by introducing a new vector
of parameters $\theta$ which is composed of the original cross-sectional
parameters and the aggregate shocks, i.e. $\theta=\left(  \beta,\nu
_{1},...,\nu_{T}\right)  $. We then denote with $f_{\theta,i}\left(
\theta,\rho\right)  $ the $i$-th moment used in the identification of the
parameters in $\theta$ and with $g_{\rho,t}\left(  \beta,\rho\right)  $ the
$t$-th moment used in the identification of the time-series parameters. Our
proposed estimator based on a combination of cross-sectional data and a long
time-series of aggregate data can then be written as the solution $\left(
\hat{\theta},\hat{\rho}\right)  $ to the following system of equations:
\begin{align}
\sum_{i=1}^{n}f_{\theta,i}\left(  \hat{\theta},\hat{\rho}\right)   &
=0,\label{Moment Cond y}\\
\sum_{t=\tau_{0}+1}^{\tau_{0}+\tau}g_{\rho,t}\left(  \hat{\beta},\hat{\rho
}\right)   &  =0. \label{Moment Cond z}%
\end{align}

Using those equations, the standard errors for $\hat{\theta}$ and $\hat{\rho}$
can be calculated using the following five steps.

\begin{enumerate}
\item Let $\phi=\left(  \theta^{\prime},\rho^{\prime}\right)  ^{\prime}$ be
the vector of parameters.

\item Let
\[
\mathbf{A}=\left[
\begin{array}
[c]{cc}%
\hat{A}_{f,\theta} & \hat{A}_{f,\rho}\\
\hat{A}_{g,\theta} & \hat{A}_{g,\rho}%
\end{array}
\right]  ,
\]
be the matrix of first order derivatives of the moments with respect to the
parameters, with
\begin{align*}
\hat{A}_{f,\theta}  &  = n^{-1}\sum_{i=1}^{n}\frac{\partial f_{\theta
,i}\left(  \hat{\theta},\hat{\rho}\right)  }{\partial\theta^{\prime}%
},\text{\ \ \ \ \ \ \ \ \ \ \ \ \ \ } \hat{A}_{f,\rho} = n^{-1}\sum_{i=1}%
^{n}\frac{\partial f_{\theta,i}\left(  \hat{\theta},\hat{\rho}\right)
}{\partial\rho^{\prime}},\\
\hat{A}_{g,\theta}  &  = \tau^{-1}\sum_{t=\tau_{0}+1}^{\tau_{0}+\tau}%
\frac{\partial g_{\rho,t}\left(  \hat{\beta},\hat{\rho}\right)  }%
{\partial\theta^{\prime}}, \text{\ \ \ \ \ \ \ \ \ \ \ } \hat{A}_{g,\rho} =
\tau^{-1}\sum_{t=\tau_{0}+1}^{\tau_{0}+\tau}\frac{\partial g_{\rho,t}\left(
\hat{\beta},\hat{\rho}\right)  }{\partial\rho^{\prime}}.
\end{align*}

\item Let
\[
\hat{\Omega}_{f}=\frac{1}{n}\sum_{i=1}^{n}f_{\theta,i}\left(  \hat{\theta
},\hat{\rho}\right)  f_{\theta,i}\left(  \hat{\theta},\hat{\rho}\right)
^{\prime}%
\]
and
\[
\hat{\Omega}_{g}=\frac{1}{n}\sum_{i=1}^{n}g_{\rho,t}\left(  \hat{\theta}%
,\hat{\rho}\right)  g_{\rho,t}\left(  \hat{\theta},\hat{\rho}\right)
^{\prime}.
\]

\item Let
\[
W=\left[
\begin{array}
[c]{cc}%
\frac{1}{n}\hat{\Omega}_{f} & 0\\
0 & \frac{1}{\tau}\hat{\Omega}_{g}%
\end{array}
\right]
\]

\item Calculate
\[
\mathbf{V}=\mathbf{A}^{-1}W\left(  \mathbf{A}^{\prime}\right)  ^{-1}%
\]
and use the square roots of the diagonal elements as the standard errors of
the estimator. For instance, if one is interested in the 95\% confidence
interval of the first component of $\phi$, it can be written as $\hat{\phi
}_{1}\pm1.96\sqrt{\mathbf{V}_{1,1}}$.
\end{enumerate}

The theoretical results in our companion paper as well as more detailed
calculations in the appendix reveal a few important points. The matrix
$\mathbf{V}$ in general is a function of aggregate shocks realized during the
observation periods of the cross-sectional sample. This randomness affects the
standard errors for both the cross-sectional and time series parameters. As a
result, caution needs to be exercised when comparing standard errors across
different observation periods or samples. On the other hand, pivotal
statistics such as t-ratios or confidence intervals have standard
distributional properties and can be compared across different samples. A
similar word of caution applies to sample descriptive statistics such as
simple sample averages obtained from short panels. These averages in general
are functions of realized values of aggregate shocks even when the
cross-sectional sample size is large. As a result, descriptive statistics are
expected to change in response to changes of the aggregate shock. Comparison
of these descriptive measures across different time periods or data sets thus
needs to be done with caution. The deep structural parameters estimated in
this paper, however, are typically thought to be fixed. As long as these
parameters are estimated consistently, their point estimators are not affected
by variation from aggregate shocks in large enough samples.\textbf{ }

In Appendix \ref{SE-formula-GE}, which is available upon request, we show for
the interested reader how the standard error formulas can be derived for the
portfolio example of Section \ref{portfolio-example} and the general
equilibrium model of Section \ref{GE-Model}. The application of the formulas
to the two examples highlights two features that determine the properties of
the asymptotic distribution of the proposed estimator. In the simple portfolio
example, there is a one-directional relationship between the cross-sectional
and time-series submodels. As a consequence, the cross-sectional parameters
can be estimated without knowledge of the time-series parameters. In addition,
agents form expectations for the main variable, end-of-period wealth, that do
not depend on the current realization of the aggregate shock. These two
features imply that the asymptotic distribution has a simple form that is
independent of the aggregate shocks. If one of these two conditions is not
satisfied, as mentioned above, the limiting distribution has a more
complicated form that depends on aggregate shocks. The more complex general
equilibrium example illustrates this point. In that case, the relationship
between the two sub-models is bidirectional, implying that there is no
recursive structure that can be used to first estimate the cross-sectional
parameters without knowledge of their time-series counterparts. As a
consequence, the asymptotic distribution depends on the aggregate variables
needed for the estimation of the cross-sectional parameters. Moreover, agents
use the current realization of the aggregate shock to form expectations about
future events. Since these expectations are used in their decision making
process, the aggregate shocks affect the limiting distribution of the
estimator by entering the variance-covariance matrix.

\section{Monte Carlo Results\label{Section: Montecarlo Results}}

In this section we report two sets of results. We first present Monte Carlo
results for the general equilibrium model with the objective of illustrating
how the estimation and inference approach introduced in this paper can be
applied in practice, and with the additional objective of documenting the
ability of our standard error formulas to produce the correct coverage
probabilities for the parameters of interest. We then provide evidence on the
magnitude of the bias that can be generated if the econometrician ignores
aggregate shocks.

To perform the Monte Carlo simulations and determine the size of the bias, we
have to set the 7 parameters of the general equilibrium model at particular
values. The most consequential parameter value is the one assigned to the
variance of the aggregate shocks $\omega^{2}$ since, as shown in Section
\ref{GE-Model}, it determines the magnitude of the bias if the econometrician
ignores the aggregate shocks. We chose the size of $\omega^{2}$ using the
estimated variance of the aggregate shocks used by Kydland and Prescott
(1982). They use an estimated variance for the quarterly U.S. cyclical output
that is equal to 0.000165. Differently from Kydland and Prescott (1982), in
our model capital is assumed to be fixed. As a consequence, the variation in
aggregate shocks affects exclusively labor demand. To account for this feature
of our model, we divided the variance estimated in Kydland and Prescott (1982)
by the square of the labor share in the economy.\footnote{The derivation of
the short-run labor demand function for a Cobb-Douglas production function
shows that this is the correct adjustment.} Since in the U.S. the labor share
is approximately 1/3, we divide 0.000165 by 1/9 to obtain 0.00149. In
addition, in our model only one of the two sectors is affected by the
aggregate shocks. To make the estimated variance consistent with our model, we
have therefore to multiply it by the square of 2 (the two sectors). With this
additional adjustment, we have a quarterly variance for the aggregate shock of
0.006. Our model has only two periods, one in which people engage in education
and one in which those individuals work. We assume that each period is
composed of 20 years and we multiply the quarterly variance of 0.006 by 4
quarters and 20 years, obtaining the aggregate variance we use in the
simulations, 0.48.

The values assigned to the variances of the productivity shocks $\sigma
_{F}^{2}$ and $\sigma_{R}^{2}$ are also important for the outcome of the Monte
Carlo exercise, since they determine the size of the individual-level
uncertainty relative to the size of the aggregate uncertainty. We chose those
variances using the estimated variance of the productivity shocks reported in
Macurdy (1982). Macurdy (1982) estimates a variance for the residuals of
yearly wages in the U.S. that is between 0.062 and 0.054. To derive our
measures of the micro variances, we multiply the upper bound of the yearly
variance estimated by Macurdy by 20 years (one of our periods), obtaining
1.2.\footnote{If we use the lower bound, the bias increases.} Lastly, in our
model the micro shocks in sector $F$ have a smaller variance than the shocks
in sector $R$. To account for this, we set $\sigma_{F}^{2}=1$ and $\sigma
_{R}^{2}=1.4$. The mean variance of the micro shocks is therefore 1.2, which
corresponds to the estimate obtained using the results in Macurdy (1982).

The remaining parameters are set equal to the following values. The mean of
the log of the risk aversion parameter $\mu$ is set equal to 0.2, which
corresponds to a mean risk aversion parameter of approximately 2. The
parameter measuring the persistence of the aggregate shock $\rho$ is initially
set equal to 0.75. We then evaluate how the results change when it is first
increased to 0.9 and then reduced to 0.5. The constant $\alpha_{0}$ and slope
$\alpha_{1}$ of the labor demand functions are chosen to be equal to 7 and -1,
respectively. The parameter characterizing the preferences for consumption
$\sigma$ is set equal to 0.6.

In the Monte Carlo exercise we consider 9 different specifications depending
on the size of the cross-section sample and length of the time-series sample.
Specifically, we simulate the model and estimate the parameters using the
following sample sizes for the cross-section: 2,500, 5,000, and 10,000
individuals; and the following lengths for the time-series: 25, 50, and 100
periods. In all cases we generate 5000 simulated data sets for the general
equilibrium model. The Monte Carlo results obtained using the method proposed
in this paper are presented in Table
\ref{Table: Montecarlo Results, Proposed Method}. The bias generated by
ignoring the aggregate shocks is reported in Table
\ref{Table: Montecarlo Results, Bias}. We only report results for the
parameters $\mu$, $\rho$, and $\omega^{2}$. All the other parameters are
estimated using the same estimators in the correct and misspecified model. The
estimates are therefore identical in the two models. Moreover, they are
estimated precisely and without significant bias in all Monte Carlo specifications.

We start by discussing the performance of the proposed approach. In the second
column of Table \ref{Table: Montecarlo Results, Proposed Method}, we report
the selected parameter estimates and in the third column the coverage
probability for those parameters of a confidence interval with 90\% nominal
coverage probability.\footnote{To perform the Monte Carlo exercise we have to
deal with a technical issue. The estimation of the risk aversion parameter
$\mu$ in the general equilibrium model requires the computation of
$\log\left(  1-\Theta_{t}\right)  $ where
\[
\Theta_{t}\equiv\frac{\log\left(  \frac{n_{t+1}^{F}}{n_{t+1}^{R}}\right)
+\frac{\sigma_{R}^{2}-\sigma_{F}^{2}}{2}+\varrho\log\nu_{t}}{\frac
{\sigma\left(  \sigma_{R}^{2}-\sigma_{F}^{2}+\omega^{2}\right)  }{2\alpha_{1}%
}}.
\]
In the model, $\Theta_{t}$ is always smaller than 1 and, hence, $\log\left(
1-\Theta_{t}\right)  $ is always well defined. In the estimation of $\mu$,
however, the true parameters included in $\Theta_{t}$ are replaced with their
estimated values. In some of the Monte Carlo repetitions, the randomness of
the estimated parameters generates values of $\Theta_{t}$ that are greater
than 1, which implies that $\log\left(  1-\Theta_{t}\right)  $ is not well
defined. A similar problem arises when we estimate the misspecified model. The
results reported in this Section are obtained by dropping all simulations for
which $\Theta_{t}\geq1$. In Appendix \ref{Censored versus Truncated Results},
we report the results obtained by using all the Monte Carlo runs and by
setting $\Theta=0.99$ in all cases in which $\Theta\geq1$.} Table
\ref{Table: Montecarlo Results, Proposed Method} documents that the accuracy
of the estimates increases with the length of the time-series. When the length
of the time-series increases from 25 to 100 the estimated persistence
parameter $\rho$ goes from $0.700$, $0.050$ lower than the true parameter, to
about $0.735$, just $0.015$ lower than the true parameter. The size of the
cross-section has no effect on the estimated value of $\rho$. A similar
pattern characterizes the estimates of the variance of the aggregate shocks,
except that in this case the size of the cross-section has a small effect on
the estimation results. For a cross-section of 10,000 individuals, an increase
from 25 to 100 periods produces a decline in the estimated $\omega^{2}$ from
$0.504$, $0.024$ higher than the true parameter, to $0.486$, just $0.06$ above
the true value. Similar trends characterize the estimates of $\omega^{2}$ for
cross-sections of 2,500 and 5,000, except that the accuracy of the estimates
improves slightly with larger cross-sections.

In the estimation of the risk aversion parameter $\mu$, we replace the other
parameters that enter the educational decision
(\ref{lem-sector-choice gamma geq 1}) with their estimated values. The small
biases in the estimation of $\rho$ and $\omega^{2}$ will therefore affect the
estimation of $\mu$, and generate patterns that are similar to the ones
observed for $\rho$ and $\omega^{2}$ when we increase the length of the
time-series and the size of the cross-section. For instance, with a
cross-section of 10,000 individuals, when we increase the time-series from 25
to 100 periods the estimated $\mu$ increases from $0.166$, $0.036$ below the
true parameter, to $0.188$, just $0.012$ below the true value. To confirm that
the small bias in the estimation of $\mu$ is generated by the small biases
that characterize the other parameters, we have also estimated $\mu$ using the
educational decision and the true value of the other parameters. We will refer
to this estimator as the infeasible estimator. The estimated values obtained
using this estimator, which by construction varies only with the length of the
time series, are reported in Table
\ref{Table: Montecarlo Results, Proposed Method}. They are always identical to
the true parameter, which confirms that the small bias in the estimation of
$\mu$ is generated by the small bias introduced by the other parameters. These
results indicate that it is important to use a long time-series when
estimating a model with aggregates shocks to reduce the noise introduced by
the estimation of the other parameters. A long time-series of aggregate
variables should therefore be preferred to a panel of data, since available
panels have a short time dimension.

We now describe the estimation of the risk aversion parameter using only
cross-sectional data. As discussed in Section \ref{GE-Model}, the parameter
$\mu$ requires both cross-section and time-series variation to be consistently
estimated. If the econometrician uses only cross-sectional data, the estimated
$\mu$ will be biased. In Table \ref{Table: Montecarlo Results, Bias} we report
the estimated $\mu$ and the corresponding bias only for the three time-series,
since the results are nearly identical across cross-sections. The numbers
indicate that the bias is positive, extremely large, and similar for all
time-series. In all cases, $\mu$ is estimated to be about six times the true
parameter and the bias to be about five times the true value. A bias of this
magnitude can have significant consequences if the estimated parameter is used
to answer policy questions, with answers that can be considerably different
from the ones that should be obtained.

In Tables \ref{Table: Montecarlo Results, Robustness} and
\ref{Table: Montecarlo Results, Robustness Bias}, we also report the effect of
changing the persistence of the aggregate shock by increasing $\rho$ from 0.75
to 0.9 and by reducing it from 0.75 to 0.5 for the the specification with
10,000 people and 100 periods. The effect is small. When we use our proposed
method the estimated coefficients are close to the true values. But if one
ignores the aggregate shocks the bias is large and positive.

Our Monte Carlo results indicate that ignoring aggregate shocks that affect
the data can have large effects on the estimation of important parameters,
such as the coefficient of risk aversion, and on the policy evaluations which
are based on them. Our results also indicate that the estimation method we
propose performs well. Given that it is relatively straightforward to use, it
is an easy solution for dealing with the presence of aggregate shocks.

\section{Summary}

Using a general econometric framework and three examples we shown that
generally, when aggregate shocks are present, model parameters cannot be
identified using cross-sectional variation alone. Identification of those
parameters requires the combination of cross-sectional and time-series data.
When those two data sources are jointly used, there are no available formulas
for the computation of test statistics and confidence intervals. We provide
new easy-to-use formulas that account for the interaction between those data
sources. Our results are expected to be helpful for the econometric analysis
of rational expectations models involving individual decision making as well
as general equilibrium models.

\FloatBarrier

\begin{table}[tbh]
\caption{Monte Carlo Results, Parameter Estimates For Correct Model}%
\label{Table: Montecarlo Results, Proposed Method}%
\begin{threeparttable}
\centering
\begin{tabular*}
{17.3cm}[c]{@{\extracolsep{\fill}}lcc}
&  & \\[-0.5cm]\hline\hline
True Parameter & Estimate & Cov. Prob.\\\hline
\multicolumn{2}{l}{\emph{Cross-sectional Sample Size: 2,500, Time-series
Sample Size: 25}} & \\\hline
\textbf{Log Risk Aversion Mean: $\mu= 0.2$} & \textbf{0.157} & \textbf{0.902}\\
Aggregate Shock Persistence: $\rho= 0.75$ & 0.700 & 0.875\\
Variance of Aggregate Shock: $\omega^{2} = 0.48$ & 0.514 & 0.840\\\hline
\multicolumn{2}{l}{\emph{Cross-sectional Sample Size: 2,500, Time-series
Sample Size: 50}} & \\\hline
\textbf{Log Risk Aversion Mean: $\mu= 0.2$} & \textbf{0.173} & \textbf{0.922}\\
Aggregate Shock Persistence: $\rho= 0.75$ & 0.722 & 0.892\\
Variance of Aggregate Shock: $\omega^{2} = 0.48$ & 0.502 & 0.867\\\hline
\multicolumn{2}{l}{\emph{Cross-sectional Sample Size: 2,500, Time-series
Sample Size: 100}} & \\\hline
\textbf{Log Risk Aversion Mean: $\mu= 0.2$} & \textbf{0.177} & \textbf{0.929}\\
Aggregate Shock Persistence: $\rho= 0.75$ & 0.735 & 0.888\\
Variance of Aggregate Shock: $\omega^{2} = 0.48$ & 0.495 & 0.888\\\hline
\multicolumn{3}{l}{\textbf{Infeasible estimator of Log Risk Aversion Mean, Cross-section of 2,500: \ $0.1997$}}\\\hline
\multicolumn{2}{l}{\emph{Cross-sectional Sample Size: 5,000, Time-series
Sample Size: 25}} & \\\hline
\textbf{Log Risk Aversion Mean: $\mu= 0.2$} & \textbf{0.158} & \textbf{0.900}\\
Aggregate Shock Persistence: $\rho= 0.75$ & 0.700 & 0.871\\
Variance of Aggregate Shock: $\omega^{2} = 0.48$ & 0.508 & 0.838\\\hline
\multicolumn{2}{l}{\emph{Cross-sectional Sample Size: 5,000, Time-series
Sample Size: 50}} & \\\hline
\textbf{Log Risk Aversion Mean: $\mu= 0.2$} & \textbf{0.180} & \textbf{0.918}\\
Aggregate Shock Persistence: $\rho= 0.75$ & 0.722 & 0.887\\
Variance of Aggregate Shock: $\omega^{2} = 0.48$ & 0.495 & 0.868\\\hline
\multicolumn{2}{l}{\emph{Cross-sectional Sample Size: 5,000, Time-series
Sample Size: 100}} & \\\hline
\textbf{Log Risk Aversion Mean: $\mu= 0.2$} & \textbf{0.178} & \textbf{0.932}\\
Aggregate Shock Persistence: $\rho= 0.75$ & 0.736 & 0.888\\
Variance of Aggregate Shock: $\omega^{2} = 0.48$ & 0.489 & 0.888\\\hline
\multicolumn{3}{l}{\textbf{Infeasible estimator of Log Risk Aversion Mean, Cross-section of 5,000: \ $0.1998$}}\\\hline
\multicolumn{2}{l}{\emph{Cross-sectional Sample Size: 10,000, Time-series
Sample Size: 25}} & \\\hline
\textbf{Log Risk Aversion Mean: $\mu= 0.2$} & \textbf{0.166} & \textbf{0.896}\\
Aggregate Shock Persistence: $\rho= 0.75$ & 0.700 & 0.870\\
Variance of Aggregate Shock: $\omega^{2} = 0.48$ & 0.504 & 0.835\\\hline
\multicolumn{2}{l}{\emph{Cross-sectional Sample Size: 10,000, Time-series
Sample Size: 50}} & \\\hline
\textbf{Log Risk Aversion Mean: $\mu= 0.2$} & \textbf{0.183} & \textbf{0.914}\\
Aggregate Shock Persistence: $\rho= 0.75$ & 0.722 & 0.883\\
Variance of Aggregate Shock: $\omega^{2} = 0.48$ & 0.492 & 0.859\\\hline
\multicolumn{2}{l}{\emph{Cross-sectional Sample Size: 10,000, Time-series
Sample Size: 100}} & \\\hline
\textbf{Log Risk Aversion Mean: $\mu= 0.2$} & \textbf{0.188} & \textbf{0.923}\\
Aggregate Shock Persistence: $\rho= 0.75$ & 0.736 & 0.882\\
Variance of Aggregate Shock: $\omega^{2} = 0.48$ & 0.486 & 0.887\\\hline
\multicolumn{3}{l}{\textbf{Infeasible estimator of Log Risk Aversion Mean, Cross-section of 10,000: \ $0.200$}}\\\hline\hline
\end{tabular*}
\begin{tablenotes}
\scriptsize{Notes: This Table reports the Monte Carlo results for the correct model obtained using our proposed estimation method. They are derived by simulating the general equilibrium model 5000 times. The second column reports the average estimated parameter, where the average is computed over the 5000 simulations. Column 3 reports the coverage probability of a confidence interval with 90\% nominal coverage probability.}
\end{tablenotes}
\end{threeparttable}
\end{table}

\FloatBarrier

\begin{table}[tbh]
\caption{Monte Carlo Results, Risk Aversion Estimates For Misspecified Model}%
\label{Table: Montecarlo Results, Bias}%
\begin{threeparttable}
\centering
\begin{tabular*}
{17.3cm}[c]{@{\extracolsep{\fill}}lcc}
&  & \\[-0.5cm]\hline\hline
True Parameter & Estimate & Bias\\\hline
\multicolumn{2}{l}{\emph{Cross-sectional Sample Size: 2,500}} & \\\hline
\textbf{Log Risk Aversion Mean: $\mu= 0.2$} & \textbf{1.163} & \textbf{0.963}%
\\\hline
\multicolumn{2}{l}{\emph{Cross-sectional Sample Size: 5,000}} & \\\hline
\textbf{Log Risk Aversion Mean: $\mu= 0.2$} & \textbf{1.173} & \textbf{0.973}%
\\\hline
\multicolumn{2}{l}{\emph{Cross-sectional Sample Size: 10,000}} & \\\hline
\textbf{Log Risk Aversion Mean: $\mu= 0.2$} & \textbf{1.179} & \textbf{0.979}%
\\\hline\hline
\end{tabular*}
\begin{tablenotes}
\scriptsize{Notes: This Table reports the Monte Carlo results for the misspecified model obtained using only cross-sectional variation. They are derived by simulating the general equilibrium model 5000 times. The second column reports the average estimated parameter, where the average is computed over the 5000 simulations. Column 3 reports the estimation bias, which is computed as the difference between the estimated and true parameter.}
\end{tablenotes}
\end{threeparttable}
\end{table}

\begin{table}[tbh]
\caption{Monte Carlo Results, Parameter Estimates For Correct Model, Different
$\rho$'s}%
\label{Table: Montecarlo Results, Robustness}%
\begin{threeparttable}
\centering
\begin{tabular*}
{17.3cm}[c]{@{\extracolsep{\fill}}lcc}
&  & \\[-0.5cm]\hline\hline
True Parameter & Estimate & Cov. Prob.\\\hline
\multicolumn{2}{l}{\emph{Cross-sectional Sample Size: 10,000, Time-series
Sample Size: 100}} & \\\hline
\textbf{Log Risk Aversion Mean: $\mu= 0.2$} & \textbf{0.212} & \textbf{0.920}\\
Aggregate Shock Persistence: $\rho= 0.9$ & 0.883 & 0.890\\
Variance of Aggregate Shock: $\omega^{2} = 0.48$ & 0.490 & 0.890\\\hline
\multicolumn{2}{l}{\emph{Cross-sectional Sample Size: 10,000, Time-series
Sample Size: 100}} & \\\hline
\textbf{Log Risk Aversion Mean: $\mu= 0.2$} & \textbf{0.179} & \textbf{0.930}\\
Aggregate Shock Persistence: $\rho= 0.5$ & 0.490 & 0.885\\
Variance of Aggregate Shock: $\omega^{2} = 0.48$ & 0.488 & 0.889\\\hline\hline
\end{tabular*}
\begin{tablenotes}
\scriptsize{See notes at Table \ref{Table: Montecarlo Results, Proposed Method}.}
\end{tablenotes}
\end{threeparttable}
\end{table}

\begin{table}[tbh]
\caption{Monte Carlo Results, Risk Aversion Estimates For Misspecified Model,
Different $\rho$'s}%
\label{Table: Montecarlo Results, Robustness Bias}%
\begin{threeparttable}
\centering
\begin{tabular*}
{17.3cm}[c]{@{\extracolsep{\fill}}lcc}
&  & \\[-0.5cm]\hline\hline
True Parameter & Estimate & Bias\\\hline
\multicolumn{2}{l}{\emph{Cross-sectional Sample Size: 10,000, $\rho= 0.9$}} &
\\\hline
\textbf{Log Risk Aversion Mean: $\mu= 0.2$} & \textbf{1.203} & \textbf{1.003}%
\\\hline
\multicolumn{2}{l}{\emph{Cross-sectional Sample Size: 10,000, $\rho= 0.5$}} &
\\\hline
\textbf{Log Risk Aversion Mean: $\mu= 0.2$} & \textbf{1.163} & \textbf{0.963}%
\\\hline\hline
\end{tabular*}
\begin{tablenotes}
\scriptsize{See notes at Table \ref{Table: Montecarlo Results, Bias}.}
\end{tablenotes}
\end{threeparttable}
\end{table}

\newpage

\setcounter{page}{1}

\appendix{}

\begin{center}
{\LARGE Appendix -- Available Upon Request}
\end{center}

\section{Discussion for Section \ref{portfolio-example}}

\subsection{Proof of (\ref{delta-port-choice})}

The maximization problem is equivalent to
\[
\max_{\alpha}\ -e^{-\delta\left(  \alpha\left(  1+r\right)  +\left(
1-\alpha\right)  \right)  }E\left[  e^{-\delta\left(  1-\alpha\right)
u_{i,t}}\right]  .
\]
Since $-\delta\left(  1-\alpha\right)  u_{i,t}\sim N\left(  -\delta\left(
1-\alpha\right)  \mu,\delta^{2}\left(  1-\alpha\right)  ^{2}\sigma^{2}\right)
$, we have
\[
E\left[  e^{-\delta\left(  1-\alpha\right)  u_{i,t}}\right]  =e^{-\delta
\left(  1-\alpha\right)  \mu+\frac{\delta^{2}\left(  1-\alpha\right)
^{2}\sigma^{2}}{2}},
\]
and the maximization problem can be rewritten as follows:
\[
\max_{\alpha}\ -e^{-\delta\left(  \alpha\left(  1+r\right)  +\left(
1-\alpha\right)  \left(  1+\mu\right)  -\frac{\delta\left(  1-\alpha\right)
^{2}\sigma^{2}}{2}\right)  }.
\]
Taking the first order condition, we have,%
\[
0=-\delta\left(  r-\mu+\sigma^{2}\delta-\alpha\sigma^{2}\delta\right)
\]
from which we obtain the solution%
\[
\alpha=\frac{1}{\sigma^{2}\delta}\left(  r-\mu+\sigma^{2}\delta\right)  .
\]

\subsection{Euler Equation and Cross Section\label{Euler-appendix}}

Our model in Section \ref{portfolio-example} is a stylized version of many
models considered in a large literature interested in estimating the parameter
$\delta$ using cross-sectional variation. Estimators are often based on moment
conditions derived from first order conditions (FOC) related to optimal
investment and consumption decisions. We illustrate the problems facing such estimators.

Assume a researcher has a cross-section of observations for individual
consumption and returns $c_{i,t}$ and $u_{i,t}$. The population FOC of our
model\footnote{We assume $\delta\neq0$ and rescale the equation by
$-\delta^{-1}.$} takes the simple form $E\left[  e^{-\delta c_{i,t}}\left(
r-u_{i,t}\right)  \right]  =0$. A just-identified moment based estimator for
$\delta$ solves the sample analog $n^{-1}\sum_{i=1}^{n}e^{-\hat{\delta}%
c_{i,t}}\left(  r-u_{i,t}\right)  =0$. It turns out that the probability limit
of $\hat{\delta}$ is equal to $\left.  \left(  \nu_{t}-r\right)  \right/
\left(  \left(  1-\alpha\right)  \sigma_{\epsilon}^{2}\right)  $, i.e.,
$\hat{\delta}$ is inconsistent.

We now compare the population FOC a rational agent uses to form their optimal
portfolio with the empirical FOC an econometrician using cross-sectional data observes:%

\[
n^{-1}\sum_{i=1}^{n}e^{-\delta c_{i,t}}\left(  r-u_{i,t}\right)  =0.
\]
Noting that $u_{i,t}=\nu_{t}+\epsilon_{i,t}$ and substituting into the budget
constraint
\[
c_{i,t}=1+\alpha r+\left(  1-\alpha\right)  u_{i,t}=1+\alpha r+\left(
1-\alpha\right)  \nu_{t}+\left(  1-\alpha\right)  \epsilon_{i,t}%
\]
we have
\begin{align}
n^{-1}\sum_{i=1}^{n}e^{-\delta c_{i,t}}\left(  r-u_{i,t}\right)   &
=n^{-1}\sum_{i=1}^{n}e^{-\delta\left(  1+\alpha r+\left(  1-\alpha\right)
\nu_{t}\right)  -\delta\left(  1-\alpha\right)  \epsilon_{i,t}}\left(
r-\nu_{t}-\epsilon_{i,t}\right) \label{Euler1}\\
&  =e^{-\delta\left(  1+\alpha r+\left(  1-\alpha\right)  \nu_{t}\right)
}\left(  \left(  r-\nu_{t}\right)  n^{-1}\sum_{i=1}^{n}e^{-\delta\left(
1-\alpha\right)  \epsilon_{i,t}}-n^{-1}\sum_{i=1}^{n}e^{-\delta\left(
1-\alpha\right)  \epsilon_{i,t}}\epsilon_{i,t}\right)  .\nonumber
\end{align}
Under suitable regularity conditions including independence of $\epsilon
_{i,t}$ in the cross-section it follows that
\begin{equation}
n^{-1}\sum_{i=1}^{n}e^{-\delta\left(  1-\alpha\right)  \epsilon_{i,t}%
}=E\left[  e^{-\delta\left(  1-\alpha\right)  \epsilon_{i,t}}\right]
+o_{p}\left(  1\right)  =e^{\frac{\delta^{2}\left(  1-\alpha\right)
^{2}\sigma_{\epsilon}^{2}}{2}}+o_{p}\left(  1\right)  \label{Euler2}%
\end{equation}
and
\begin{equation}
n^{-1}\sum_{i=1}^{n}e^{-\delta\left(  1-\alpha\right)  \epsilon_{i,t}}%
\epsilon_{i,t}=E\left[  e^{-\delta\left(  1-\alpha\right)  \epsilon_{i,t}%
}\epsilon_{i,t}\right]  +o_{p}\left(  1\right)  =-\delta\left(  1-\alpha
\right)  \sigma_{\epsilon}^{2}e^{\frac{\delta^{2}\left(  1-\alpha\right)
^{2}\sigma_{\epsilon}^{2}}{2}}+o_{p}\left(  1\right)  . \label{Euler3}%
\end{equation}
Taking limits as $n\rightarrow\infty$ in (\ref{Euler1}) and substituting
(\ref{Euler2}) and (\ref{Euler3}) then shows that the method of moments
estimator based on the empirical FOC asymptotically solves
\begin{equation}
\left(  \left(  r-\nu_{t}\right)  +\delta\left(  1-\alpha\right)
\sigma_{\epsilon}^{2}\right)  e^{\frac{\delta^{2}\left(  1-\alpha\right)
^{2}\sigma_{\epsilon}^{2}}{2}}=0. \label{Euler4}%
\end{equation}
Solving for $\delta$ we obtain
\[
\operatorname*{plim}\hat{\delta}=\frac{\nu_{t}-r}{\left(  1-\alpha\right)
\sigma_{\epsilon}^{2}}.
\]
This estimate is inconsistent because the cross-sectional data set lacks cross
sectional ergodicity, or in other words does not contain the same information
about aggregate risk as is used by rational agents. Therefore, the empirical
version of the FOC is unable to properly account for aggregate risk and return
characterizing the risky asset. The estimator based on the FOC takes the form
of an implicit solution to an empirical moment equation, which obscures the
effects of cross-sectional non-ergodicity. A more illuminative approach uses
our modelling strategy in Section \ref{section-model}.

On the other hand, it is easily shown using properties of the Gaussian moment
generating function that the population FOC is proportional to
\begin{equation}
E\left[  e^{-\delta\left(  1-\alpha\right)  u_{i,t}}\left(  r-u_{i,t}\right)
\right]  =\left(  r-\mu+\delta\left(  1-\alpha\right)  \sigma^{2}\right)
e^{-\delta\left(  1-\alpha\right)  \mu+\frac{\delta^{2}\left(  1-\alpha
\right)  ^{2}\sigma^{2}}{2}}=0. \label{Euler5}%
\end{equation}
The main difference between (\ref{Euler2}) and (\ref{Euler3}) lies in the fact
that $\sigma_{v}^{2}$ is estimated to be $0$ in the sample and that $\nu
_{t}\neq\mu$ in general. Note that (\ref{Euler5}) implies that consistency may
be achieved with a large number of repeated cross sections, or a panel data
set with a long time series dimension. However, this raises other issues
discussed in Section \ref{section-long-panel}.

\section{Details of Section \ref{GE-Model}}

\subsection{Proof of (\ref{lem-sector-choice gamma geq 1})}

In the proof we will drop the $i$ subscripts for notational purposes. The
individual will choose education $F$ if
\[
E\left[  \left.  V_{t+1}\left(  F,\nu_{t+1},\varepsilon_{t+1}^{F}\right)
\right\vert \nu_{t}\right]  \geq E\left[  \left.  V_{t+1}\left(  R,\nu
_{t+1},\varepsilon_{t+1}^{R}\right)  \right\vert \nu_{t}\right]  .
\]
Using (\ref{equi-flex-wage}) and (\ref{equi-rigid-wage}) later in Section
\ref{sec-proof-wage-f-wage-R}, we write%
\[
V_{t+1}\left(  F,\nu_{t+1},\varepsilon_{t+1}^{F}\right)  =\dfrac{\left[
\left(  \left(  \dfrac{n_{t+1}^{F}\sigma T}{e^{\alpha_{0}}}\right)
^{1/\alpha_{1}}\sigma T\right)  ^{\sigma}\left(  \left(  1-\sigma\right)
T\right)  ^{1-\sigma}\right]  ^{1-\gamma}}{1-\gamma}\exp\left(  \sigma\left(
1/\alpha_{1}\right)  \left(  1-\gamma\right)  \varepsilon_{t+1}^{F}\right)  ,
\]
and%
\begin{align*}
V_{t+1}\left(  R,\nu_{t+1},\varepsilon_{t+1}^{R}\right)   &  =\dfrac{\left[
\left(  \left(  \dfrac{n_{t+1}^{R}\sigma T}{e^{\alpha_{0}}}\right)
^{1/\alpha_{1}}\sigma T\right)  ^{\sigma}\left(  \left(  1-\sigma\right)
T\right)  ^{1-\sigma}\right]  ^{1-\gamma}}{1-\gamma}\exp\left(  \sigma\left(
1/\alpha_{1}\right)  \left(  1-\gamma\right)  \varepsilon_{t+1}^{R}\right) \\
&  \times\left(  \nu_{t+1}^{-\sigma\left(  1/\alpha_{1}\right)  \left(
1-\gamma\right)  }\right)  .
\end{align*}
It follows that education $F$ is chosen if and only if%
\begin{align}
\left(  n_{t+1}^{F}\right)  ^{\sigma\left(  1-\gamma\right)  /\alpha_{1}}\geq
&  \left(  n_{t+1}^{R}\right)  ^{\sigma\left(  1-\gamma\right)  /\alpha_{1}%
}\nonumber\\
&  \times\frac{E\left[  \exp\left(  \sigma\left(  1/\alpha_{1}\right)  \left(
1-\gamma\right)  \varepsilon_{t+1}^{R}\right)  \right]  E_{t}\left[  \nu
_{t+1}^{-\sigma\left(  1/\alpha_{1}\right)  \left(  1-\gamma\right)  }\right]
}{E\left[  \exp\left(  \sigma\left(  1/\alpha_{1}\right)  \left(
1-\gamma\right)  \varepsilon_{t+1}^{F}\right)  \right]  }.
\label{sector-choice-3}%
\end{align}

Recall that $E\left[  \exp\left(  \varepsilon_{t}^{S}\right)  \right]  =1$ for
$S=F,R$. It follows that $\varepsilon_{t+1}^{F}\sim N\left(  -\frac{\sigma
_{F}^{2}}{2},\sigma_{F}^{2}\right)  $, and $\varepsilon_{t+1}^{R}\sim N\left(
-\frac{\sigma_{R}^{2}}{2},\sigma_{R}^{2}\right)  $, and as a consequence,%
\begin{align}
E\left[  \exp\left(  \frac{\sigma\left(  1-\gamma\right)  }{\alpha_{1}%
}\varepsilon_{t+1}^{F}\right)  \right]   &  =\exp\left(  -\frac{\sigma\left(
1/\alpha_{1}\right)  \left(  1-\gamma\right)  \sigma_{F}^{2}}{2}+\frac{\left(
\sigma\left(  1/\alpha_{1}\right)  \left(  1-\gamma\right)  \right)
^{2}\sigma_{F}^{2}}{2}\right)  ,\label{decision-alt3'}\\
E\left[  \exp\left(  \frac{\sigma\left(  1-\gamma\right)  }{\alpha_{1}%
}\varepsilon_{t+1}^{R}\right)  \right]   &  =\exp\left(  -\frac{\sigma\left(
1/\alpha_{1}\right)  \left(  1-\gamma\right)  \sigma_{R}^{2}}{2}+\frac{\left(
\sigma\left(  1/\alpha_{1}\right)  \left(  1-\gamma\right)  \right)
^{2}\sigma_{R}^{2}}{2}\right)  . \label{decision-alt3''}%
\end{align}
\textbf{ }Also, because $\log\nu_{t+1}=\rho\log\nu_{t}+\eta_{t}$, or
$\nu_{t+1}=\nu_{t}^{\rho}\exp\left(  \eta_{t}\right)  $, we can write%
\[
E_{t}\left[  \nu_{t+1}^{-\sigma\left(  1-\gamma\right)  \left(  1/\alpha
_{1}\right)  }\right]  =E_{\eta}\left[  \left(  \nu_{t}^{\rho}\exp\left(
\eta_{t}\right)  \right)  ^{-\sigma\left(  1-\gamma\right)  \left(
1/\alpha_{1}\right)  }\right]  =\nu_{t}^{-\rho\sigma\left(  1-\gamma\right)
\left(  1/\alpha_{1}\right)  }E\left[  \exp\left(  -\sigma\left(
1-\gamma\right)  \left(  1/\alpha_{1}\right)  \eta_{t}\right)  \right]  .
\]
where $E_{\eta}\left[  \cdot\right]  $ denotes the integral with respect to
$\eta_{t}$ alone. The assumption that $\eta_{t}\sim N\left(  0,\omega
^{2}\right)  $ allows us to write%
\[
E\left[  \exp\left(  -\sigma\left(  1-\gamma\right)  \left(  1/\alpha
_{1}\right)  \eta_{t}\right)  \right]  =\exp\left(  \frac{\left(
\sigma\left(  1-\gamma\right)  \left(  1/\alpha_{1}\right)  \right)  ^{2}}%
{2}\omega^{2}\right)
\]
recognizing that the expectation on the left is nothing but the moment
generating function of $N\left(  0,\omega^{2}\right)  $ evaluated at
$-\sigma\left(  1-\gamma\right)  \left(  1/\alpha_{1}\right)  $. Therefore, we
have%
\begin{equation}
E_{t}\left[  \nu_{t+1}^{-\sigma\left(  1-\gamma\right)  \left(  1/\alpha
_{1}\right)  }\right]  =\nu_{t}^{-\rho\sigma\left(  1-\gamma\right)  \left(
1/\alpha_{1}\right)  }\exp\left(  \frac{\left(  \sigma\left(  1-\gamma\right)
\left(  1/\alpha_{1}\right)  \right)  ^{2}}{2}\omega^{2}\right)
\label{decision-alt3}%
\end{equation}
Combining (\ref{decision-alt3'}), (\ref{decision-alt3''}), and
(\ref{decision-alt3}), we obtain%
\begin{align*}
&  \frac{E\left[  \exp\left(  \sigma\left(  1/\alpha_{1}\right)  \left(
1-\gamma\right)  \varepsilon_{t+1}^{R}\right)  \right]  E_{t}\left[  \nu
_{t+1}^{-\sigma\left(  1/\alpha_{1}\right)  \left(  1-\gamma\right)  }\right]
}{E\left[  \exp\left(  \sigma\left(  1/\alpha_{1}\right)  \left(
1-\gamma\right)  \varepsilon_{t+1}^{F}\right)  \right]  }\\
&  =\nu_{t}^{-\rho\sigma\left(  1-\gamma\right)  \left(  1/\alpha_{1}\right)
}\exp\left(  \frac{\left(  \sigma\left(  1-\gamma\right)  \left(  1/\alpha
_{1}\right)  \right)  ^{2}}{2}\left(  \sigma_{R}^{2}-\sigma_{F}^{2}+\omega
^{2}\right)  \right) \\
&  \times\exp\left(  -\frac{\sigma\left(  1/\alpha_{1}\right)  \left(
1-\gamma\right)  \left(  \sigma_{R}^{2}-\sigma_{F}^{2}\right)  }{2}\right)
\end{align*}

As a consequence, (\ref{sector-choice-3}) is equivalent to
\begin{align}
\left(  n_{t+1}^{F}\right)  ^{\sigma\left(  1-\gamma\right)  /\alpha_{1}}  &
\geq\left(  n_{t+1}^{R}\right)  ^{\sigma\left(  1-\gamma\right)  /\alpha_{1}%
}\nu_{t}^{-\rho\sigma\left(  1-\gamma\right)  \left(  1/\alpha_{1}\right)
}\exp\left(  \frac{\left(  \sigma\left(  1-\gamma\right)  \left(  1/\alpha
_{1}\right)  \right)  ^{2}}{2}\left(  \sigma_{R}^{2}-\sigma_{F}^{2}+\omega
^{2}\right)  \right) \nonumber\\
&  \times\exp\left(  -\frac{\sigma\left(  1/\alpha_{1}\right)  \left(
1-\gamma\right)  \left(  \sigma_{R}^{2}-\sigma_{F}^{2}\right)  }{2}\right)
\label{decision-alt1}%
\end{align}
when $1-\gamma>0$, and to
\begin{align}
\left(  n_{t+1}^{F}\right)  ^{\sigma\left(  1-\gamma\right)  /\alpha_{1}}  &
\leq\left(  n_{t+1}^{R}\right)  ^{\sigma\left(  1-\gamma\right)  /\alpha_{1}%
}\nu_{t}^{-\rho\sigma\left(  1-\gamma\right)  \left(  1/\alpha_{1}\right)
}\exp\left(  \frac{\left(  \sigma\left(  1-\gamma\right)  \left(  1/\alpha
_{1}\right)  \right)  ^{2}}{2}\left(  \sigma_{R}^{2}-\sigma_{F}^{2}+\omega
^{2}\right)  \right) \nonumber\\
&  \times\exp\left(  -\frac{\sigma\left(  1/\alpha_{1}\right)  \left(
1-\gamma\right)  \left(  \sigma_{R}^{2}-\sigma_{F}^{2}\right)  }{2}\right)
\label{decision-alt2}%
\end{align}
when $1-\gamma<0$.

Consider first the case $1-\gamma>0$. Taking logs of (\ref{decision-alt1}), we
obtain%
\begin{align*}
\frac{\sigma\left(  1-\gamma\right)  }{\alpha_{1}}\log n_{t+1}^{F}  &
\geq\frac{\sigma\left(  1-\gamma\right)  }{\alpha_{1}}\log n_{t+1}^{R}%
-\rho\frac{\sigma\left(  1-\gamma\right)  }{\alpha_{1}}\log\nu_{t}\\
&  +\frac{\left(  \sigma\left(  1-\gamma\right)  \right)  ^{2}}{2\alpha
_{1}^{2}}\left(  \sigma_{R}^{2}-\sigma_{F}^{2}+\omega^{2}\right)
-\frac{\sigma\left(  1/\alpha_{1}\right)  \left(  1-\gamma\right)  \left(
\sigma_{R}^{2}-\sigma_{F}^{2}\right)  }{2}.
\end{align*}
Dividing by $\sigma$ and multiplying by $\alpha_{1}<0$, we conclude that the
decision is equivalent to%
\[
\left(  1-\gamma\right)  \left(  \log\frac{n_{t+1}^{F}}{n_{t+1}^{R}}%
+\frac{\left(  \sigma_{R}^{2}-\sigma_{F}^{2}\right)  }{2}+\rho\log\nu
_{t}\right)  \leq\frac{\sigma\left(  1-\gamma\right)  ^{2}\left(  \sigma
_{R}^{2}-\sigma_{F}^{2}+\omega^{2}\right)  }{2\alpha_{1}}%
\]
Dividing by $\sigma\left(  1-\gamma\right)  \left(  \sigma_{R}^{2}-\sigma
_{F}^{2}+\omega^{2}\right)  >0$, we obtain%
\[
\frac{\log\frac{n_{t+1}^{F}}{n_{t+1}^{R}}+\frac{\sigma_{R}^{2}-\sigma_{F}^{2}%
}{2}+\rho\log\nu_{t}}{\sigma\left(  \sigma_{R}^{2}-\sigma_{F}^{2}+\omega
^{2}\right)  }\leq\frac{1-\gamma}{2\alpha_{1}}.
\]

Multiplying by $2\alpha_{1}<0$, we obtain%
\[
\frac{\log\frac{n_{t+1}^{F}}{n_{t+1}^{R}}+\frac{\sigma_{R}^{2}-\sigma_{F}^{2}%
}{2}+\rho\log\nu_{t}}{\frac{\sigma\left(  \sigma_{R}^{2}-\sigma_{F}^{2}%
+\omega^{2}\right)  }{2\alpha_{1}}}\geq1-\gamma
\]
or
\[
\gamma\geq1-\frac{\log\left(  \frac{n_{t+1}^{F}}{n_{t+1}^{R}}\right)
+\frac{\sigma_{R}^{2}-\sigma_{F}^{2}}{2}+\rho\log\nu_{t}}{\frac{\sigma\left(
\sigma_{R}^{2}-\sigma_{F}^{2}+\omega^{2}\right)  }{2\alpha_{1}}}%
\]
which proves inequality (\ref{lem-sector-choice gamma geq 1}) for the
$1-\gamma>0$ case.

Consider now the case $1-\gamma<0$. Taking logs of (\ref{decision-alt2}), we
obtain%
\begin{align*}
\frac{\sigma\left(  1-\gamma\right)  }{\alpha_{1}}\log n_{t+1}^{F}  &
\leq\frac{\sigma\left(  1-\gamma\right)  }{\alpha_{1}}\log n_{t+1}^{R}%
-\rho\frac{\sigma\left(  1-\gamma\right)  }{\alpha_{1}}\log\nu_{t}\\
&  +\frac{\left(  \sigma\left(  1-\gamma\right)  \right)  ^{2}}{2\alpha
_{1}^{2}}\left(  \sigma_{R}^{2}-\sigma_{F}^{2}+\omega^{2}\right)
-\frac{\sigma\left(  1/\alpha_{1}\right)  \left(  1-\gamma\right)  \left(
\sigma_{R}^{2}-\sigma_{F}^{2}\right)  }{2}%
\end{align*}
Dividing by $\sigma$ and multiplying by $\alpha_{1}<0$, we conclude that the
decision is equivalent to%
\[
\left(  1-\gamma\right)  \left(  \log\frac{n_{t+1}^{F}}{n_{t+1}^{R}}%
+\frac{\sigma_{R}^{2}-\sigma_{F}^{2}}{2}+\rho\log\nu_{t}\right)  \geq
\frac{\sigma\left(  1-\gamma\right)  ^{2}\left(  \sigma_{R}^{2}-\sigma_{F}%
^{2}+\omega^{2}\right)  }{2\alpha_{1}}%
\]
Dividing by by $\sigma\left(  1-\gamma\right)  \left(  \sigma_{R}^{2}%
-\sigma_{F}^{2}+\omega^{2}\right)  <0$, we obtain%
\[
\frac{\log\frac{n_{t+1}^{F}}{n_{t+1}^{R}}+\frac{\sigma_{R}^{2}-\sigma_{F}^{2}%
}{2}+\rho\log\nu_{t}}{\sigma\left(  \sigma_{R}^{2}-\sigma_{F}^{2}+\omega
^{2}\right)  }\leq\frac{\left(  1-\gamma\right)  }{2\alpha_{1}}%
\]
Multiplying by $2\alpha_{1}<0$, we obtain%
\[
\frac{\log\frac{n_{t+1}^{F}}{n_{t+1}^{R}}+\frac{\sigma_{R}^{2}-\sigma_{F}^{2}%
}{2}+\rho\log\nu_{t}}{\frac{\sigma\left(  \sigma_{R}^{2}-\sigma_{F}^{2}%
+\omega^{2}\right)  }{2\alpha_{1}}}\geq1-\gamma
\]
or%
\[
\gamma\geq1-\frac{\log\frac{n_{t+1}^{F}}{n_{t+1}^{R}}+\frac{\sigma_{R}%
^{2}-\sigma_{F}^{2}}{2}+\rho\log\nu_{t}}{\frac{\sigma\left(  \sigma_{R}%
^{2}-\sigma_{F}^{2}+\omega^{2}\right)  }{2\alpha_{1}}}%
\]
which proves inequality (\ref{lem-sector-choice gamma geq 1}) for the
$1-\gamma<0$ case as well.

\subsection{Proof of (\ref{wage-F}) and (\ref{wage-R}%
)\label{sec-proof-wage-f-wage-R}}

Note that individual heterogeneity is completely summarized by the vector
$\chi_{t}\equiv\left(  \varepsilon_{t}^{F},\varepsilon_{t}^{R},\gamma\right)
$. This means that the labor supply for each type $\chi$ of workers can be
written $h_{t}^{F}\left(  \chi\right)  $ and $h_{t}^{R}\left(  \chi\right)  $.
We assume that the measure of individuals such that $\left(  \varepsilon
_{t}^{F},\varepsilon_{t}^{R},\gamma\right)  \in A$ for some $A\subset R^{3}$
is given by $N_{t}\int_{A}G\left(  d\chi\right)  $, where $G$ is a joint CDF.
For simplicity, we assume that $G$ is such that the first and second
components are independent of each other. Recall that we also assume that
$\int\exp\left(  \varepsilon_{t}\right)  G\left(  d\chi\right)  =1$.

We can rewrite (\ref{sector-choice}) as follows:
\begin{align}
&  E\left[  \left.  \dfrac{\left[  \left(  \sigma w_{t+1}^{F}\left(  \nu
_{t+1}\right)  \exp\left(  \varepsilon_{t+1}^{F}\right)  T\right)  ^{\sigma
}\left(  \left(  1-\sigma\right)  T\right)  ^{1-\sigma}\right]  ^{1-\gamma}%
}{1-\gamma}\right\vert \nu_{t}\right] \nonumber\\
&  \geq E\left[  \left.  \dfrac{\left[  \left(  \sigma w_{t+1}^{R}\left(
\nu_{t+1}\right)  \exp\left(  \varepsilon_{t+1}^{R}\right)  T\right)
^{\sigma}\left(  \left(  1-\sigma\right)  T\right)  ^{1-\sigma}\right]
^{1-\gamma}}{1-\gamma}\right\vert \nu_{t}\right]  . \label{sector-choice-2}%
\end{align}
As a consequence, education $F$ is chosen if%
\begin{equation}
\psi\left(  \gamma,\nu_{t}\right)  \equiv E\left[  \left.  \frac{\left[
\left(  w_{t+1}^{F}\left(  \nu_{t+1}\right)  \exp\left(  \varepsilon_{t+1}%
^{F}\right)  \right)  \right]  ^{\sigma\left(  1-\gamma\right)  }}{1-\gamma
}\right\vert \nu_{t}\right]  -E\left[  \left.  \frac{\left[  w_{t+1}%
^{R}\left(  \nu_{t+1}\right)  \exp\left(  \varepsilon_{t+1}^{R}\right)
\right]  ^{\sigma\left(  1-\gamma\right)  }}{1-\gamma}\right\vert \nu
_{t}\right]  \geq0 \label{education inequality}%
\end{equation}
Specifically, an individual chooses $F$ if $\psi\left(  \gamma,\nu_{t}\right)
>0$. We can now introduce the equilibrium condition for education $F$. It
takes the following form:%
\begin{align*}
H_{t+1}^{D,F}  &  =N_{t+1}\int_{E=F}h_{t+1}^{F}\left(  \chi\right)  G\left(
d\chi\right)  =N_{t+1}\sigma\mathcal{T}\int_{\psi\left(  \gamma,\nu
_{t}\right)  \geq0}\exp\left(  \varepsilon_{t+1}^{F}\right)  G\left(
d\chi\right) \\
H_{t+1}^{D,R}  &  =N_{t+1}\int_{E=R}h_{t+1}^{F}\left(  \chi\right)  G\left(
d\chi\right)  =N_{t+1}\sigma\mathcal{T}\int_{\psi\left(  \gamma,\nu
_{t}\right)  <0}\exp\left(  \varepsilon_{t+1}^{R}\right)  G\left(
d\chi\right)
\end{align*}
Using independence between $\gamma$ and $\varepsilon$ as well as $\int%
\exp\left(  \varepsilon_{t}^{F}\right)  G\left(  d\chi\right)  =1$, we can
write%
\begin{align}
\int_{\psi\left(  \gamma,\nu_{t}\right)  \geq0}\exp\left(  \varepsilon
_{t+1}^{F}\right)  G\left(  d\chi\right)   &  =\left(  \int_{\psi\left(
\gamma,\nu_{t}\right)  \geq0}G\left(  d\chi\right)  \right)  \left(  \int%
\exp\left(  \varepsilon_{t+1}^{F}\right)  G\left(  d\chi\right)  \right)
\nonumber\\
&  =\int_{\psi\left(  \gamma,\nu_{t}\right)  \geq0}G\left(  d\chi\right)
\nonumber\\
&  =\text{Fraction of workers in Sector }F \label{frac-flexible}%
\end{align}
so we can write $H_{t}^{D,F}=n_{t}^{F}\sigma T$, where $n^{F}$ is the measure
of individuals that chose education $F$\textbf{.} Taking logs, we have:
\[
\log H_{t}^{D,F}=\log n_{t}^{F}+\log\sigma+\log T,
\]
Substituting for $H_{t}^{D,F}$, we obtain the following equilibrium
condition:
\[
\alpha_{0}+\alpha_{1}\log w_{t}^{F}=\log n_{t}^{F}+\log\sigma+\log T,
\]
Solving for $\log w_{t}^{F}$, we have the log equilibrium wage:
\begin{equation}
\left(  z_{t}^{F}\equiv\right)  \quad\log w_{t}^{F}=\dfrac{\log n_{t}^{F}%
+\log\sigma+\log T-\alpha_{0}}{\alpha_{1}}. \label{equi-flex-wage}%
\end{equation}
This wage is for the unit of effective labor. Because the worker $i$ provides
$\sigma\exp\left(  \varepsilon_{t}\right)  T$ of effective labor, his recorded
earning is $\sigma\exp\left(  \varepsilon_{t}\right)  T\exp\left(  \dfrac{\log
n_{t}^{F}+\log\sigma+\log T-\alpha_{0}}{\alpha_{1}}\right)  $. Because the
individual works for $\sigma T$ hours, his wage for the labor is $\exp\left(
\varepsilon_{t}\right)  \exp\left(  \dfrac{\log n_{t}^{F}+\log\sigma+\log
T-\alpha_{0}}{\alpha_{1}}\right)  $; we will assume that the cross section
\textquotedblleft error\textquotedblright\ consist of $n$ i.i.d. copies of
$\varepsilon_{t}$, i.e.,the observed log equilibrium individual wage follows:%
\[
\log w_{it}^{F}=\dfrac{\log n_{t}^{F}+\log\sigma+\log T-\alpha_{0}}{\alpha
_{1}}+\varepsilon_{it}^{F}.
\]

Because of the normalization $E\left[  \exp\left(  \varepsilon_{it}%
^{R}\right)  \right]  =1$, the second equality in (\ref{frac-flexible}) also
applies to the $R$ sector, and as a consequence, the equilibrium condition for
education $R$ has the following form:
\[
H_{t}^{D,R}=n_{t}^{R}\sigma T,
\]
where $n^{R}$ is the measure of individuals that chose education $R$.
Substituting for $H_{t}^{D,R}$ and solving for $\log w_{t}^{R}$, we obtain the
following equilibrium wage for $R$:
\begin{equation}
\left(  z_{t}^{R}\equiv\right)  \quad\log w_{t}^{R}=\dfrac{\log n_{t}^{R}%
+\log\sigma+\log T-\alpha_{0}-\log\nu_{t}}{\alpha_{1}}.
\label{equi-rigid-wage}%
\end{equation}
By the same reasoning, the observed log equilibrium wage would look like%
\[
\log w_{it}^{R}=\dfrac{\log n_{t}^{R}+\log\sigma+\log T-\alpha_{0}-\log\nu
_{t}}{\alpha_{1}}+\varepsilon_{it}^{R}.
\]

\section{Long Panels?\label{section-long-panel}}

Our proposal requires access to two data sets, a cross-section (or short
panel) and a long time series of aggregate variables. One may wonder whether
we may obtain an estimator with similar properties by exploiting panel data
sets in which the time series dimension of the panel data is large enough.

One obvious advantage of combining two sources of data is that time series
data may contain variables that are unavailable in typical panel data sets.
For example the inflation rate potentially provides more information about
aggregate shocks than is available in panel data. We argue with a toy model
that even without access to such variables, the estimator based on the two
data sets is expected to be more precise, which suggests that the advantage of
data combination goes beyond availability of more observable variables.

Consider the alternative method based on one long panel data set, in which
both $n$ and $T$ go to infinity. Since the number of aggregate shocks $\nu
_{t}$ increases as the time-series dimension $T$ grows, we expect that the
long panel analysis can be executed with tedious yet straightforward arguments
by modifying ideas in Hahn and Kuersteiner (2002), Hahn and Newey (2004) and
Gagliardini and Gourieroux (2011), among others.

We will now illustrate a potential problem with the long panel approach with a
simple artificial example. Suppose that the econometrician is interested in
the estimation of a parameter $\gamma$ that characterizes the following system
of linear equations:
\begin{align*}
q_{i,t}  &  =x_{i,t}\frac{\gamma}{\omega}+\nu_{t}+\varepsilon_{i,t}\qquad
i=1,\ldots,n;\;t=1,\ldots,T,\\
\nu_{t}  &  =\omega\nu_{t-1}+u_{t}.
\end{align*}
The variables $q_{i,t}$ and $x_{i,t}$ are observed and it is assumed that
$x_{i,t}$ is strictly exogenous in the sense that it is independent of the
error term $\varepsilon_{i,t}$, including all leads and lags. For simplicity,
we also assume that $u_{t}$ and $\varepsilon_{i,t}$ are normally distributed
with zero mean and that $\varepsilon_{i,t}$ is i.i.d. across both $i$ and $t$.
We will denote by $\delta$ the ratio $\left.  \gamma\right/  \omega$.

In order to estimate $\gamma$ based on the panel data $\left\{  \left(
q_{i,t},x_{i,t}\right)  ,i=1,\ldots,n;\;t=1,\ldots,T\right\}  $, we can adopt
a simple two-step estimator of $\gamma$. In a first step, the parameter
$\delta$ and the aggregate shocks $\nu_{t}$ are estimated using an Ordinary
Least Square (OLS) regression of $q_{i,t}$ on $x_{i,t}$ and time dummies. In
the second step, the time-series parameter $\omega$ is estimated by regressing
$\widehat{\nu}_{t}$ on $\widehat{\nu}_{t-1}$, where $\widehat{\nu}_{t}$,
$t=1,\ldots,T$, are the aggregate shocks estimated in the first step using the
time dummies. An estimator of $\gamma$ can then be obtained as
$\widehat{\delta}\widehat{\omega}$.

The following remarks are useful to understand the properties of the estimator
$\widehat{\gamma}=\widehat{\delta}\widehat{\omega}$. First, even if $\nu_{t}$
were observed, for $\widehat{\omega}$ to be a consistent estimator of $\omega$
we would need $T$ to go to infinity, under which assumption we have
$\widehat{\omega}=\omega+O_{p}\left(  T^{-1/2}\right)  $. This implies that it
is theoretically necessary to assume that our data source is a
\textquotedblleft long\textquotedblright\ panel, i.e., $T\rightarrow\infty$.
Similarly, $\hat{\nu}_{t}$ is a consistent estimator of $\nu_{t}$ only if $n$
goes to infinity. As a consequence, we have $\hat{\nu}_{t}=\nu_{t}%
+O_{p}\left(  n^{-1/2}\right)  $. This implies that it is in general
theoretically necessary to assume that $n\rightarrow\infty$.\footnote{For
$\widehat{\omega}$ to have the same distribution as if $\nu_{t}$ were
observed, we need $n$ to go to infinity faster than $T$ or equivalently that
$T=o\left(  n\right)  $. See Heckman and Sedlacek (1985, p. 1088).} Moreover,
if $n$ and $T$ both go to infinity, $\widehat{\delta}$ is a consistent
estimator of $\delta$ and $\widehat{\delta}=\delta+O_{p}\left(  n^{-1/2}%
T^{-1/2}\right)  $. All this implies that
\[
\widehat{\gamma}=\widehat{\delta}\widehat{\omega}=\left(  \delta+O_{p}\left(
\frac{1}{\sqrt{nT}}\right)  \right)  \left(  \omega+O_{p}\left(  \frac
{1}{\sqrt{T}}\right)  \right)  =\delta\omega+O_{p}\left(  \frac{1}{\sqrt{T}%
}\right)  =\gamma+O_{p}\left(  \frac{1}{\sqrt{T}}\right)  .
\]

The $O_{p}\left(  n^{-1/2}T^{-1/2}\right)  $ estimation noise of
$\widehat{\delta}$, which is dominated by the $O_{p}\left(  T^{-1/2}\right)  $
error from estimating $\widehat{\omega}$, is the term that would arise if
$\omega$ were not estimated. The term reflects typical findings in long panel
analysis (i.e., large $n$, large $T$), where the standard errors are inversely
proportional to the square root of the number $n\times T$ of observations. The
fact that the estimation error of $\widehat{\gamma}$ is dominated by the
$O_{p}\left(  T^{-1/2}\right)  $ term indicates that the number of
observations is effectively equal to $T$, i.e., the long panel should be
treated as a time series problem for all practical purposes.

This conclusion has two interesting implications. First, the sampling noise
due to cross-section variation should be ignored and the \textquotedblleft
standard\textquotedblright\ asymptotic variance formulae should generally be
avoided in panel data analysis when aggregate shocks are present. We note that
Lee and Wolpin's (2006, 2010) standard errors use the standard formula that
ignores the $O_{p}\left(  T^{-1/2}\right)  $ term. Second, since in most cases
the time-series dimension $T$ of a panel data set is relatively small, despite
the theoretical assumption that it grows to infinity, estimators based on
panel data will generally be more imprecise than may be expected from the
\textquotedblleft large\textquotedblright\ number $n\times T$ of
observations.\footnote{This raises an interesting point. Suppose there is an
aggregate time series data set available with which consistent estimation of
$\gamma$ is feasible at the standard rate of convergence. Also suppose that
the number of time series observations, say $\tau$, is a lot larger than $T$.
In that case we conjecture that the panel data analysis is strictly dominated
by the time series analysis from an efficiency point of view.}

\section{Asymptotic Distribution and Standard Error Formulas for
Examples\label{SE-formula-GE}}

In this section, we discuss how the discussion in Section
\ref{Standard Errors} applies to the general equilibrium model. We also
present characterizations of the asymptotic distributions for the examples in
Sections \ref{portfolio-example} and \ref{GE-Model}.\textbf{ }

\subsection{Standard Error Formula Applied to the General Equilibrium Model}

Recall our assumption that the (repeated) cross-sectional data include $n$
i.i.d. observations $\left(  w_{i,t},c_{i,t}^{\ast},l_{i,t}^{\ast}%
,F_{i,t}\right)  $ for \textit{working} individuals from two periods $t=1,2$.
Here, $F_{i,t}$ denotes a dummy variable that is equal to one if the agent
chooses $S=F$ in the \textit{previous} period.\textbf{ }Recall that we use%
\begin{align*}
\frac{1}{\bar{n}_{1}}\sum_{i=1}^{\bar{n}_{1}}\log w_{i,1}^{F}-\frac{1}{\bar
{n}_{2}}\sum_{i=1}^{\bar{n}_{2}}\log w_{i,2}^{F}  &  =\dfrac{1}%
{\widehat{\alpha}_{1}}\left(  \log n_{1}^{F}-\log n_{2}^{F}\right) \\
\frac{1}{\bar{n}_{1}}\sum_{i=1}^{\bar{n}_{1}}\frac{c_{i,1}^{\ast}}%
{l_{i,1}^{\ast}}  &  =w_{1}^{F}\frac{\widehat{\sigma}}{1-\widehat{\sigma}}%
\end{align*}
as well as%
\begin{equation}
\widehat{\log\nu_{t}}=\widehat{\alpha}_{1}\left(  \log w_{t}^{F}-\log
w_{t}^{R}\right)  -\left(  \log n_{t}^{F}-\log n_{t}^{R}\right)  .
\label{pi2equation for estimation of as'}%
\end{equation}
The parameters $\varrho$ and $\omega^{2}$ can then be consistently estimated
by the time-series regression of the following equation:
\begin{equation}
\widehat{\log\nu_{t+1}}=\varrho\widehat{\log\nu_{t}}+\eta_{t}.
\label{pi2equation for estimation of rho'}%
\end{equation}
In addition to these equations, we will use the cross section variances of
$\log w_{i,1}^{F}$ and $\log w_{i,1}^{R}$ to estimate $\sigma_{F}^{2}$ and
$\sigma_{R}^{2}$. We also have the log likelihood from a sample of $n$
individuals (cross section) is \textbf{ }%
\[
\sum_{i=1}^{n}\left\{  F_{i,2}\log\left[  1-\Phi\left(  \log\left(
1-\Theta\right)  -\mu\right)  \right]  +\left(  1-F_{i,2}\right)  \log\left[
\Phi\left(  \log\left(  1-\Theta\right)  -\mu\right)  \right]  \right\}
\]
where $\Theta$ is constant across $i$ and given by
\begin{equation}
\Theta\equiv\dfrac{\log\left(  \frac{n_{2}^{F}}{n_{2}^{R}}\right)
+\frac{\sigma_{R}^{2}-\sigma_{F}^{2}}{2}+\varrho\log\nu_{1}}{\frac
{\sigma\left(  \sigma_{R}^{2}-\sigma_{F}^{2}+\omega^{2}\right)  }{2\alpha_{1}%
}}. \label{pi2cap-theta-def}%
\end{equation}
The moments employed in the estimation of $\alpha_{1}$ and $\sigma$ take the
following form:%
\begin{align*}
\frac{1}{\bar{n}_{1}}\sum_{i=1}^{\bar{n}_{1}}\log w_{i,1}^{F}-\frac{1}{\bar
{n}_{2}}\sum_{i=1}^{\bar{n}_{2}}\log w_{i,2}^{F}  &  =\dfrac{1}%
{\widehat{\alpha}_{1}}\left(  \log n_{1}^{F}-\log n_{2}^{F}\right) \\
\frac{1}{\bar{n}_{1}}\sum_{i=1}^{\bar{n}_{1}}\frac{c_{i,1}^{\ast}}%
{l_{i,1}^{\ast}}  &  =w_{1}^{F}\frac{\widehat{\sigma}}{1-\widehat{\sigma}}%
\end{align*}
To simplify notation we introduce two redundant parameters $\delta_{1}$ and
$\delta_{2}$%
\[
\frac{1}{\bar{n}_{1}}\sum_{i=1}^{\bar{n}_{1}}\log w_{i,1}^{F}=\widehat{\delta
}_{1},\quad\frac{1}{\bar{n}_{2}}\sum_{i=1}^{\bar{n}_{2}}\log w_{i,2}%
^{F}=\widehat{\delta}_{2}%
\]
and understand
\begin{equation}
\widehat{\alpha}_{1}=\frac{\log n_{1}^{F}-\log n_{2}^{F}}{\widehat{\delta}%
_{1}-\widehat{\delta}_{2}}. \label{pi2alpha-alt}%
\end{equation}
Given that our asymptotics are based on $n\rightarrow\infty$, we need to
express moments in terms of $n$:%
\begin{align*}
\sum_{i=1}^{n}F_{i,1}\left(  \log w_{i,1}^{F}-\delta_{1}\right)   &  =0,\\
\sum_{i=1}^{n}F_{i,2}\left(  \log w_{i,2}^{F}-\delta_{2}\right)   &  =0,\\
\sum_{i=1}^{n}F_{i,1}\left(  \frac{c_{i,1}^{\ast}}{l_{i,1}^{\ast}}-w_{1}%
^{F}\frac{\sigma}{1-\sigma}\right)   &  =0.
\end{align*}
For the estimation of $\sigma_{F}^{2}=\sigma_{\varepsilon}^{2}$, we use the
fact that the second moment is the sum of the variance and the square of the
first moment and let%
\[
\sum_{i=1}^{n}F_{i,1}\left(  \left(  \log w_{i,1}^{F}\right)  ^{2}-\left(
\sigma_{F}^{2}+\delta_{1}^{2}\right)  \right)  =0.
\]
Likewise, for the estimation of $\sigma_{R}^{2}$,
\begin{align*}
\sum_{i=1}^{n}\left(  1-F_{i,1}\right)  \left(  \log w_{i,1}^{R}-\delta
_{3}\right)   &  =0,\\
\sum_{i=1}^{n}\left(  1-F_{i,1}\right)  \left(  \left(  \log w_{i,1}%
^{R}\right)  ^{2}-\left(  \sigma_{R}^{2}+\delta_{3}^{2}\right)  \right)   &
=0.
\end{align*}
For the estimation of the parameters $\rho$ and $\omega^{2}$, the OLS
estimator of $\varrho$ and the corresponding estimator for $\omega^{2}$
solve:
\[
\frac{1}{\tau}\sum_{t=1}^{\tau}\widehat{\log\nu_{t}}\left(  \widehat{\log
\nu_{t+1}}-\widehat{\varrho}\widehat{\log\nu_{t}}\right)  =0
\]
and
\[
\frac{1}{\tau}\sum_{t=1}^{\tau}\left(  \widehat{\log\nu_{t+1}}%
-\widehat{\varrho}\widehat{\log\nu_{t}}\right)  ^{2}=\widehat{\omega}^{2}.
\]
Replacing for $\widehat{\log\nu_{t+1}}$ and $\widehat{\log\nu_{t}}$ using
equation(\ref{pi2equation for estimation of as'}), as well as
(\ref{pi2alpha-alt}), we obtain the following two moment conditions:%
\begin{align*}
\sum_{t=1}^{\tau}  &  \left(
\begin{array}
[c]{c}%
\frac{\log n_{1}^{F}-\log n_{2}^{F}}{\delta_{1}-\delta_{2}}\left(  \log
w_{t}^{F}-\log w_{t}^{R}\right) \\
-\left(  \log n_{t}^{F}-\log n_{t}^{R}\right)
\end{array}
\right)  \times\\
&  \left(  \left(
\begin{array}
[c]{c}%
\frac{\log n_{1}^{F}-\log n_{2}^{F}}{\delta_{1}-\delta_{2}}\left(  \log
w_{t+1}^{F}-\log w_{t+1}^{R}\right) \\
-\left(  \log n_{t+1}^{F}-\log n_{t+1}^{R}\right)
\end{array}
\right)  -\varrho\left(
\begin{array}
[c]{c}%
\frac{\log n_{1}^{F}-\log n_{2}^{F}}{\delta_{1}-\delta_{2}}\left(  \log
w_{t}^{F}-\log w_{t}^{R}\right) \\
-\left(  \log n_{t}^{F}-\log n_{t}^{R}\right)
\end{array}
\right)  \right)  =0,
\end{align*}%
\[
\sum_{t=1}^{\tau}\left(  \left(  \left(
\begin{array}
[c]{c}%
\frac{\log n_{1}^{F}-\log n_{2}^{F}}{\delta_{1}-\delta_{2}}\left(  \log
w_{t+1}^{F}-\log w_{t+1}^{R}\right) \\
-\left(  \log n_{t+1}^{F}-\log n_{t+1}^{R}\right)
\end{array}
\right)  -\varrho\left(
\begin{array}
[c]{c}%
\frac{\log n_{1}^{F}-\log n_{2}^{F}}{\delta_{1}-\delta_{2}}\left(  \log
w_{t}^{F}-\log w_{t}^{R}\right) \\
-\left(  \log n_{t}^{F}-\log n_{t}^{R}\right)
\end{array}
\right)  \right)  ^{2}-\omega^{2}\right)  =0.
\]
For the rest of the parameters, we note that $F_{i,2}$ is chosen with
probability $1-\Phi\left(  \log\left(  1-\Theta\right)  -\mu\right)  $ for
\[
\Theta=\dfrac{\log\left(  \frac{n_{2}^{F}}{n_{2}^{R}}\right)  +\frac
{\sigma_{R}^{2}-\sigma_{F}^{2}}{2}+\varrho\log\nu_{1}}{\frac{\sigma\left(
\sigma_{R}^{2}-\sigma_{F}^{2}+\omega^{2}\right)  }{2\alpha_{1}}},
\]
so $\mu$ can be estimated by Probit MLE, where the FOC can be shown to be
\[
0=\sum_{i=1}^{n}\left\{  F_{i,2}-\left[  1-\Phi\left(  \log\left(
1-\Theta\right)  -\mu\right)  \right]  \right\}
\]
where\textbf{ }%
\begin{align*}
\Theta &  =\dfrac{\log\left(  \frac{n_{2}^{F}}{n_{2}^{R}}\right)
+\frac{\sigma_{R}^{2}-\sigma_{F}^{2}}{2}+\varrho\log\nu_{1}}{\frac
{\sigma\left(  \sigma_{R}^{2}-\sigma_{F}^{2}+\omega^{2}\right)  }{2\alpha_{1}%
}}\\
&  =\dfrac{\log\left(  \frac{n_{2}^{F}}{n_{2}^{R}}\right)  +\frac{\sigma
_{R}^{2}-\sigma_{F}^{2}}{2}+\varrho\left(  \frac{\log n_{1}^{F}-\log n_{2}%
^{F}}{\delta_{1}-\delta_{2}}\left(  \log w_{1}^{F}-\log w_{1}^{R}\right)
-\left(  \log n_{1}^{F}-\log n_{1}^{R}\right)  \right)  }{\frac{\sigma\left(
\sigma_{R}^{2}-\sigma_{F}^{2}+\omega^{2}\right)  }{2}\frac{\delta_{1}%
-\delta_{2}}{\log n_{1}^{F}-\log n_{2}^{F}}}%
\end{align*}
Here, we used the fact that
\begin{align*}
\log\nu_{1}  &  =\alpha_{1}\left(  \log w_{1}^{F}-\log w_{1}^{R}\right)
-\left(  \log n_{1}^{F}-\log n_{1}^{R}\right) \\
\alpha_{1}  &  =\frac{\log n_{1}^{F}-\log n_{2}^{F}}{\delta_{1}-\delta_{2}}%
\end{align*}

Based on the previous discussion, we can now present moments in the form of
(\ref{Moment Cond y}) and (\ref{Moment Cond z}). In our case, $\log\nu_{1}$ is
estimated with the aid of aggregate variables, so we have $\beta
=\theta=\left(  \mu,\delta_{1},\delta_{2},\sigma,\delta_{3},\sigma_{F}%
^{2},\sigma_{R}^{2}\right)  ^{\prime}$ and $\rho=\left(  \varrho,\omega
^{2}\right)  ^{\prime}$. We see that the cross sectional moments are%
\begin{align*}
\frac{1}{n}\sum_{i=1}^{n}F_{i,1}\left(  \log w_{i,1}^{F}-\delta_{1}\right)
&  =0,\\
\frac{1}{n}\sum_{i=1}^{n}F_{i,2}\left(  \log w_{i,2}^{F}-\delta_{2}\right)
&  =0,\\
\frac{1}{n}\sum_{i=1}^{n}F_{i,1}\left(  \frac{c_{i,1}^{\ast}}{l_{i,1}^{\ast}%
}-w_{1}^{F}\frac{\sigma}{1-\sigma}\right)   &  =0,\\
\frac{1}{n}\sum_{i=1}^{n}\left\{  F_{i,2}-\left[  1-\Phi\left(  \log\left(
1-\Theta\right)  -\mu\right)  \right]  \right\}   &  =0,
\end{align*}
\textbf{ }and%
\begin{align*}
\sum_{i=1}^{n}F_{i,1}\left(  \left(  \log w_{i,1}^{F}\right)  ^{2}-\left(
\sigma_{F}^{2}+\delta_{1}^{2}\right)  \right)   &  =0,\\
\sum_{i=1}^{n}\left(  1-F_{i,1}\right)  \left(  \log w_{i,1}^{R}-\delta
_{3}\right)   &  =0,\\
\sum_{i=1}^{n}\left(  1-F_{i,1}\right)  \left(  \left(  \log w_{i,1}%
^{R}\right)  ^{2}-\left(  \sigma_{R}^{2}+\delta_{3}^{2}\right)  \right)   &
=0,
\end{align*}
where
\[
\Theta=\dfrac{\log\left(  \frac{n_{2}^{F}}{n_{2}^{R}}\right)  +\frac
{\sigma_{R}^{2}-\sigma_{F}^{2}}{2}+\varrho\left(  \frac{\log n_{1}^{F}-\log
n_{2}^{F}}{\delta_{1}-\delta_{2}}\left(  \log w_{1}^{F}-\log w_{1}^{R}\right)
-\left(  \log n_{1}^{F}-\log n_{1}^{R}\right)  \right)  }{\frac{\sigma\left(
\sigma_{R}^{2}-\sigma_{F}^{2}+\omega^{2}\right)  }{2}\frac{\delta_{1}%
-\delta_{2}}{\log n_{1}^{F}-\log n_{2}^{F}}},
\]
and the time series moments are%
\begin{align*}
\frac{1}{\tau}\sum_{t=1}^{\tau}\log\nu_{t}\left(  \log\nu_{t+1}-\varrho\log
\nu_{t}\right)   &  =0,\\
\frac{1}{\tau}\sum_{t=1}^{\tau}\left(  \left(  \log\nu_{t+1}-\varrho\log
\nu_{t}\right)  ^{2}-\omega^{2}\right)   &  =0,
\end{align*}
where%
\[
\log\nu_{t}=\frac{\log n_{1}^{F}-\log n_{2}^{F}}{\delta_{1}-\delta_{2}}\left(
\log w_{t}^{F}-\log w_{t}^{R}\right)  -\left(  \log n_{t}^{F}-\log n_{t}%
^{R}\right)  .
\]
Letting
\begin{equation}
f_{\theta,i}\left(  \theta,\rho\right)  =\left[
\begin{array}
[c]{c}%
F_{i,1}\left(  \log w_{i,1}^{F}-\delta_{1}\right) \\
F_{i,1}\left(  \frac{c_{i,1}^{\ast}}{l_{i,1}^{\ast}}-w_{1}^{F}\frac{\sigma
}{1-\sigma}\right) \\
F_{i,1}\left(  \left(  \log w_{i,1}^{F}\right)  ^{2}-\left(  \sigma_{F}%
^{2}+\delta_{1}^{2}\right)  \right) \\
\left(  1-F_{i,1}\right)  \left(  \log w_{i,1}^{R}-\delta_{3}\right) \\
\left(  1-F_{i,1}\right)  \left(  \left(  \log w_{i,1}^{R}\right)
^{2}-\left(  \sigma_{R}^{2}+\delta_{3}^{2}\right)  \right) \\
F_{i,2}\left(  \log w_{i,2}^{F}-\delta_{2}\right) \\
F_{i,2}-\left[  1-\Phi\left(  \log\left(  1-\Theta\right)  -\mu\right)
\right]
\end{array}
\right]  , \label{pi2f-for-us}%
\end{equation}
and%
\begin{equation}
g_{\rho,t}\left(  \beta,\rho\right)  =\left[
\begin{array}
[c]{c}%
\log\nu_{t}\left(  \log\nu_{t+1}-\varrho\log\nu_{t}\right) \\
\left(  \log\nu_{t+1}-\varrho\log\nu_{t}\right)  ^{2}-\omega^{2}%
\end{array}
\right]  \label{pi2g-for-us}%
\end{equation}
we can compute%
\[
\hat{\Omega}_{f}=\frac{1}{n}\sum_{i=1}^{n}f_{\theta,i}f_{\theta,i}^{\prime}%
\]
and
\[
\hat{\Omega}_{g}=\tau^{-1}\sum_{t=1}^{\tau}g_{\rho,t}g_{\rho,t}^{\prime}%
\]
and
\begin{equation}
\hat{W}=\left[
\begin{array}
[c]{cc}%
\frac{1}{n}\hat{\Omega}_{f} & 0\\
0 & \frac{1}{\tau}\hat{\Omega}_{g}%
\end{array}
\right]  . \label{pi2W-for-us}%
\end{equation}

We are now ready to describe the five steps required in the computation of
test statistics and confidence intervals for the general equilibrium model. As
a first step, let $\theta=\beta=\left(  \mu,\delta_{1},\delta_{2}%
,\sigma,\delta_{3},\sigma_{F}^{2},\sigma_{R}^{2}\right)  ^{\prime}$ and
$\rho=\left(  \varrho,\omega^{2}\right)  ^{\prime}$. Observe that the
aggregate shock is not in the set of estimated parameters, since the general
equilibrium model implies that $\log\nu_{t}=\alpha_{1}\left(  \log w_{t}%
^{F}-\log w_{t}^{R}\right)  -\left(  \log n_{t}^{F}-\log n_{t}^{R}\right)  $.
In the second, third, and fourth steps compute the matrices $\mathbf{A}$,
$\hat{\Omega}_{f}$, $\hat{\Omega}_{g}$, and $W$ using the vectors of moments
$f_{\theta,i}$ and $g_{\rho,t}$ derived above. In the last step, calculate the
variance matrix $V=\mathbf{A}^{-1}W\left(  \mathbf{A}^{\prime}\right)  ^{-1}$
and form related t-ratios and confidence intervals.

\subsection{Limiting Distributions}

We first consider the portfolio choice problem in Section
\ref{portfolio-example}. In this example, the time series log likelihood is
given by
\[
\tau^{-1}\sum_{s=1}^{\tau}\log\left(  \phi\left(  \left(  \nu_{t}-\mu\right)
/\sigma_{\nu}\right)  /\sigma_{\nu}\right)
\]
where $\phi$ is the PDF of $N\left(  0,1\right)  $. The likelihood is
maximized that $\hat{\mu}=\tau^{-1}\sum_{s=1}^{\tau}\nu_{t}$ and $\hat{\sigma
}_{\nu}^{2}=\tau^{-1}\sum_{s=1}^{\tau}\left(  \nu_{t}-\hat{\mu}\right)  ^{2}.$
The cross-sectional likelihood is given by
\[
n^{-1}\sum_{i=1}^{n}\log\left(  \phi\left(  \left(  u_{i1}-\nu_{1}\right)
/\sigma_{\epsilon}\right)  /\sigma_{\epsilon}\right)  +n^{-1}\sum_{i=1}%
^{n}\log\left(  \phi\left(  \left(  \alpha_{i1}-\alpha\right)  /\sigma
_{e}\right)  /\sigma_{e}\right)
\]
where $\alpha=\left(  \delta\left(  \sigma_{\epsilon}^{2}+\sigma_{\nu}%
^{2}\right)  +r-\mu\right)  /\delta\left(  \sigma_{\epsilon}^{2}+\sigma_{\nu
}^{2}\right)  $. For given values of $\mu,r,$and $\sigma_{\nu}^{2}$ there is a
one-to-one mapping between the parameters $\left(  \delta,\sigma_{\epsilon
}^{2},\sigma_{e}^{2},\nu_{1}\right)  $ and $\left(  \alpha,\sigma_{\epsilon
}^{2},\sigma_{e}^{2},\nu_{1}\right)  .$ Maximizing the likelihood with respect
to $\left(  \delta,\sigma_{\epsilon}^{2},\sigma_{e}^{2},\nu_{1}\right)  $ is
thus equivalent to maximizing the likelihood with respect to $\left(
\alpha,\sigma_{\epsilon}^{2},\sigma_{e}^{2},\nu_{1}\right)  $ and then solving
for $\left(  \delta,\sigma_{\epsilon}^{2},\sigma_{e}^{2},\nu_{1}\right)  $.
The maximizer for $\left(  \alpha,\sigma_{\epsilon}^{2},\sigma_{e}^{2},\nu
_{1}\right)  $ is the standard MLE of the normal distribution for mean and
variance, $\hat{\nu}_{1}=n^{-1}\sum_{i=1}^{n}u_{i1}$, $\hat{\alpha}=n^{-1}%
\sum_{i=1}^{n}\alpha_{i1}$, $\hat{\sigma}_{\epsilon}^{2}=n^{-1}\sum_{i=1}%
^{n}\left(  u_{i1}-\hat{\nu}_{1}\right)  ^{2}$ and $\hat{\sigma}_{e}%
=n^{-1}\sum_{i=1}^{n}\left(  \alpha_{i1}-\hat{\alpha}\right)  ^{2}$. The
limiting distributions of these estimators are given by
\[
\tau^{1/2}\left(
\begin{array}
[c]{c}%
\hat{\mu}-\mu\\
\hat{\sigma}_{\nu}^{2}-\sigma_{\nu}^{2}%
\end{array}
\right)  \rightarrow_{d}N\left(  0,\left[
\begin{array}
[c]{cc}%
\sigma_{\nu}^{2} & 0\\
0 & 2\sigma_{\nu}^{2}%
\end{array}
\right]  \right)  ,
\]
and
\[
n^{1/2}\left(
\begin{array}
[c]{c}%
\hat{\alpha}-\alpha\\
\hat{\sigma}_{\epsilon}^{2}-\sigma_{\epsilon}^{2}\\
\hat{\sigma}_{e}^{2}-\sigma_{e}^{2}\\
\hat{\nu}_{1}-\nu_{1}%
\end{array}
\right)  \rightarrow_{d}N\left(  0,\left[
\begin{array}
[c]{cccc}%
\sigma_{e}^{2} & 0 & 0 & 0\\
0 & 2\sigma_{\epsilon}^{2} & 0 & 0\\
0 & 0 & 2\sigma_{e}^{2} & 0\\
0 & 0 & 0 & \sigma_{\epsilon}^{2}%
\end{array}
\right]  \right)  .
\]
From the results in Hahn, Kuersteiner, and Mazzocco (2016) the convergence of
the two vectors is joint, with asymptotic independence between cross-section
and time series parameters, and stable with respect to $\nu_{1}$. However,
because of the particularly simple nature of the model the limiting
distributions are conventional Gaussian limits with fixed variances. To obtain
the limiting distribution of $\hat{\delta}$ one now simply applies the delta
method and the continuous mapping theorem. More specifically, we have
$\hat{\delta}=\left(  \hat{\mu}-r\right)  /\left(  \left(  \hat{\sigma
}_{\epsilon}^{2}+\hat{\sigma}_{\nu}^{2}\right)  \left(  1-\hat{\alpha}\right)
\right)  $ and
\begin{align}
n^{-1/2}\left(  \hat{\delta}-\delta\right)   &  =\frac{\mu-r}{\left(
\sigma_{\epsilon}^{2}+\sigma_{\nu}^{2}\right)  \left(  1-\alpha\right)  ^{2}%
}n^{1/2}\left(  \hat{\alpha}-\alpha\right)  -\frac{\mu-r}{\left(
\sigma_{\epsilon}^{2}+\sigma_{\nu}^{2}\right)  ^{2}\left(  1-\alpha\right)
}n^{1/2}\left(  \hat{\sigma}_{\epsilon}^{2}-\sigma_{\epsilon}^{2}\right)
\label{delta_expansion}\\
&  +\frac{1}{\left(  \sigma_{\epsilon}^{2}+\sigma_{\nu}^{2}\right)  \left(
1-\alpha\right)  }\sqrt{\frac{n}{\tau}}\tau^{1/2}\left(  \hat{\mu}-\mu\right)
-\frac{\mu-r}{\left(  \sigma_{\epsilon}^{2}+\sigma_{\nu}^{2}\right)
^{2}\left(  1-\alpha\right)  }\sqrt{\frac{\tau}{n}}\tau^{1/2}\left(
\hat{\sigma}_{\nu}^{2}-\sigma_{\nu}^{2}\right)  +o_{p}\left(  1\right)
,\nonumber
\end{align}
leading to a limiting distribution of $\hat{\delta}$ given by
\[
n^{-1/2}\left(  \hat{\delta}-\delta\right)  \rightarrow_{d}N\left(
0,\frac{2\left(  1-\alpha\right)  ^{2}\left(  \mu-r\right)  ^{2}\left(
\sigma_{\epsilon}^{2}+\kappa\sigma_{\nu}^{2}\right)  +\left(  \sigma
_{\epsilon}^{2}+\sigma_{\nu}^{2}\right)  ^{2}\left(  \left(  \mu-r\right)
^{2}\sigma_{e}^{2}+\left(  1-\alpha\right)  ^{2}\kappa\sigma_{\nu}^{2}\right)
}{\left(  1-\alpha\right)  ^{4}\left(  \sigma_{\epsilon}^{2}+\sigma_{\nu}%
^{2}\right)  ^{4}}\right)
\]
where $\kappa=\lim\frac{n}{\tau}$ and the variance formula uses the fact that
the four components in (\ref{delta_expansion}) are asymptotically independent.
The formula for the variance is indicative of the fact that first step
estimation of the time series parameters can be ignored if $\tau$ is much
larger than $n,$ such that $\kappa$ is close to zero. However, this is an
unlikely scenario given that cross-sectional samples tend to be quite large.

We now consider the general equilibrium example. It is useful to analyze the
form of the limiting distribution of a set of GMM estimators based on $f$ and
$g.$ Define the empirical moment functions as
\[
h_{n}\left(  \theta,\rho\right)  =n^{-1}\sum_{i=1}^{n}f_{\theta,i}\left(
\theta,\rho\right)  ,\text{ }k_{\tau}\left(  \beta,\rho\right)  =\tau^{-1}%
\sum_{t=\tau_{0}+1}^{\tau_{0}+\tau}g_{\rho,t}\left(  \beta,\rho\right)  .
\]
and the moment based criterion functions $F_{n}\left(  \theta,\rho\right)
=-h_{n}\left(  \theta,\rho\right)  ^{\prime}\hat{\Omega}_{y}^{-1}h_{n}\left(
\theta,\rho\right)  $ and $G_{\tau}\left(  \beta,\rho\right)  =-k_{\tau
}\left(  \beta,\rho\right)  ^{\prime}\hat{\Omega}_{\nu}^{-1}k_{\tau}\left(
\beta,\rho\right)  .$The estimators then are defined as the solution $\left(
\hat{\theta},\hat{\rho}\right)  $ to
\begin{align*}
\frac{\partial F_{n}\left(  \hat{\theta},\hat{\rho}\right)  }{\partial\theta}
&  =0\\
\frac{\partial G_{\tau}\left(  \hat{\beta},\hat{\rho}\right)  }{\partial\rho}
&  =0.
\end{align*}
Because the GMM estimators are exactly identified in our example these
equations reduce to
\begin{align*}
h_{n}\left(  \hat{\theta},\hat{\rho}\right)   &  =0\\
k_{\tau}\left(  \hat{\beta},\hat{\rho}\right)   &  =0.
\end{align*}
We focus on the just identified case and refer the reader to our companion
paper Hahn, Kuersteiner and Mazzocco (2016) for a general treatment. The
limiting distribution of $\hat{\theta},\hat{\rho}$ depends on the joint
limiting distribution of $h_{n}\left(  \theta_{0},\rho_{0}\right)  $ and
$k_{\tau}\left(  \beta_{0},\rho_{0}\right)  .$

Recall $\log w_{it}^{F}=\alpha_{1}^{-1}\left(  \log n_{t}^{F}+\log\sigma+\log
T-\alpha_{0}\right)  +\varepsilon_{it}^{F}$ such that
\[
\delta_{1}=\alpha_{1}^{-1}\left(  \log n_{1}^{F}+\log\sigma+\log T-\alpha
_{0}\right)  -\frac{\sigma_{F}^{2}}{2}.
\]
Similarly, let $\delta_{2}=\alpha_{1}^{-1}\left(  \log n_{2}^{F}+\log
\sigma+\log T-\alpha_{0}\right)  -\sigma_{F}^{2}/2,$%
\[
\delta_{3}=\alpha_{1}^{-1}\left(  \log n_{1}^{R}+\log\sigma+\log T-\alpha
_{0}-\log\nu_{1}\right)  -\frac{\sigma_{R}^{2}}{2}%
\]
and define $p\left(  \Theta\right)  =\Phi\left(  \log\left(  1-\Theta\right)
-\mu\right)  .$ Let $\mathcal{C}$ be the $\sigma$-field generated by $\log
n_{1}^{R},\log n_{1}^{F},\log n_{2}^{F}$ and $\log\nu_{1}$ such that $\Theta,$
$w_{1}^{F},$ $\delta_{1},\delta_{2}$ and $\delta_{3}$ are measurable with
respect to $\mathcal{C}.$ Based on the theory in our companion paper, the
moment functions converge jointly and stably to independent mixed Gaussian
limits
\[
n^{1/2}h_{n}\left(  \theta_{0},\rho_{0}\right)  \rightarrow_{d}\Omega
_{f}^{1/2}\xi_{h}\sim N\left(  0,\Omega_{f}\right)  \text{ }\left(
\mathcal{C}\text{-stably}\right)
\]
where $\xi_{h}\sim N\left(  0,I\right)  $ and is independent of any
$\mathcal{C}$-measurable random variable,%
\begin{align*}
\Omega_{f,1}  &  =\left[
\begin{array}
[c]{ccc}%
p\left(  \bar{\Theta}_{1}\right)  \sigma_{F}^{2} & p\left(  \bar{\Theta}%
_{1}\right)  \frac{w_{1}^{F}\sigma}{1-\sigma}\sigma_{F}^{2} & 2\delta
_{1}\sigma_{F}^{2}\\
p\left(  \bar{\Theta}_{1}\right)  \frac{w_{1}^{F}\sigma}{1-\sigma}\sigma
_{F}^{2} & p\left(  \bar{\Theta}_{1}\right)  \left(  \frac{w_{1}^{F}\sigma
}{1-\sigma}\right)  ^{2}\left(  e^{\sigma_{F}^{2}}-1\right)  & p\left(
\bar{\Theta}_{1}\right)  \frac{w_{1}^{F}\sigma}{1-\sigma}\left(  2\delta
_{1}+1\right)  \sigma_{\epsilon}^{2}\\
2\delta_{1}\sigma_{F}^{2} & p\left(  \bar{\Theta}_{1}\right)  \frac{w_{1}%
^{F}\sigma}{1-\sigma}\left(  2\delta_{1}+1\right)  \sigma_{F}^{2} & p\left(
\bar{\Theta}_{1}\right)  \left(  2\sigma_{\epsilon}^{4}+4\delta_{1}^{2}%
\sigma_{F}^{2}\right)
\end{array}
\right]  ,\\
\Omega_{f,2}  &  =\left[
\begin{array}
[c]{cc}%
\left(  1-p\left(  \bar{\Theta}_{1}\right)  \right)  \sigma_{R}^{2} &
2\delta_{3}\sigma_{R}^{2}\\
2\delta_{3}\sigma_{R}^{2} & \left(  1-p\left(  \bar{\Theta}_{1}\right)
\right)  \left(  2\sigma_{R}^{4}+4\delta_{3}^{2}\sigma_{R}^{2}\right)
\end{array}
\right]  ,\\
\Omega_{f,3}  &  =\left[
\begin{array}
[c]{cc}%
p\left(  \bar{\Theta}_{1}\right)  \sigma_{\epsilon}^{2} & 0\\
0 & p\left(  \bar{\Theta}_{2}\right)  \left(  1-p\left(  \bar{\Theta}%
_{2}\right)  \right)
\end{array}
\right]
\end{align*}
and
\[
\Omega_{f}=\left[
\begin{array}
[c]{ccc}%
\Omega_{f,1} & 0 & 0\\
0 & \Omega_{f,2} & 0\\
0 & 0 & \Omega_{f,3}%
\end{array}
\right]  .
\]
Here, we let
\[
\bar{\Theta}_{t}\equiv\dfrac{\log\left(  \frac{n_{t}^{F}}{n_{t}^{R}}\right)
+\frac{\left(  \pi_{2}^{2}-1\right)  \sigma_{\varepsilon}^{2}}{2}+\varrho
\log\nu_{t-1}}{\frac{\sigma\left(  \sigma_{R}^{2}-\sigma_{F}^{2}+\omega
^{2}\right)  }{2\alpha_{1}}}%
\]
for clarity.\textbf{ }For the time series sample it is straight forward to see
that under suitable regularity conditions%
\[
\tau^{1/2}k_{\tau}\left(  \beta_{0},\rho_{0}\right)  \rightarrow_{d}\Omega
_{g}^{1/2}\xi_{k}\sim N\left(  0,\Omega_{g}\right)  \text{ }\left(
\mathcal{C}\text{-stably}\right)
\]
where $\xi_{k}\sim N\left(  0,I\right)  $ and independent of any $\mathcal{C}%
$-measurable random variable and
\[
\Omega_{g}=\left[
\begin{array}
[c]{cc}%
\frac{\omega^{4}}{1-\varrho_{0}^{2}} & 0\\
0 & 2\omega^{4}%
\end{array}
\right]  .
\]
The results in Hahn, Kuersteiner and Mazzocco (2016) imply that $\xi_{h}$ and
$\xi_{k}$ are independent Gaussian random variables conditional on
$\mathcal{C}$. The explicit formulas make clear that in this model the
limiting variance does depend on macro variables including common shocks and
other observables. Since these variables remain random in the limit as $n$ and
$\tau$ tend to infinity, the resulting limiting distribution is mixed Gaussian
and the convergence to the limit is joint with the macro variables or
$\mathcal{C}$-stable. The later is important because the influence matrix $A,$
as we show below, also depends on these same macro variables.

Next compute the limits%
\begin{align*}
A_{f,\theta}  &  =\operatorname*{plim}n^{-1}\sum_{i=1}^{n}\frac{\partial
f_{\theta,i}\left(  \theta_{0},\rho_{0}\right)  }{\partial\theta^{\prime}%
},\text{\ \ \ \ \ \ \ \ \ \ \ \ \ \ }A_{f,\rho}=\operatorname*{plim}n^{-1}%
\sum_{i=1}^{n}\frac{\partial f_{\theta,i}\left(  \theta_{0},\rho_{0}\right)
}{\partial\rho^{\prime}},\\
A_{g,\theta}  &  =\operatorname*{plim}\tau^{-1}\sum_{t=\tau_{0}+1}^{\tau
_{0}+\tau}\frac{\partial g_{\rho,t}\left(  \beta_{0},\rho_{0}\right)
}{\partial\theta^{\prime}},\text{\ \ \ \ \ \ \ \ \ \ \ }A_{g,\rho
}=\operatorname*{plim}\tau^{-1}\sum_{t=\tau_{0}+1}^{\tau_{0}+\tau}%
\frac{\partial g_{\rho,t}\left(  \beta_{0},\rho_{0}\right)  }{\partial
\rho^{\prime}}.
\end{align*}
First, letting $\dot{p}\left(  \Theta\right)  =\phi\left(  \log\left(
1-\Theta\right)  -\mu\right)  $ where $\phi$ is the PDF of $N\left(
0,1\right)  $, \renewcommand{\baselinestretch}{1.5}{\footnotesize
\[
A_{f,\theta}=\left[
\begin{array}
[c]{ccccccc}%
0 & -p\left(  \bar{\Theta}_{1}\right)  & 0 & 0 & 0 & 0 & 0\\
0 & 0 & 0 & -\frac{-w_{1}^{F}}{\left(  1-\sigma\right)  ^{2}}p\left(
\bar{\Theta}_{1}\right)  & 0 & 0 & 0\\
0 & -2\delta_{1}p\left(  \bar{\Theta}_{1}\right)  & 0 & 0 & 0 & -p\left(
\bar{\Theta}_{1}\right)  & 0\\
0 & 0 & 0 & 0 & -\left(  1-p\left(  \bar{\Theta}_{1}\right)  \right)  & 0 &
0\\
0 & 0 & 0 & 0 & -2\delta_{3}\left(  1-p\left(  \bar{\Theta}_{1}\right)
\right)  & 0 & -\left(  1-p\left(  \bar{\Theta}_{1}\right)  \right) \\
0 & 0 & -p\left(  \bar{\Theta}_{2}\right)  & 0 & 0 & 0 & 0\\
-\dot{p}\left(  \bar{\Theta}_{2}\right)  & -\frac{\dot{p}\left(  \bar{\Theta
}_{2}\right)  }{1-\bar{\Theta}_{2}}\frac{\partial\bar{\Theta}_{2}}%
{\partial\delta_{1}} & -\frac{\dot{p}\left(  \bar{\Theta}_{2}\right)  }%
{1-\bar{\Theta}_{2}}\frac{\partial\bar{\Theta}_{2}}{\partial\delta_{2}} &
-\frac{\dot{p}\left(  \bar{\Theta}_{2}\right)  }{1-\bar{\Theta}_{2}}%
\frac{\partial\bar{\Theta}_{2}}{\partial\sigma} & 0 & -\frac{\dot{p}\left(
\bar{\Theta}_{2}\right)  }{1-\bar{\Theta}_{2}}\frac{\partial\bar{\Theta}_{2}%
}{\partial\sigma_{F}^{2}} & -\frac{\dot{p}\left(  \bar{\Theta}_{2}\right)
}{1-\bar{\Theta}_{2}}\frac{\partial\bar{\Theta}_{2}}{\partial\sigma_{R}^{2}}%
\end{array}
\right]  ,
\]
\renewcommand{\baselinestretch}{1.5}}{\normalsize Next, consider the two
cross-derivative terms where the first one is given by%
\[
A_{f,\rho}=\left[
\begin{array}
[c]{cc}%
0 & 0\\
0 & 0\\
0 & 0\\
0 & 0\\
0 & 0\\
0 & 0\\
-\frac{\dot{p}\left(  \bar{\Theta}\right)  }{1-\bar{\Theta}}\frac{\partial
\bar{\Theta}}{\partial\varrho} & -\frac{\dot{p}\left(  \bar{\Theta}\right)
}{1-\bar{\Theta}}\frac{\partial\bar{\Theta}}{\partial\omega^{2}}%
\end{array}
\right]  .
\]
Next note that
\[
\log\nu_{t}=\frac{\log n_{1}^{F}-\log n_{2}^{F}}{\delta_{1}-\delta_{2}}\left(
\log w_{t}^{F}-\log w_{t}^{R}\right)  -\left(  \log n_{t}^{F}-\log n_{t}%
^{R}\right)
\]
such that $\partial\log\nu_{t}/\partial\theta$ is non-zero for elements
$\delta_{1}$ and $\delta_{2}$. For $\log\nu_{t}\left(  \log\nu_{t+1}%
-\varrho\log\nu_{t}\right)  $ the derivative $\left(  \partial\log\nu
_{t}/\partial\theta\right)  \left(  \log\nu_{t+1}-\varrho_{0}\log\nu
_{t}\right)  $ has zero expectation because $\left(  \log\nu_{t+1}-\varrho
_{0}\log\nu_{t}\right)  =\eta_{t}$. For $\left(  \log\nu_{t+1}-\varrho\log
\nu_{t}\right)  ^{2}-\omega^{2}$ we obtain partial derivatives equal to
$2\eta_{t}\left(  \partial\log\nu_{t+1}/\partial\theta-\varrho\partial\log
\nu_{t}/\partial\theta\right)  $. Since $\eta_{t}$ is orthogonal to all data
in $\log\nu_{t}$ it follows that $E\left[  \eta_{t}\left(  \partial\log
\nu_{t+1}/\partial\theta-\varrho\partial\log\nu_{t}/\partial\theta\right)
\right]  =E\left[  \eta_{t}\partial\log\nu_{t+1}/\partial\theta\right]  $.
Under suitable regularity conditions it then follows that sample averages
converge to these expectations, leading to%
\[
A_{g,\theta}=\left[
\begin{array}
[c]{ccccccc}%
0 & E\left[  \log\nu_{t}\left(  \frac{\partial\log\nu_{t+1}}{\partial
\delta_{1}}-\varrho\frac{\partial\log\nu_{t}}{\partial\delta_{1}}\right)
\right]  & E\left[  \log\nu_{t}\left(  \frac{\partial\log\nu_{t+1}}%
{\partial\delta_{2}}-\varrho\frac{\partial\log\nu_{t}}{\partial\delta_{2}%
}\right)  \right]  & 0 & 0 & 0 & 0\\
0 & 2E\left[  \eta_{t}\frac{\partial\log\nu_{t+1}}{\partial\delta_{1}}\right]
& 2E\left[  \eta_{t}\frac{\partial\log\nu_{t+1}}{\partial\delta_{2}}\right]  &
0 & 0 & 0 & 0
\end{array}
\right]  .
\]
Finally, straight forward calculations show that under suitable regularity
conditions ensuring a law of large numbers for an autoregressive process the
limits in $A_{g,\rho}$ are given by
\[
A_{g,\rho}=\left[
\begin{array}
[c]{cc}%
-\frac{\omega^{2}}{1-\varrho^{2}} & 0\\
0 & -1
\end{array}
\right]  .
\]
The limiting distribution of $\hat{\theta}$ is a consequence of Hahn,
Kuersteiner and Mazzocco (2016), Theorem 2 and Corollary 2. Using the notation
developed here we have
\[
\sqrt{n}\left(  \hat{\theta}-\theta_{0}\right)  \overset{d}{\rightarrow
}-A^{f,\theta}\Omega_{f}^{1/2}\xi_{h}-\sqrt{\kappa}A^{g,\rho}\Omega_{g}%
^{1/2}\xi_{k}\text{ (}\mathcal{C}\text{-stably)}%
\]
where
\begin{align*}
A^{f,\theta}  &  =A_{f,\theta}^{-1}+A_{f,\theta}^{-1}A_{f,\rho}\left(
A_{g,\rho}-A_{g,\theta}A_{f,\theta}^{-1}A_{f,\rho}\right)  ^{-1}A_{g,\theta
}A_{f,\theta}^{-1}\\
A^{g,\rho}  &  =-A_{f,\theta}^{-1}A_{f,\rho}\left(  A_{g,\rho}-A_{g,\theta
}A_{f,\theta}^{-1}A_{f,\rho}\right)  ^{-1}.
\end{align*}
The limiting distribution of $\hat{\theta}$ is mixed Gaussian $N\left(
0,\Omega_{\theta}\right)  $, with random weight matrix $\Omega_{\theta
}=A^{f,\theta}\Omega_{f}A^{f,\theta\prime}+\kappa A^{g,\rho}\Omega
_{g}A^{g,\rho\prime}$ where we have shown how the elements of $A$ and
$\Omega_{f}$ depend on macro variables and unobserved macro shocks. Similarly,
the limiting distribution of $\hat{\rho}$ is also mixed Gaussian and can be
derived in a similar fashion. }

\section{Proof of (\ref{education inequality misspecified})\label{sec-misspecification}}

Suppose that our econometrician tries to estimate $\mu$ using
only cross-section data sets misspecifies the model and assumes that the
difference in the labor demand functions of the two types of firms is not due
to the aggregate shock, but to different intercepts, i.e.,
\begin{align*}
\log H_{t+1}^{D,F}  &  =\alpha_{0}+\alpha_{1}\log w_{t+1}^{F}\\
\log H_{t+1}^{D,R}  &  =\alpha_{0}^{\prime}+\alpha_{1}\log w_{t+1}^{R}%
\end{align*}
with $\alpha_{0}\neq\alpha_{0}^{\prime}$. The equilibrium wages are then
\begin{align}
\log w_{t+1}^{F}  &  =\dfrac{\log n_{t+1}^{F}+\log\sigma+\log T-\alpha_{0}%
}{\alpha_{1}},\nonumber\\
\log w_{t+1}^{R}  &  =\dfrac{\log n_{t+1}^{R}+\log\sigma+\log T-\alpha
_{0}^{\prime}}{\alpha_{1}}, \label{equi-wages}%
\end{align}
and as a consequence, equation (\ref{sector-choice-3}) is changed to%
\begin{align*}
&  \left(  n_{t+1}^{F}\right)  ^{\sigma\left(  1-\gamma\right)  /\alpha_{1}%
}\left[  \left(  \left(  \dfrac{1}{e^{\alpha_{0}}}\right)  ^{1/\alpha_{1}%
}\right)  ^{\sigma}\right]  ^{1-\gamma}E\left[  \exp\left(  \sigma\left(
1/\alpha_{1}\right)  \left(  1-\gamma\right)  \varepsilon_{t+1}^{F}\right)
\right] \\
&  \geq\left(  n_{t+1}^{R}\right)  ^{\sigma\left(  1-\gamma\right)
/\alpha_{1}}\left[  \left(  \left(  \dfrac{1}{e^{\alpha_{0}^{\prime}}}\right)
^{1/\alpha_{1}}\right)  ^{\sigma}\right]  ^{1-\gamma}E\left[  \exp\left(
\sigma\left(  1/\alpha_{1}\right)  \left(  1-\gamma\right)  \varepsilon
_{t+1}^{R}\right)  \right]  .
\end{align*}
Note that%
\begin{align*}
E\left[  \exp\left(  \sigma\left(  1/\alpha_{1}\right)  \left(  1-\gamma
\right)  \varepsilon_{t+1}^{F}\right)  \right]   &  =\exp\left(  -\frac
{\sigma\left(  1/\alpha_{1}\right)  \left(  1-\gamma\right)  }{2}%
\sigma_{\varepsilon}^{2}\right)  \exp\left(  \frac{\left(  \sigma\left(
1-\gamma\right)  \left(  1/\alpha_{1}\right)  \right)  ^{2}}{2}\sigma_{F}%
^{2}\right)  ,\\
E\left[  \exp\left(  \sigma\left(  1/\alpha_{1}\right)  \left(  1-\gamma
\right)  \varepsilon_{t+1}^{R}\right)  \right]   &  =\exp\left(  -\frac
{\sigma\left(  1/\alpha_{1}\right)  \left(  1-\gamma\right)  }{2}\pi_{2}%
^{2}\sigma_{\varepsilon}^{2}\right)  \exp\left(  \frac{\left(  \sigma\left(
1-\gamma\right)  \left(  1/\alpha_{1}\right)  \right)  ^{2}}{2}\sigma_{R}%
^{2}\right)  ,
\end{align*}
and
\begin{align*}
\left[  \left(  \left(  \dfrac{1}{e^{\alpha_{0}}}\right)  ^{1/\alpha_{1}%
}\right)  ^{\sigma}\right]  ^{1-\gamma}\exp\left(  -\frac{\sigma\left(
1/\alpha_{1}\right)  \left(  1-\gamma\right)  }{2}\sigma_{F}^{2}\right)   &
=\exp\left(  -\sigma\left(  1/\alpha_{1}\right)  \left(  1-\gamma\right)
\widetilde{\alpha}_{0}\right)  ,\\
\left[  \left(  \left(  \dfrac{1}{e^{\alpha_{0}^{\prime}}}\right)
^{1/\alpha_{1}}\right)  ^{\sigma}\right]  ^{1-\gamma}\exp\left(  -\frac
{\sigma\left(  1/\alpha_{1}\right)  \left(  1-\gamma\right)  }{2}\sigma
_{R}^{2}\right)   &  =\exp\left(  -\sigma\left(  1/\alpha_{1}\right)  \left(
1-\gamma\right)  \widetilde{\alpha}_{0}^{\prime}\right)  ,
\end{align*}
where
\[
\widetilde{\alpha}_{0}=\alpha_{0}+\frac{1}{2}\sigma_{F}^{2}=\alpha
_{0}-E\left[  \varepsilon_{t+1}^{F}\right]  ,\quad\widetilde{\alpha}%
_{0}^{\prime}=\alpha_{0}^{\prime}+\frac{1}{2}\sigma_{R}^{2}=\alpha_{0}%
^{\prime}-E\left[  \varepsilon_{t+1}^{R}\right]  .
\]
Therefore, the econometrician will conclude that $F$ is chosen if%
\begin{multline*}
\left(  n_{t+1}^{F}\right)  ^{\sigma\left(  1-\gamma\right)  /\alpha_{1}}%
\exp\left(  -\sigma\left(  1/\alpha_{1}\right)  \left(  1-\gamma\right)
\widetilde{\alpha}_{0}\right) \\
\geq\left(  n_{t+1}^{R}\right)  ^{\sigma\left(  1-\gamma\right)  /\alpha_{1}%
}\exp\left(  -\sigma\left(  1/\alpha_{1}\right)  \left(  1-\gamma\right)
\widetilde{\alpha}_{0}^{\prime}\right)  \exp\left(  \frac{\left(
\sigma\left(  1-\gamma\right)  \left(  1/\alpha_{1}\right)  \right)  ^{2}}%
{2}\left(  \sigma_{R}^{2}-\sigma_{F}^{2}\right)  \right)
\end{multline*}
when $1-\gamma>0$, and to%
\begin{multline*}
\left(  n_{t+1}^{F}\right)  ^{\sigma\left(  1-\gamma\right)  /\alpha_{1}}%
\exp\left(  -\sigma\left(  1/\alpha_{1}\right)  \left(  1-\gamma\right)
\widetilde{\alpha}_{0}\right) \\
\leq\left(  n_{t+1}^{R}\right)  ^{\sigma\left(  1-\gamma\right)  /\alpha_{1}%
}\exp\left(  -\sigma\left(  1/\alpha_{1}\right)  \left(  1-\gamma\right)
\widetilde{\alpha}_{0}^{\prime}\right)  \exp\left(  \frac{\left(
\sigma\left(  1-\gamma\right)  \left(  1/\alpha_{1}\right)  \right)  ^{2}}%
{2}\left(  \sigma_{R}^{2}-\sigma_{F}^{2}\right)  \right)
\end{multline*}
when $1-\gamma<0$. This implies that $F$ is chosen if%
\begin{equation}
\gamma\geq1-\frac{\log\left(  \frac{n_{t+1}^{F}}{n_{t+1}^{R}}\right)  +\left(
\widetilde{\alpha}_{0}^{\prime}-\widetilde{\alpha}_{0}\right)  }{\frac
{\sigma\left(  \pi_{2}^{2}-1\right)  \sigma_{\varepsilon}^{2}}{2\alpha_{1}}}.
\label{education-choice-F-misspecification}%
\end{equation}
Note that%
\[
\widetilde{\alpha}_{0}^{\prime}-\widetilde{\alpha}_{0}=\alpha_{0}^{\prime
}-\alpha_{0}+\frac{1}{2}\left(  \sigma_{R}^{2}-\sigma_{F}^{2}\right)  .
\]

We now argue that $\alpha_{0}^{\prime}-\alpha_{0}$ above should
be understood to be equal to $\log v_{t+1}$. Note that the econometrician can
estimate $\alpha_{1}$ consistently using equation
(\ref{equation for estimation of a1}), which is based on cross-section
variation. The econometrician can also estimate $\alpha_{0}^{\prime}%
-\alpha_{0}$ consistently by $\widehat{\alpha}_{1}\left(  \log w_{t+1}%
^{F}-\log w_{t+1}^{R}\right)  -\left(  \log n_{t+1}^{F}-\log n_{t+1}%
^{R}\right)  $. Comparing with (\ref{equation for estimation of as}), we
conclude that the econometrician's estimator is exactly equal to our earlier
estimator of $\log\nu_{t+1}$. This is a natural consequence of the nature of
the econometrician's misspecification, who assumes that the difference in the
equilibrium wages in (\ref{equi-wages}) reflects the difference of intercepts
of the labor demand functions. However, this assumption is incorrect and the
difference of the intercepts is due to the aggregate shock, i.e, $\alpha
_{0}^{\prime}=\alpha_{0}+\log v_{t+1}$.

It follows that the econometrician's conclusion
(\ref{education-choice-F-misspecification}) above can be equivalently written
with $\alpha_{0}^{\prime}-\alpha_{0}$ replaced by $\log v_{t+1}$, which
establishes (\ref{education inequality misspecified}).

\section{Censored versus Truncated Results\label{Censored versus Truncated Results}}

As mentioned in the main text, to perform the Monte Carlo
exercise we have to deal with a technical issue. The estimation of the risk
aversion parameter $\mu$ in the general equilibrium model requires the
computation of $\log\left(  1-\Theta\right)  $ where
\[
\Theta\equiv\dfrac{\log\left(  \frac{n_{2}^{F}}{n_{2}^{R}}\right)
+\frac{\sigma_{R}^{2}-\sigma_{F}^{2}}{2}+\varrho\log\nu_{1}}{\frac
{\sigma\left(  \sigma_{R}^{2}-\sigma_{F}^{2}+\omega^{2}\right)  }{2\alpha_{1}%
}}%
\]
In the model, $\Theta$ is always smaller than 1 and, hence, $\log\left(
1-\Theta\right)  $ is always well defined. In the estimation of $\mu$,
however, the true parameters included in $\Theta$ are replaced with their
estimated values. In some of the Monte Carlo repetitions, the randomness of
the estimated parameters generates values of $\Theta$ that are greater than 1,
which implies that $\log\left(  1-\Theta\right)  $ is not well define. We deal
with this issue by presenting two sets of results. A first set in which we
only use Monte Carlo runs in which $\Theta<1$. We will refer to these results
as the \textquotedblleft truncated\textquotedblleft\ results. A second set in
which we set $\Theta=0.99$ if $\Theta>1$ and report our findings using all the
Monte Carlo runs. We will refer to the second set as the \textquotedblright
censored\textquotedblleft\ set. With the results, we also report the number of
simulations in which $\Theta>1$. An examination of the probability of choosing
education $F$ clarifies that the censored set tends to bias the estimates of
$\mu$ downward: by setting $\Theta$ closer to 1, the MLE estimator of $\mu$
tends to minus infinity. The truncated set may therefore provide a more
accurate description of the true bias. But the censored set is also
informative because it documents the potential effect of replacing the true
parameters of the model with their estimates in the estimation of parameters
that are affected by both cross-sectional and time-series variation.

This issue is even more significant when the risk aversion
parameter is estimated using the misspecified model. In that case, $\Theta$
can be greater than 1 for two different reasons. First, as in the general
equilibrium model, the true parameters are replaced by their estimated
counterparts. Second, $\Theta$ is misspecified and, hence, there is no reason
to expect that it satisfies the theoretical restriction $\Theta<1$. We
therefore expect the downward bias for the misspecified model in the censored
results and the number of cases in which $\Theta>1$ to be larger than in the
general equilibrium model.

Tables \ref{Table: Montecarlo Results, Proposed Method, Appendix}
and \ref{Table: Montecarlo Results, Bias, Appendix} compare the results
obtained using the censored sample with the results obtained using the
truncated sample. There are three patterns worth highlighting. First, when the
censored sample is used, as expected, the average of the estimated risk
aversion parameter obtained employing our proposed method is always lower.
Second, with our proposed method the number of cases in which $\Theta>1$
decreases with the length of the time-series, since the persistence and the
variance of the aggregate shocks are estimated more precisely. This suggests
that it is important to employ a long time-series of aggregate data to avoid
situations in which the estimated parameters are incompatible with the
structure of the model. Lastly, as expected, when we use the misspecified
model, the number of cases in which $\Theta>1$ is much larger and the
misspecification bias goes from being positive to being negative.

\begin{table}[ptbh]
\caption{Monte Carlo Results, Parameter Estimates For Correct Model}%
\label{Table: Montecarlo Results, Proposed Method, Appendix}%
\begin{threeparttable}
\centering
\begin{tabular*}
{17.3cm}[c]{@{\extracolsep{\fill}}lcccc}
&  &  &  & \\[-0.5cm]\hline\hline
& \multicolumn{2}{c}{\emph{Censored Results}} &
\multicolumn{2}{c}{\emph{Truncated Results}}\\\hline
True Parameter & Estimate & Cov. Prob. & Estimate & N. Cases\\\hline
\multicolumn{4}{l}{\emph{Cross-sectional Sample Size: 2,500, Time-series Sample
Size: 25}} & \\\hline
\textbf{Log Risk Aversion Mean: $\mu= 0.2$} & \textbf{0.053} & \textbf{0.896} &
\textbf{0.157} & 120/5000\\\hline
\multicolumn{4}{l}{\emph{Cross-sectional Sample Size: 2,500, Time-series Sample
Size: 50}} & \\\hline
\textbf{Log Risk Aversion Mean: $\mu= 0.2$} & \textbf{0.114} & \textbf{0.918} &
\textbf{0.173} & 73/5000\\\hline
\multicolumn{4}{l}{\emph{Cross-sectional Sample Size: 2,500, Time-series Sample
Size: 100}} & \\\hline
\textbf{Log Risk Aversion Mean: $\mu= 0.2$} & \textbf{0.129} & \textbf{0.926} &
\textbf{0.177} & 59/5000\\\hline
\multicolumn{4}{l}{\emph{Cross-sectional Sample Size: 5,000, Time-series Sample
Size: 25}} & \\\hline
\textbf{Log Risk Aversion Mean: $\mu= 0.2$} & \textbf{0.070} & \textbf{0.893} &
\textbf{0.158} & 102/5000\\\hline
\multicolumn{4}{l}{\emph{Cross-sectional Sample Size: 5,000, Time-series Sample
Size: 50}} & \\\hline
\textbf{Log Risk Aversion Mean: $\mu= 0.2$} & \textbf{0.130} & \textbf{0.915} &
\textbf{0.180} & 61/5000\\\hline
\multicolumn{4}{l}{\emph{Cross-sectional Sample Size: 5,000, Time-series Sample
Size: 100}} & \\\hline
\textbf{Log Risk Aversion Mean: $\mu= 0.2$} & \textbf{0.148} & \textbf{0.930} &
\textbf{0.178} & 38/5000\\\hline
\multicolumn{4}{l}{\emph{Cross-sectional Sample Size: 10,000, Time-series
Sample Size: 25}} & \\\hline
\textbf{Log Risk Aversion Mean: $\mu= 0.2$} & \textbf{0.082} & \textbf{0.889} &
\textbf{0.166} & 95/5000\\\hline
\multicolumn{4}{l}{\emph{Cross-sectional Sample Size: 10,000, Time-series
Sample Size: 50}} & \\\hline
\textbf{Log Risk Aversion Mean: $\mu= 0.2$} & \textbf{0.139} & \textbf{0.910} &
\textbf{0.183} & 54/5000\\\hline
\multicolumn{4}{l}{\emph{Cross-sectional Sample Size: 10,000, Time-series
Sample Size: 100}} & \\\hline
\textbf{Log Risk Aversion Mean: $\mu= 0.2$} & \textbf{0.159} & \textbf{0.923} &
\textbf{0.188} & 36/5000\\\hline\hline
\end{tabular*}
\begin{tablenotes}
\scriptsize{Notes: This Table reports the Monte Carlo results for the correct model obtained using our proposed estimation method. They are derived by simulating the general equilibrium model 5000 times. The second column reports the average estimated parameter, where the average is computed over the 5000 simulations, when we use all the Monte Carlo runs and set $\Theta_{t}=0.99$ in all cases in which $\Theta_{t} \geq 1$. Column 3 reports the corresponding coverage probability of a confidence interval with 90\% nominal coverage probability. Columns 4 reports the average estimated parameter when we drop all simulations for which $\Theta_{t} \geq 1$. Column 5 reports the number of case in which $\Theta_{t} \geq 1$.}
\end{tablenotes}
\end{threeparttable}
\end{table}

\begin{table}[ptbh]
\caption{Monte Carlo Results, Parameter Estimates For Misspecified Model}%
\label{Table: Montecarlo Results, Bias, Appendix}%
\begin{threeparttable}
\centering
\begin{tabular*}
{17.3cm}[c]{@{\extracolsep{\fill}}lcccc}
&  &  &  & \\[-0.5cm]\hline\hline
& \multicolumn{2}{c}{\emph{Censored Results}} &
\multicolumn{2}{c}{\emph{Truncated Results}}\\\hline
True Parameter & Estimate & Bias & Estimate & N. Cases\\\hline
\multicolumn{4}{l}{\emph{Cross-sectional Sample Size: 2,500}} & \\\hline
\textbf{Log Risk Aversion Mean: $\mu= 0.2$} & \textbf{-0.990} & \textbf{-1.190} &
\textbf{1.163} & 1170/1000\\
\multicolumn{4}{l}{\emph{Cross-sectional Sample Size: 5,000}} & \\\hline
\textbf{Log Risk Aversion Mean: $\mu= 0.2$} & \textbf{-0.996} & \textbf{-1.196} &
\textbf{1.173} & 1180/5000\\
\multicolumn{4}{l}{\emph{Cross-sectional Sample Size: 10,000}} & \\\hline
\textbf{Log Risk Aversion Mean: $\mu= 0.2$} & \textbf{-0.997} & \textbf{-1.197} &
\textbf{1.179} & 1185/5000\\\hline\hline
\end{tabular*}
\begin{tablenotes}
\scriptsize{Notes: This Table reports the Monte Carlo results for the misspecified model obtained using only cross-sectional variation. They are derived by simulating the general equilibrium model 5000 times. The second column reports the average estimated parameter, where the average is computed over the 5000 simulations, when we use all the Monte Carlo runs and set $\Theta_{t}=0.99$ in all cases in which $\Theta_{t} \geq 1$. Column 3 reports the corresponding coverage probability of a confidence interval with 90\% nominal coverage probability. Columns 4 reports the average estimated parameter when we drop all simulations for which $\Theta_{t} \geq 1$. Column 5 reports the number of case in which $\Theta_{t} \geq 1$.}
\end{tablenotes}
\end{threeparttable}
\end{table}

\end{document}